\documentclass{aa}
\usepackage{graphicx}
\usepackage[varg]{txfonts}
\usepackage[
colorlinks,citecolor=blue,linkcolor=blue,urlcolor=blue]{hyperref}
\usepackage{amsmath}
\usepackage[usenames]{xcolor}
\usepackage{comment}
\usepackage{multirow}
\usepackage{chngpage}
\usepackage{lscape}
\usepackage{url}

\newcommand{\udef}{\stackrel{\mathrm{def}}{=}}
\usepackage{forest}
\usepackage{tikz-qtree}
\usetikzlibrary{shadows,trees}

\definecolor{Wildstrawberry}{rgb}{1.0, 0.26, 0.64}
\newcommand{\newtext}[1]{{{#1}}}

\newcommand{\kms}{{\mathrm{km\ s^{-1}}}}
\newcommand{\Msun}{{\mathrm{M}_\odot}}

\DeclareRobustCommand{\Eqref}[1]{Eq.~\ref{#1}}
\DeclareRobustCommand{\Figref}[1]{Fig.~\ref{#1}}
\DeclareRobustCommand{\Tabref}[1]{Table~\ref{#1}}
\DeclareRobustCommand{\Secref}[1]{Sec.~\ref{#1}}

\defcitealias{tauris:98}{TT98}
\defcitealias{ramirez-agudelo:15}{R-A15}

\begin{document}

\title{Massive Runaway and Walkaway Stars}
\subtitle{A study of the kinematical imprints of the physical processes \\ governing the evolution and explosion of their binary progenitors}

\author{M.~Renzo\inst{1} \and E.~Zapartas\inst{1} \and S.~E.~de~Mink\inst{1} \and
 Y.~G\"otberg\inst{1} \and S.~Justham\inst{2,3} \and
 R.~J.~Farmer\inst{1} \and R.~G.~Izzard\inst{4,5}
 \and S.~Toonen\inst{1}
 \and H.~Sana\inst{6}.} 

\institute{{Astronomical Institute Anton Pannekoek, University of
    Amsterdam, 1098 XH Amsterdam, The Netherlands}
  \and{School of Astronomy \& Space Science, University of the Chinese
    Academy of Sciences, Beijing 100012, China}
  \and{National Astronomical Observatories, Chinese Academy of
    Sciences, Beijing 100012, China}
  \and{Astrophysics Research Group, Faculty of Engineering and
    Physical Sciences, University of Surrey, Guildford, Surrey, GU27XH, United Kingdom}
    \and{Institute of Astronomy, University of Cambridge, Madingley Road, Cambridge CB30HA, UK}
  \and{Institute of Astronomy, KU Leuven, Celestijnenlaan 200 D, B-3001 Leuven, Belgium}}

\offprints{M.~Renzo, \href{mailto:m.renzo@uva.nl}{m.renzo@uva.nl}}
\date{}
\abstract{We perform an extensive numerical study of
  the evolution of massive binary systems to predict the peculiar
  velocities that stars obtain when their companion collapses and
  disrupts the system. Our aim is to (i) identify which predictions are robust against model uncertainties and assess their implications, (ii) investigate which physical processes leave a clear imprint and may therefore be constrained observationally and (iii) provide a suite of publicly available model predictions\newtext{, to allow for the use of kinematic
    constraints from the \emph{Gaia} mission}.  
We find that $22_{-8}^{+26}$\% of all massive binary systems merge
prior to the first core-collapse in the system. Of the remainder,
$86_{-9}^{+11}\%$ become unbound because of the
core-collapse. Remarkably, this rarely produce runaway stars
(\newtext{observationally defined as}
stars with velocities above $30\,\kms$). These are outnumbered by more
than an order of magnitude by slower unbound companions, or ``walkaway
stars’’. This is a robust outcome of our simulations and is due to the
reversal of the mass ratio prior to the explosion and widening of the
orbit, as we show analytically and numerically. \newtext{For stars more massive
than $15\,M_\odot$,} we estimate \newtext{that}
$10^{+5}_{-8}$\% \newtext{are} walkaways and only $0.5^{+1.0}_{-0.4}$\% \newtext{are} runaways, nearly all of which have
accreted mass from their companion. Our findings are consistent with
earlier studies, however the low runaway fraction we find is in
tension with observed fractions \newtext{of about}
10\%. \newtext{Thus, astrometric data on presently single massive stars can
  potentially constrain the physics of massive binary evolution}.
Finally, we show
that the high end of the mass distributions of runaway stars is very
sensitive to the assumed black hole natal kicks and propose this as a
potentially stringent test for the explosion mechanism. We also
discuss companions remaining bound which can evolve into X-ray
 and gravitational wave sources.}

\keywords{stars: kinematics, binaries, massive, supernovae }
\maketitle{}

\section{Introduction}
\label{sec:intro}

Stars with initial mass larger than
about $7.5\,M_\odot$ are the progenitors of black holes (BH) and
neutron stars (NS). These stars play an important role in shaping galaxies
through their radiative, chemical, mechanical feedback
(e.g., \citealt{larson:74, ceverino:09}).
Most young, unevolved, massive stars have a nearby companion with
which they form a close binary system
\citep[e.g.,][]{sana:12,
  chini:12,kobulnicky:14,almeida:17}. 
Binary systems that remain bound throughout the entire evolution of both
stars can give rise to many exotic phenomena, including X-ray binaries
\citep[e.g., ][]{gott:71,Bolton1972, Webster+1972,
  van-den-Heuvel+1972}, binary neutron stars
\citep[e.g.,][]{hulse:75,wijers:92}, gamma-ray bursts
\citep[e.g.,][]{izzard:04b,becerra:16, LVC:17:grb}, and gravitational wave events
\citep{LVC:16b, BNSmerger:17}.  
However, only a small fraction of massive stars born \newtext{in}
binary systems are expected to stay together their entire lives. The
majority of systems is disrupted by the first core collapse event,
which can separate the newly formed compact object from its former
companion star
\citep[e.g.,][]{tauris:98, belczynski:99,
  belczynski:08, eldridge:11}. 

Single NSs can be detected as pulsars or magnetars. Many of
them are observed to have large proper motions (e.g.,
\citealt{gunn:70,lyne:94,hobbs:05}, but see also \citealt{verbunt:17}).
Conversely, single BHs are only detectable under special
circumstances, for example through lensing events when passing in
front of a background star \citep[e.g.,][]{wyrzykowski:16}, or if they
accrete gas from the ambient medium
\citep[e.g.,][]{fender:13, gaggero:17}. Therefore, to probe the
population of stellar-mass black holes we are limited to X-ray or
gravitational wave observations that can only target the rare
cases that remain bound to their companion and are close enough for
the BH to accrete or merge. To learn more about the black holes that
form in less special cases, it is worth \newtext{investigating} the imprints
they may leave on their former companion star. 

The primary focus of this study is the population of unbound main sequence companions,
which can be identified observationally because of their peculiar
spatial velocities compared to the surrounding population. Their velocities can either be detected as proper
motions (i.e.,\ their motion in the plane of the sky measured directly
from the displacement of the star) or as radial velocities (i.e.,\ their
motion perpendicular to the plane of the sky, measured from the
Doppler shift of the spectral lines). Large spatial velocities have
been inferred for a significant sub-population of young massive stars
\citep[e.g.][]{blaauw:61, Cruz-Gonzalez+1974, gies:86, gies:87,
  hoogerwerf:00, hoogerwerf:01, tetzlaff:11, boubert:18}.  

\citet[][]{blaauw:61} introduced the term ``runaway stars'' for those 
in the fast tail of the velocity distribution for a given spectral type. The typical threshold
adopted to define the tail of this distribution for O and B-type stars is $v \gtrsim 30\,\mathrm{km \ s^{-1}}$
\citep[e.g.,][]{blaauw:56, gies:86, dedonder:97, hoogerwerf:00, hoogerwerf:01, dray:05, eldridge:11}, although sometimes other values
have been considered \citep[e.g., $40\,\mathrm{km\ s^{-1}}$
in][]{blaauw:61, dewit:05, boubert:18}. However, as we will argue based on simulations presented in this work, the majority of unbound companions is expected to exhibit velocities well below these thresholds.  
We will refer to these slow unbound former companions resulting from
disrupted binary systems as ``walkaway stars'' \citep[to our knowledge first coined
by][]{de-Mink+2012} to distinguish them from the faster counterparts.

\begin{figure}[tbp]
  \centering
  \includegraphics[width=0.5\textwidth]{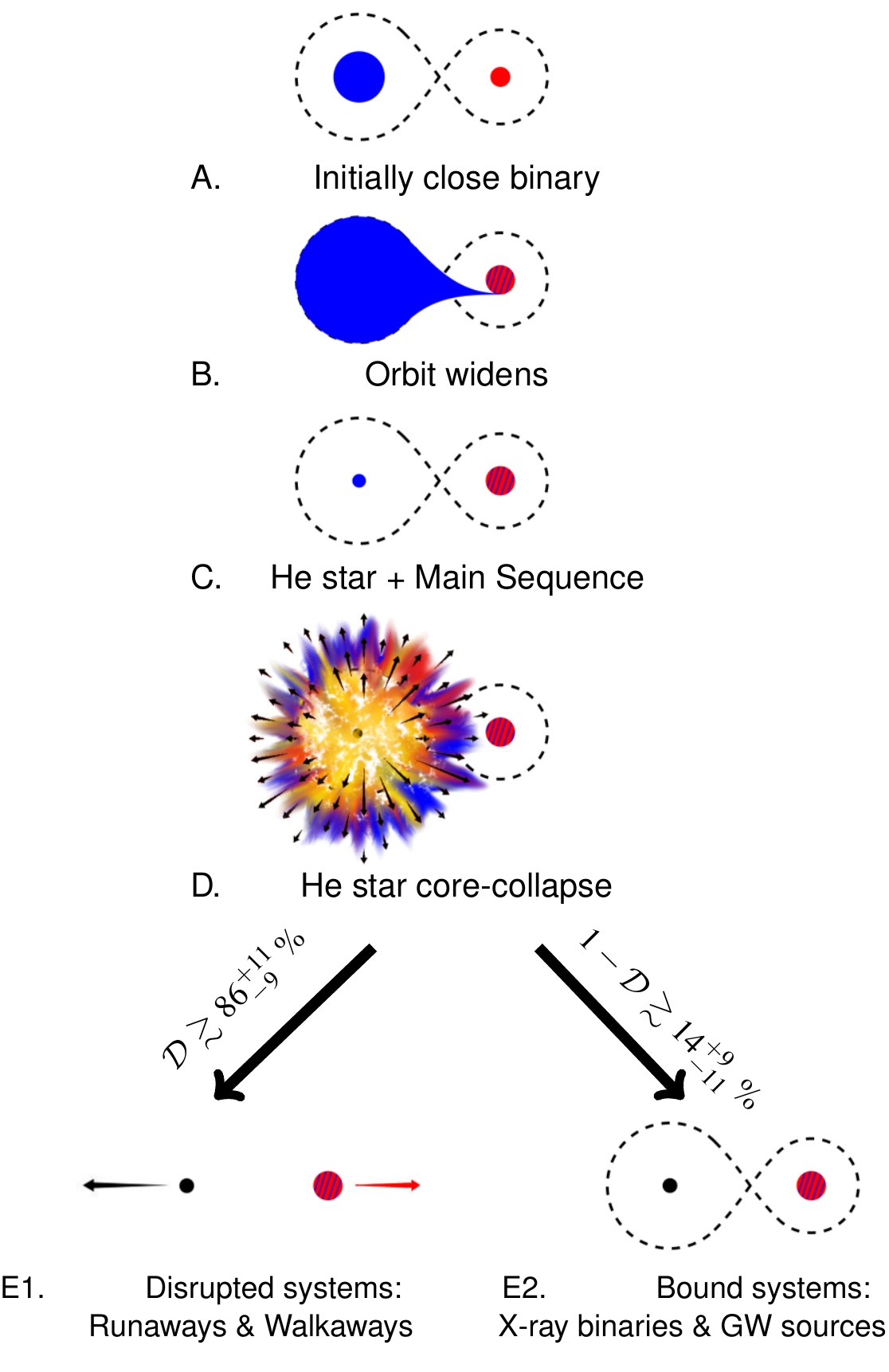}
  \caption{Schematic depiction of the evolution of a close massive
    binary through stable mass transfer. The
    evolution forks at the first core-collapse event in the system: the vast
    majority of the systems is disrupted and each produces a runaway or
    walkaway star which can travel far from its birth location. The
    systems remaining bound are possible progenitors of X-ray binaries
    and gravitational wave sources. The fraction of binaries disrupted
    $\mathcal{D}$ comes from the simulations presented in this study.
}
  \label{fig:cartoon}
\end{figure}

A proposed explanation for the large spatial velocity of runaways is that
they originate from disrupted binary systems \citep[]{zwicky:57,
  blaauw:61, boersma:61},
which naturally explains the lower number of companions they have,
compared to typical massive stars \citep[e.g.,][]{blaauw:61, gies:86, sana:14}. \Figref{fig:cartoon} sketches the
typical evolution of a \newtext{massive} binary system. Most binaries are disrupted
at the time of the first core-collapse.

An alternative mechanism to produce stars with peculiar spatial
velocities is dynamical ejection from a star cluster \citep[e.g.,][]{poveda:67, leonard:91
}. Dynamical interactions with
a supermassive BH can also disrupt a binary, but the ejection
velocities achieved in this scenario are typically much higher
\citep[$\gtrsim$$10^3\,\mathrm{km\ s^{-1}}$,][]{hills:88}.

Both mechanisms, i.e. the ``binary
disruption scenario'' and the ``dynamical ejection scenario'', are
expected to act in nature, but their relative importance is not not well
constrained \citep[e.g.,][]{hoogerwerf:00,hoogerwerf:01,guseinov:05}.
\citet{hoogerwerf:01} analyzed the properties of a sample of 56 nearby
runaway stars and 9 radio pulsars, and traced back the runaways to
their most likely parent stellar group. From the sub-sample for which a
clear identification of the parent group was possible, they estimated that the disruption of binaries is
responsible for roughly two thirds of observed runaways \citep[see
also][]{gies:86, gies:87, stone:91, hoogerwerf:00, dincel:15,
  boubert:17b}. However, this claim could not be confirmed \newtext{in
  the re-analysis of the same sample by
\cite{jilinski:10}, who found that most of the runaways in the sample
were bound spectroscopic binaries}.

Unbound stars resulting from the disruption of a binary are of
potential interest for several topical questions in
astrophysics. Their kinematics and stellar properties bear imprints of the uncertain physical processes that govern the evolution of their
binary progenitor systems, including the expected phase of mass
transfer between the two stars (phase B. in
\Figref{fig:cartoon}). Detailed knowledge of the spatial velocity of
runaway stars can improve the accuracy of wind mass-loss rate
determinations relying on their bow shocks \citep[e.g.][]{gull:79,
  kobulnicky:18}. Of particular interest is the
question \newtext{of} whether they can provide any unique constraints on the physics of
core-collapse, in particular the natal kick on the compact objects
that they produce. Such kicks are expected either from asymmetries in
the explosion and/or the neutrino emission
\citep[][]{shklovskii:70,
  wongwathanarat:13,
  janka:13,janka:17}, and they determine which systems disrupt and
eject the companion star and which systems remain bound and thus have a
potential as future X-ray and gravitational wave sources. 

These unbound stars have also been considered for their potential
importance as non-canonical sources of stellar feedback
\citep[e.g.,][]{larson:74, ceverino:09}. 
They can travel long distances and end their lives tens to hundreds of
parsecs away from their birth location. Because of their motion, their
ionizing photons are
less likely to be absorbed by their birth clouds. Therefore, in the context of the
re-ionization of the early universe, the ionizing radiation of stars
ejected from a binary are more likely to
escape and become available for ionization of hydrogen in the
interGalactic medium \citep[e.g.][]{conroy:12,kimm:14,ma:16}. Furthermore they are
expected to explode in lower density regions, which may \newtext{change} the
impact they have as sources of turbulence in the ISM \citep[e.g.][]{gatto:15}. 

The present and upcoming \emph{Gaia} data releases provide an important
observational motivation for this study \citep[e.g.,][]{perryman:01}. The determination of
distances, proper motions and radial velocities by the \emph{Gaia} satellite
is expected to drastically increase the available sample of massive
stars with precisely known velocities
\citep{
  Gaia-Collaboration+2016, Gaia-Collaboration+2016a}.

In this study, we present a systematic theoretical study of the
kinematical signatures of stars ejected from \newtext{massive} binary systems and how
these depend on the uncertain physical processes governing binary
stellar evolution. For this purpose we employ a rapid binary
population synthesis code, which we have updated to account for the main relevant processes that affect the disruption of the binary systems (see \Secref{sec:methods}).  

We first provide insight in our simulations by presenting the results
for an individual system in \Secref{sec:example}.
We present in \Secref{sec:analytics} an analytic estimate of the
typical 
velocity of ejected companions expected from the
disruption of binaries. Then, in \Secref{sec:fiducial_res} we describe the results of a fiducial simulation of a
full population of binary stars, for which we adopt realistic
input assumptions provided by detailed simulations and/or observations
when available.  In addition, we present in
\Secref{sec:param_variation} an extensive grid of models where we vary the most relevant uncertain assumptions. This allows us to study (i) which predictions are robust against the model uncertainties and (ii) which uncertain physical processes leave a clear imprint on the observables that can be used to test and constrain them. 

Various earlier studies have discussed the evolution and
interaction of populations of massive binary systems. Examples
include, but are not limited to, \citet[][]{vanbeveren:82, dedonder:97, Fryer+1997, Fryer+2001, fryer:12,
  Belczynski+2012b, repetto:12, Fragos+2013a, Grudzinska+2015,
  boubert:18}. Most
of these studies focus on the minor fraction of systems
that remain bound and are the progenitors of X-ray binaries and/or
gravitational wave sources. A few studies focus specifically on
disrupted systems and unbound stars. Examples are
\cite{vanrensbergen:96,dedonder:97, dray:05, eldridge:11, bray:16, boubert:17b,
  zapartas:17b, boubert:18}. We expand on these studies by using updated physical
assumptions, and by focusing on which physical processes could be
constrained using the Galactic population of unbound stars. A
comparison to previously published results is given in Appendix~\ref{sec:literature}. 

Our first main result is that the majority of disrupted binaries
ejects a slow-moving walkaway star ($v < 30\,\mathrm{km\ s^{-1}}$), rather than a faster runaway. This result is robust against
variations of uncertain parameters in the model, as we discuss in \Secref{sec:param_variation}.
This would imply that the more easily detected runaway stars only
reveal a small subset of the population of unbound stars that are
former companions of disrupted binary systems.  We discuss the implications in \Secref{sec:discussion}.

Our second main result is that the kinematic properties and absolute
number of unbound companions depend sensitively on assumptions
concerning the  natal kicks of BHs. We show the imprints this leaves on
the mass distribution and discuss whether future observations can be
used to constrain these processes.

Section~\ref{sec:bound} briefly describes the population of systems remaining bound after the first
core-collapse.

We also find that the fraction of runaways among massive stars
predicted by our simulations is much lower than the observed value in
all our parameter variations. This finding agrees with the results of
\cite{eldridge:11}, but is in potential contrast
with the observational results of \cite{hoogerwerf:01}, and needs further
investigation.

Finally, we discuss how our results \newtext{could be used} to
include the effect of unbound binaries in models for stellar
feedback. We will provide our numerical results upon publication. 

\section{Binary population synthesis calculations}
\label{sec:methods}

We carry out population synthesis calculations of isolated binaries with the rapid binary evolution code \texttt{binary\_c}
\citep[][]{izzard:04,izzard:06,izzard:09,demink:13,schneider:15,izzard:18}. This
code is based on the algorithms by \cite{tout:97,hurley:00,hurley:02}, which rely on the analytic fits to the
single stellar evolution models from \cite{pols:98}.

First, we compute a fiducial population using observationally
\newtext{favored} assumptions for the free parameters that describe the initial
conditions and physical assumptions.
Then, we check the robustness of our results (or
equivalently, to which assumptions they are most sensitive) by varying
the free parameters that are most relevant for the
velocity distribution of disrupted binaries. To limit
computational costs, we explore variations in each parameter
one-by-one while keeping the other parameters fixed to our
fiducial choices, following the approach of
\cite{Fragos+2013a,demink:13,demink:14,zapartas:17,zapartas:17b,
  belczynski:17}. Effectively, \newtext{\texttt{binary\_c} treats}
each parameter as independent from the others, therefore this approach does not account for possible
correlations between either initial distributions \citep[e.g., the one between initial period and mass ratio
suggested by][]{moe:17} or uncertain
physical processes (e.g., core spin and natal kick amplitude). Varying
two (or more) parameters simultaneously, or changing the algorithmic
representation of the uncertain physical processes, might possibly result in
larger variations than those we present in
\Secref{sec:param_variation}.

\subsection{Initial distributions and parameters}

Each binary system in our calculations is characterized by a zero age
main sequence (ZAMS)
mass for the primary\footnote{Throughout this study, we define the primary star
  to be the initially more massive star, even
  if it becomes the less massive star in the binary during the evolution.}
$M_1^\mathrm{ZAMS}$, initial mass ratio $q\udef M_2/M_1$, and initial orbital
period $P^\mathrm{ZAMS}$. \newtext{\texttt{binary\_c} builds} a grid in this parameter space and
weighs each system according to the initial distributions
described below. \newtext{We present our results in terms of
  probability ``per binary system''. This can be converted in a
  probability per unit stellar mass by dividing it by the mean mass of
a binary system in the population ($0.42\,M_\odot$ for our fiducial
assumptions, assuming there are no binaries for $M_1^\mathrm{ZAMS}\leq2\,M_\odot$, see below).}

We select $N_{M_1}$ primary stars with ZAMS mass $M_1^\mathrm{ZAMS}$
at logarithmically spaced intervals in the range $7.5\,M_\odot\leq M_1^\mathrm{ZAMS}
\leq 100\,M_\odot$. We weigh each primary star with an initial mass function (IMF) with
slope $\alpha=-2.3$ 
\citep[][]{kroupa:01}. In our model
variations, we explore values of $\alpha$ of -1.9
\citep[][]{schneider:18} and -3.

For each primary star of mass $M_1^\mathrm{ZAMS}$, we select $N_q$ secondaries with mass
$M_2^\mathrm{ZAMS} = qM_1^\mathrm{ZAMS}$, taken at regular intervals
in mass ratio $q$ between 0.1 and 1 assuming a
flat distribution \citep[slope $\kappa=0$, e.g.,][]{kouwenhoven:05}. We consider in
\Secref{sec:param_variation} also variations with $\kappa=\pm1$.

Finally, for each pair of masses $(M_1^\mathrm{ZAMS},M_2^\mathrm{ZAMS})$ we choose $N_p$ different periods
$P^\mathrm{ZAMS}$ equally spaced in $\log_{10}(P^\mathrm{ZAMS}/\mathrm{days})$
between 0.15 and 5.5. We weigh the birth probability of each binary system with a
mass-dependent distribution for the initial orbital period: if $M_1<15\,M_\odot$ we assume
a flat distribution in $\log_{10}(P^\mathrm{ZAMS}/\mathrm{days})$
\citep[][]{opik:24,kobulnicky:07}; while for $M_1\geq15\,M_\odot$ we assume a
powerlaw distribution in $\log_{10}(P^\mathrm{ZAMS}/\mathrm{days})$ with exponent
$\pi=-0.55$ \citep[][]{sana:12}. We explore values of $\pi=0\,$and\,$\pi=-1$
(for all values of $M_1^\mathrm{ZAMS}$)
in our model variations. We chose our upper-limit on the initial period, $10^{5.5}$\,days, to include
wide systems in which the stars effectively evolve as single stars \citep[][]{demink:15}.

To limit the dimensions of the parameter space, we assume
all orbits to be initially circular (i.e., eccentricity
$e=0$). Because we focus on the population of stars resulting from the
disruption of post-interaction binaries this assumption is not
critical \citep[][]{demink:15}: orbits are expected to
circularize
because of tides before or during mass transfer
(\citealt{belczynski:99,hurley:02}, but see also \citealt{eldridge:09}
for arguments against circularization of the orbit in post-main
sequence mass transfer). 

In our fiducial model, we assume a canonical metallicity (i.e., mass fraction of
elements heavier than helium) $Z=0.02$. This value is slightly above
the most recent determination for solar neighborhood
\cite{asplund:09}, but it should generally describe young population
of massive stars in the Milky-Way. The models from \cite{pols:98}, which serve as
input for our computations, adopt the
isotopic mixture of \cite{anders:89}. In
\Secref{sec:param_variation}, we also consider $Z=0.0002$, $0.0047$, $0.008$,
and $0.03$.

We initialize the stellar rotation rate with a mass-dependent
equatorial velocity according to \cite{hurley:00}, assuming
alignment between the stellar spins and the orbital angular momentum. In one of our
parameter variations, we draw the initial spin velocity randomly from
the distribution given by \citealt{ramirez-agudelo:15} (R-A15, see last
line in \Tabref{tab:parameters}), but maintain the assumption of
alignment. However, changing the initial rotation rate has almost no
impact on our results because binary interaction processes overwrite it \citep[][]{demink:13}.  

We build a grid in the initial parameter space, and initialize a
binary system in each cell of such grid. 
Each system is assigned a ``probability''
which can then be multiplied by the
amount of time spent by the system in that cell to obtain a quantity comparable to
observed number counts. See Appendix~\ref{app:data} for more details.

For each set of parameters, the total size of our model grid is
$N_{M_1}\times N_q \times N_p = 50 \times 50 \times 100 = 250\,000$ 
binary systems\footnote{This number is then multiplied by the number
  of natal kicks that we draw for each core-collapse, see \Secref{sec:SN_kick}.}.  Increasing this resolution to
\newtext{$N_{M_1}\times N_q \times N_p = 100 \times 100 \times 200$} introduced no
significant variation in our results.

\subsection{Physical assumptions}

We follow the evolution of each binary system until the
first core-collapse (CC) event or until they merge. 
Stellar winds can have a dramatic effect on the evolution of single
stars \citep[e.g.,][]{renzo:17} and impact significantly the orbital
evolution in a binary. We include wind mass loss in our single star models as implemented by
\cite{demink:13}, i.e.\ we use a combination of mass loss rates from
\cite{vink:00,vink:01} for hot hydrogen-rich stars,
\cite{nieuwenhuijzen:90} for cool stars, 
 including a ``luminous blue variable'' enhancement as in
 \citealt{hurley:00}, and from \cite{hamann:98} reduced by a factor of
 10 for
 Wolf-Rayet stars. We also include the wind mass loss
 enhancement for fast rotating stars following \cite{maeder:00}.
 
We include the effects
of tides on the spins and orbital angular momentum using the
algorithm from \cite{hurley:02} based on the calculations of
\cite{zahn:77} and \cite{hut:81}. 

When the radius of \newtext{one} star exceeds the Roche radius,
(calculated using \citealt{eggleton:83} fitting formula)
we use the algorithm of \cite{claeys:14} to determine the
mass transfer rate (see their Eq.~10).

\newtext{In our fiducial simulation, we assume a variable mass
  transfer efficiency.} We limit the accretion rate of the
accretor to 10 times its Kelvin-Helmholtz timescale, i.e., we assume a
mass-transfer efficiency $\beta_\mathrm{RLOF}
=\beta_\mathrm{th}$. Larger rates are
likely to drive the accretor out of thermal equilibrium\footnote{We do
not model the internal structure of the stars, therefore, we do not
follow its possible bloating during mass transfer, which could
potentially enhance its mass loss and related spin down.} and lead to
unstable mass transfer \citep[][]{neo:77,hurley:02}. For a
large portion of the parameter space, $\beta_\mathrm{th}$ results in a
rather conservative mass transfer \citep[][]{schneider:15}. Since the
efficiency of mass transfer is a major uncertainty in binary evolution
\citep[][]{demink:07}, we also consider parameter
variations with $\beta=0,\,0.5,$ and 1, respectively, \newtext{which
  bracket the range of physical possibilities}. 
 
Matter that is not accreted is assumed to leave the system with
the specific angular momentum of the orbit of the accretor
\citep[e.g.,][]{soberman:97,vandenheuvel:17}, $h = \gamma_\mathrm{RLOF}
J_\mathrm{orb}/(M_\mathrm{don}+M_\mathrm{acc})$,
with $\gamma_\mathrm{RLOF}=M_\mathrm{don}/M_\mathrm{acc}$, where
\newtext{$J_\mathrm{orb}$ is the orbital angular momentum, $M_\mathrm{don}$ and
$M_\mathrm{acc}$  the donor and accretor masses, respectively}. In our
model variations, we also explore a scenario where the mass that is not
accreted leaves the binary from the outer Lagrangian point L2 and is assumed to form a circumbinary disk \citep[i.e.,
$\gamma_\mathrm{RLOF}=\sqrt{2}(M_\mathrm{don}+M_\mathrm{acc})^2/(M_\mathrm{acc}M_\mathrm{don})\equiv
\gamma_\mathrm{disk}$,][]{artymowicz:94}. We further consider the
assumptions that the mass that is not accreted leaves the system with the
orbital specific angular momentum, i.e.~$\gamma=1$, which is the standard model of \cite{belczynski:08, dominik:12, dominik:13}.

If, by the time the first star fills its Roche lobe, the mass
ratio of the accretor to the donor is smaller than a certain threshold
($M_\mathrm{acc}/M_\mathrm{don}<q_\mathrm{crit}$), we assume that the
system enters a common envelope phase. The threshold value
$q_\mathrm{crit}$ is uncertain and
depends on the evolutionary stage of the donor. In our fiducial simulation $q_\mathrm{crit,A}=0.65$ for a
main sequence (MS) donor \citep[][]{demink:07}, $q_\mathrm{crit,B}=0.4$ for a Hertzsprung gap
donor \citep[][]{hurley:02}, \newtext{and $q_\mathrm{crit,RSG}=0.25$ for
core-helium burning and red supergiant donors \citep[][]{claeys:14}}.
We also consider values of $q_\mathrm{crit,A}=0.25,\,0.8$,
$q_\mathrm{crit,B}=1,\,0.5,\,0$, \newtext{and $q_\mathrm{crit,RSG}=1.0$}
in our model variations.

We treat common envelope evolution using the $\alpha_{\rm CE}
\lambda$-formalism \citep[][]{webbink:84, livio:88, dekool:90, hurley:02}. In all
variations, we use an analytic fit to the $\lambda_{g}$ values of
\cite{dewi:00} for the binding energy parameter $\lambda$. Those values do not include the energy stored in thermal
motions and the ionization state of the material within the
envelope. It is unknown what fraction of that internal energy is
useful in unbinding the envelope \citep[see, e.g.,][and references
therein]{dewi:00, ivanova:13}.  We assume $\alpha_{\rm CE} = 1$ for
our fiducial simulation, i.e., perfectly efficient use of the
liberated gravitational potential energy from the orbit but without
additional energy sources.  We explore variations with  $\alpha_{\rm
  CE} = 0.1$, i.e., inefficient use of the orbital energy to eject the
common envelope, and $\alpha_{\rm CE} = 10$, intended as indicative of
a fairly extreme case of additional energy input. \newtext{We also
  test the combination of $\alpha_\mathrm{CE}=0.1$ and $10$ with
  $q_\mathrm{crit, RSG}=1.0$, to test the influence of the efficient
  use of energy in common envelope ejection in cases when a common
  envelope is more likely.} For
binaries that interact while neither star has a well-defined core-envelope structure, such as
MS stars and helium-MS stars, we assume that the common envelope
phase leads to a merger as described in \cite{demink:13,
  schneider:15}.

\subsubsection{Natal kick}
\label{sec:SN_kick}

When a CC event happens in a binary system three different physical
ingredients contribute to the possible disruption of the system and
the ejection of the companion star.

\begin{enumerate}
\item[i.] The orbit is modified by the sudden change in gravitational potential, because
of the mass lost through ejecta \citep[the so-called ``Blaauw
kick'',][]{zwicky:57,blaauw:61, boersma:61}. However, this is not the
dominant effect, because the typical evolution of a massive close binary
system involves mass transfer through Roche lobe overflow (RLOF) before the first CC (cf.~\Figref{fig:cartoon}). RLOF
removes the envelope of the initially more massive star, limiting
the amount of mass that can be ejected by the SN. The
``Blaauw kick'' alone rarely unbinds the system \citep[][]{huang:63,tutukov:73,leonard:94}.
\item[ii.] The SN shock can interact with the companion star.
  \citep[e.g.,][]{wheeler:75,liu:15, rimoldi:16, hirai:18}. The shock can dynamically
  remove mass from the companion (stripping), heat the envelope of the
  secondary, thus enhancing its own mass
  loss (ablation), and deposit mass and momentum on the secondary. The injection of energy can inflate the companion and make
  it look redder for a duration comparable to its thermal timescale. Our
  treatment of accretion only changes the mass of the companion, without
  checking for structural readjustments of the star. However, the
  interaction between the ejecta and the secondary is typically a small effect \citep[e.g.,][]{liu:15}.
\item[iii.]  The natal kick of the compact object changes 
  its kinetic energy and momentum. This kick is caused by asymmetries
  in the SN ejecta and/or neutrino flux at the explosion, possibly
  seeded by the late core and shell burning phases of stellar
  evolution \citep[see e.g.,][]{
    wongwathanarat:13,
    holland-ashford:17,grefenstette:17,katsuda:18}. It
is typically parametrized using a kick velocity $\mathbf{v}_k$, drawn
from a distribution of amplitude $v_k\equiv |\mathbf{v}_k|$ and directions (see below). The natal
kick is the dominant reason for the disruption of binaries.
\end{enumerate}

We have updated the treatment of binary disruptions by a CC event in
\texttt{binary\_c}, following \citealt{tauris:98} (TT98 hereafter). To re-calculate the post-CC
orbital parameters, their algorithm
assumes instantaneous
loss of the ejecta, because the ejecta velocity $v_\mathrm{ej}$ is much larger than the
orbital velocity $v_\mathrm{orb}$ ($v_\mathrm{orb} \lesssim 10^2\,\mathrm{km \
  s^{-1}} \ll v_\mathrm{ej}\sim10^4\,\mathrm{km \
  s^{-1}}$). This algorithm considers the motion of the compact
object within the pre-CC orbit, and the interaction between
the ejecta and the companion star. 
We set the
relative change of mass 
of the secondary due to stripping, ablation, and
accretion of SN ejecta following a fit to the three-dimensional
hydrodynamical simulations
of \cite{liu:15} for the impact of the ejecta on a
$M_2=3.5\,M_\odot$ star, which is the most massive companion they
considered. We also assume an efficiency of momentum
transfer\footnote{$\eta=0.5$ in the algorithm from \citetalias{tauris:98}.} from the
ejecta to the companion star of 0.5 (\citetalias{tauris:98}). The contribution of 
the interactions between the SN ejecta and the secondary star to the total natal kick is
typically small, $\lesssim10\,\mathrm{km \ s^{-1}}$
(\citetalias{tauris:98}; \citealt{liu:15}; \citealt{hirai:18}).

In our fiducial run, the kick direction is assumed to be isotropically
distributed in the frame of the collapsing star \citep[][]{wongwathanarat:13,
  bear:17}, although  \cite{johnston:05} and \cite{kaplan:08} suggest that there
is weak evidence for kick-spin alignment in well observed pulsars
such as the Crab or Vela. \cite{johnston:05} also underlined that the
direction of the pulsar spin does not necessarily match the direction
of the pre-CC spin of the stellar core, if significant torques
develop during the CC \citep[e.g.][]{kazeroni:16}. In our model
variations, we also consider kicks constrained in a cone with an
opening angle of
$\alpha=10\,\mathrm{degrees}$ oriented along the spin of the exploding
star (which we assume to be perpendicular to the orbital plane), and
kicks constrained at angles from the orbital plane $90 - \alpha
\leq 45\,\mathrm{degrees}$. The kick amplitude $|\mathbf{v}_k| \equiv
v_k$ is drawn from a Maxwellian distribution with one-dimensional root
mean squared dispersion $\sigma_\mathrm{kick}=265\,\mathrm{km \  s^{-1}}$ \citep[][]{hobbs:05}. Such
distribution is motivated by the observation of the \newtext{proper motions} of pulsars in
\cite{hobbs:05} (see also \citealt{lyne:94}). 
We also compute
populations with $\sigma_\mathrm{kick}=0,\,300,\,1000\,\mathrm{km\
  s^{-1}}$ in our model variations.

The value of $v_k$ drawn is then reduced to consider the amount of matter that 
falls back after the successful launch of the SN shock,
i.e.~$v_k \rightarrow v_k(1-f_b)$, were $f_b$ is the fallback
fraction taken from\footnote{We correct for a missing parenthesis in
  their equations, which can be found by dimensional analysis \citep[][]{belczynski:17}.} Eq.~16 in \cite{fryer:12}, corresponding to their
``rapid SN engine'' which reproduces the NS-BH mass gap between
$\sim2-5\,M_\odot$ (e.g., \citealt{farr:11}, however see also \citealt{wyrzykowski:16}
regarding the existence of this mass gap). 
The inclusion of fallback
also determines the mass of the compact remnant obtained after each
CC event, and we set the mass boundary between NSs and BHs to
$2.5\,M_\odot$.

The fallback fraction is highly uncertain in BH formation. The algorithm we use
here assumes a large amount of fallback, implying close to zero
BH kick amplitudes, sometimes referred to as a ``BH momentum kick'' \citep[][]{belczynski:08,stevenson:17}. Whether
this is realistic is subject of debate in the literature. Evidence for
non-zero BH natal kicks comes from the observed
Galactic latitude of BH X-ray binaries
(e.g., \citealt{fragos:09,repetto:12,repetto:15,repetto:17}, although
see also \citealt{mandel:16}), the
possibility of retrograde BH spin \citep[e.g.,][]{morningstar:14}, eccentric orbits in these systems
\citep[e.g.][]{remillard:06}, the
small number of Wolf-Rayet star-BH binaries \citep[e.g.,][]{dray:05} and the gravitational wave constraints on
BH spins \citep[e.g.,][]{oshaughnessy:17,wysocki:18}. Recently,
multi-dimensional calculations of fallback in CC resulting in the
formation of a BH \newtext{found} that large natal kicks might be possible \citep[][]{chan:18}. Therefore, we
also consider a model without the rescaling of the kick amplitude, but still
including fallback for the remnant mass calculation. This variation is
effectively equivalent to a so-called ``velocity kick'' for the
BHs. To test an ``intermediate'' BH kick, we run a simulation using
$\sigma_\mathrm{kick}=100\,\mathrm{km\ s^{-1}}$ and no fallback
down scaling of the kick amplitude for
the CC resulting in BH formation, and ou\newtext{r} fiducial kick
($\sigma_\mathrm{kick}=265\,\mathrm{km\ s^{-1}}$, including fallback scaling)
for CC resulting in NS formation.

Another debated issue is whether the CC of low-mass iron
(or oxygen-neon-magnesium) cores produces small natal
kicks, possibly resulting in a bi-modal kick distribution
\citep[][]{katz:75, arzoumanian:02, pfahl:02,
  podsiadlowski:04, 
  knigge:11, 
  beniamini:16, tauris:17,
  verbunt:17, verbunt:17b}. To
explore the possibility of low kicks for low mass collapsing cores, we
run one model variation reducing the natal kick for NSs less massive
than $1.35\,M_\odot$ \citep[][]{schwab:10,knigge:11} drawing
the natal kicks for these NSs from a Maxwellian with
$\sigma_\mathrm{kick}=30\,\mathrm{km\ s^{-1}}$ creating a
double peaked distribution. For more massive NSs and BHs the kick is drawn from a
Maxwellian with $\sigma_\mathrm{kick}=265\,\mathrm{km\ s^{-1}}$.

In all our simulations, we draw $N_\mathrm{kick}=20$ intrinsic natal
kick directions and amplitudes for each CC in our population, effectively increasing the size of our sample by this
factor. Convergence tests showed no significant variations
of our results increasing $N_\mathrm{kick}$ to 50.

\subsection{Caveats}
\label{sec:caveats}
In addition to the limitations discussed  earlier in this section there are a few more caveats that should be kept in mind when comparing our simulations with observations (aside from the possible effects of biases). 

We only consider runaways and walkaways resulting from the disruption of binary systems at the first stellar collapse. Observational samples of runaway stars also include another sub-population, coming from cluster ejection \citep[e.g.,][]{poveda:67, leonard:91
}.  Similarly, walkaway stars produced by the disruption of binaries will add to slow-moving stars generated by the dissolution of clusters which did not go through binary interactions, see e.g.~\cite{allison:12}. In \Secref{sec:discussion}, we consider how spectroscopic measurements may allow to disentangle binary products from the outcome of purely dynamical processes.
   
All the calculations presented here implicitly assume a constant star
formation history (SFH) for a duration longer than the longest stellar
lifetime of interest.  However, a recent change in the SFH of a
region will affect the ratio of runaways and walkaway stars over the
number of normal stars. For example, a recent increase in the SFH leads to an increase of systems that did not yet have time to interact. For a binary to be disrupted, a CC event is needed, and it can only happen after the lifetime of the primary star is over. This  introduces a delay corresponding to at least the lifetime of the most massive stars \citep[$\sim$3\,Myr,][]{zapartas:17}, between a SFH event and the first binary disruption.  A recent increase in the SFH can therefore decrease the fraction of runaway and walkaway stars in a population. 

Finally we warn \newtext{about} the possible effects of stochasticity
\citep[][]{justham:12, eldridge:12}.  We have aimed to simulate
sufficient binary systems such that our results are converged, but it
should  be kept in mind that the fastest runaways come from relatively
rare channels. In a relatively low mass stellar population one will not fully sample these rare events and stochastic effects can be large.

\section{Example of the evolution of a binary system}
\label{sec:example}

\begin{figure*}[!htbp]
  \centering  
\includegraphics[width=\textwidth]{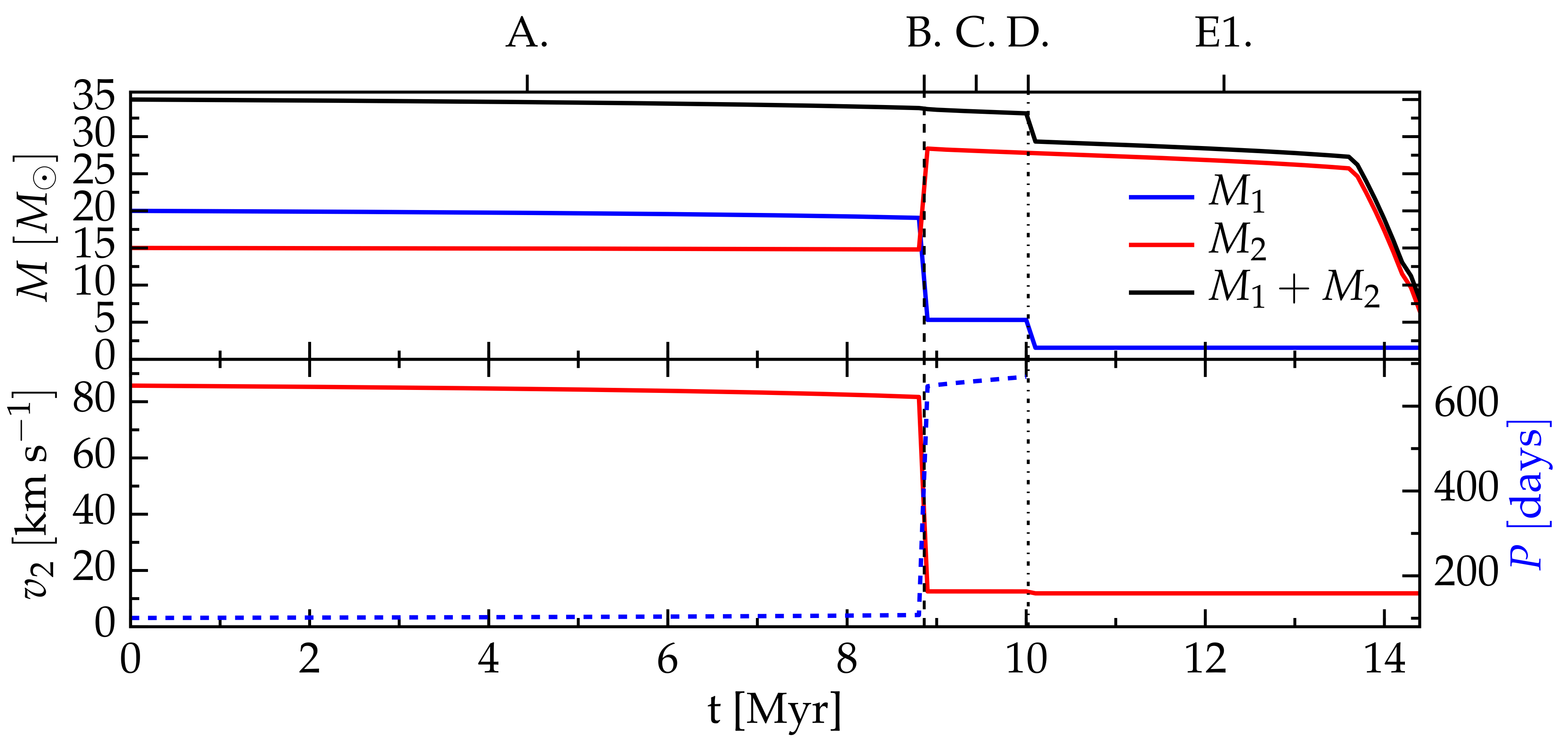}  
  \caption{Evolution of an example binary system with 
    $M_1^\mathrm{ZAMS}=20\,M_\odot$, $M_2^\mathrm{ZAMS}=15\,M_\odot$
    with $P^\mathrm{ZAMS}=100\,\mathrm{days}$, described in
    \Secref{sec:example} (see also the cartoon in \Figref{fig:cartoon}). The top panel shows the
    primary, secondary, and total mass. The bottom panel shows the
    orbital velocity of the secondary (left
    y-axis) and period of the binary (right y-axis). The labels
    indicate the following phases: A. the MS of the primary, B. case B
    RLOF, emphasized by the vertical dashed line, C. the binary is
    made of a He star (possibly looking like a Wolf-Rayet \newtext{star}) and a
    rejuvenated MS secondary; D. (and the vertical dot-dashed line)
    marks the explosion of the primary in a SNIb/Ic which unbinds the
    NS and its companion. During the phase E1. the secondary is a slow
    moving walkaway star, traveling for a duration of about half the
    main sequence lifetime of the primary.}
  \label{fig:example}
\end{figure*}

We first describe the typical evolution of a binary system that
produces an unbound companion as a result of the disruption of the
system by the first CC event. The aim is to provide the reader with
some insight in\newtext{to} our simulations. As a representative example, we consider
a  $M_1^\mathrm{ZAMS}=20\,M_\odot$ primary star,
a $M_2^\mathrm{ZAMS}=15\,M_\odot$ secondary with
$P^\mathrm{ZAMS}=100$\,days (corresponding to a separation
$a_\mathrm{ZAMS}\simeq300\,R_\odot$). \Figref{fig:example} shows the
evolution of the masses, orbital period, and orbital velocity of the
secondary (and ejection \newtext{velocity},
after the CC of the primary) 
as a function of time. The labels on the top axis correspond to the phases depicted in the cartoon shown in \Figref{fig:cartoon}.  

\subsection{Binary evolution until first CC}
\label{sec:example_evol}

Our example system remains detached during the MS evolution
of the primary (phase A.~in \Figref{fig:example}). Because of wind mass loss, the period increases by
about $7\%$.

After $\sim$8.8\,Myr, the primary leaves
the MS, and starts expanding on a thermal timescale as it burns hydrogen
in a shell. Shortly after, at the vertical dashed line B.~in \Figref{fig:example}, the primary fills its Roche lobe
initiating case B
RLOF \citep[][]{kippenhahn:67}.
For a system which has initial mass ratio close to one
($q_\mathrm{ZAMS}=0.75$ in this example),
RLOF is almost conservative in our calculations, i.e.~nearly all the
mass lost by the primary is accreted by the companion (cf.~\citealt{schneider:15}). The
secondary star accretes  $\Delta M_2\simeq14\,M_\odot$, and becomes the
more massive star, now with a mass of about $28\,M_\odot$ (red line in the top
panel of \Figref{fig:example}). It is also spun up to
near to breakup rotation \citep[][]{packet:81,demink:09}, and it is rejuvenated because of the increased
mass and consequent growth of its convective core
\citep[][]{hellings:83,schneider:16}. The primary  becomes
a $\sim$$5.3\,M_\odot$ helium star \citep[][]{gotberg:17, gotberg:18}, and
the orbits widens during mass transfer, reaching a period
longer than 600\,days (phase C.~in \Figref{fig:example}). 

After about $10$\,Myr, the primary reaches CC (dot-dashed vertical line
D.~in \Figref{fig:example}). Just before the core-collapse
of the primary, the orbit is circular with a period of
 about 700\,days. The secondary has a mass of about $28\,M_\odot$ and a pre-SN orbital velocity of
about $v_2 = (M_1/(M_1+M_2))v_\mathrm{orb} \simeq 12.5\,\mathrm{km\ s^{-1}}$.

If the system is disrupted at the first CC event, the final spatial velocity of the
secondary is nearly equal to its pre-explosion orbital velocity
\citep[e.g.,][]{blaauw:61,eldridge:11}. For comparison, similar
initial masses but a shorter initial period of $P^\mathrm{ZAMS}=7$\,days would experience a very similar
evolutionary path, but it would only widen to a pre-CC period of
$\sim$$45$\,days, resulting in a pre-CC orbital velocity of the
secondary slightly larger than $30\,\mathrm{km\  s^{-1}}$.

Because of the rejuvenating effect of
mass accretion, the secondary remains on the MS for another $\sim$$3.6\,\mathrm{Myr}$ after
the CC of the primary.

\subsection{Effects of the natal kick on the post-CC orbit}
\label{sec:example_kick}

In our example system, the CC of
the primary results in a stripped-envelope SN of type\footnote{We do not attempt to distinguish between stripped SN types (IIb, Ib or Ic depending on
the presence or lack of He lines in the spectrum).} \newtext{IIb/}Ib/Ic 
\citep[][]{filippenko:97}, which forms a NS of about
$M_\mathrm{NS}=1.6\,M_\odot$, with a fallback fraction $f_b=0.13$. The SN ejecta mass is $\Delta M_\mathrm{SN} = M_1^\mathrm{pre-CC}-M_\mathrm{NS}
\simeq 3.7\,M_\odot$, where $M_1^\mathrm{pre-CC}$ is the primary mass just
before the CC, and $M_\mathrm{NS}$ is the mass of the resulting
NS. Because the ejecta mass is less than half of the total pre-CC mass
of the system, $M_1^\mathrm{pre-CC}+M_2^\mathrm{pre-CC}\simeq33\,M_\odot$,
the loss of the SN ejecta alone is insufficient to
unbind the system \citep[][]{blaauw:61}. Whether the system is
disrupted depends on the amplitude and direction of the natal kick
imparted to the newly formed NS.

We distinguish three cases: (i) a large kick with $v_\mathrm{k}\gg v_\mathrm{orb}$ results in disruption of the binary regardless of the
direction, (ii) an intermediate kick $v_\mathrm{k} \simeq v_\mathrm{orb}$ disrupts the system only if the kick
orientation is favorable, while (iii) a small kick, $v_\mathrm{k}
\ll v_\mathrm{orb}$, never unbinds the binary, although it
can still modify its orbit (e.g., \citealt{brandt:95, kalogera:96}, \citetalias{tauris:98}).

To illustrate each case, we consider here three different fixed natal kick
amplitudes $v_k\simeq 870, 230$, and $43\,\mathrm{km\ s^{-1}}$ (respectively
corresponding to $1000,265$, and $50\,\mathrm{km \ s^{-1}}$ before the
rescaling of the kick amplitude with the fallback fraction). For comparison, the pre-CC orbital
velocity in our example system is $v_\mathrm{orb}^\mathrm{pre-CC}\simeq
100\,\mathrm{km\ s^{-1}}$ (corresponding to
$v_2^\mathrm{pre-CC}\simeq12.5\,\mathrm{km\ s^{-1}}$). For each kick amplitude $v_k$, we draw 10\,000 different directions,
isotropically distributed in the frame of the exploding star \citep[e.g.,][]{wongwathanarat:13}.

For large natal kicks, our example system is disrupted in 99.9\% of
all the directions drawn. In the remaining 0.1\%, the
compact object is shot in\newtext{to} the envelope of the secondary (see
\Secref{sec:bound} for a brief discussion).
In the case of disruption, the ejected star acquires a velocity
$v_\mathrm{dis}\simeq v_2^\mathrm{pre-SN}=12.5\,\mathrm{km\
  s^{-1}}$. With $v_\mathrm{dis} < 30\,\mathrm{km\
  s^{-1}}$, it is below the threshold to
be \newtext{observationally} classified as runaway, 
and thus it is
an example of what we refer to as a walkaway star.

In the case of intermediate kick amplitudes (here we use
$230\,\mathrm{km\ s^{-1}}$), the angle $\theta$ between the kick direction and the
pre-CC orbital velocity of the primary, given by $\cos\theta = \mathbf{v_k}\cdot\mathbf{v}_1^\mathrm{pre-SN}/(v_k
\ v_1^\mathrm{pre-SN})$, determines whether the kick disrupts the
system. Kicks that are roughly aligned with the pre-CC orbital
velocity, i.e.\ $0 \lesssim \theta\lesssim\pi/4$, successfully disrupt the
binary. Conversely, kicks that are roughly oriented opposite to the
pre-CC orbital velocity of the exploding star just tend to slow the
orbital motion: the system will remain bound on an eccentric
orbit. In our example, we find that about $84\%$ of the kick
directions result in disruptions and produce a walkaway
star with $v_\mathrm{dis}\simeq v_2^\mathrm{pre-SN}$.

With small kick amplitudes (here $v_k\lesssim50\,\mathrm{km \ s^{-1}}$),
our example system is never disrupted. The post-SN eccentricity, separation, and systemic velocity of the
binary vary with the kick direction \citep[e.g.,][]{brandt:95,
  kalogera:96, tauris:98}. We discuss the population of binaries surviving the
first CC (and hosting a compact object) in \Secref{sec:bound}.

In wide pre-CC systems, small kicks have a larger relative impact on the
velocity of the unbound secondary. The reason is that in wide pre-CC
orbits, smaller kicks result in a longer time for the compact object
to exit the orbit and exert a gravitational pull on the other
star. Conversely, for short pre-SN period systems (unlike the example
considered here) the effects of large kicks 
are more important. This is because the time for the compact object to
exit the pre-CC orbit is short, but larger kicks are more
likely to result in the disruption of the system than small kicks.

\section{Analytic  estimates}
\label{sec:analytics}

\begin{figure*}[htb]
  \centering
  \includegraphics[width=0.33\textwidth]{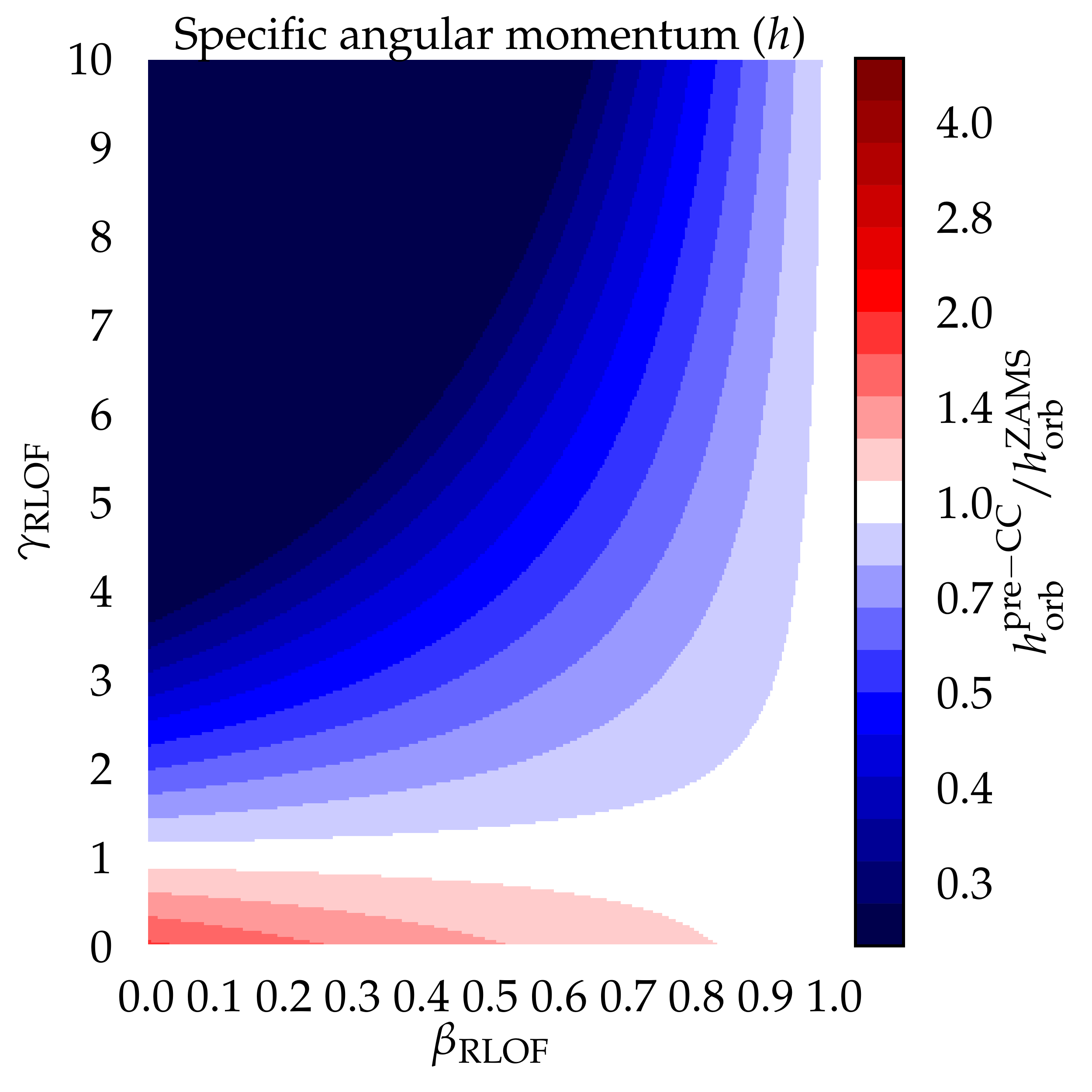}
  \includegraphics[width=0.33\textwidth]{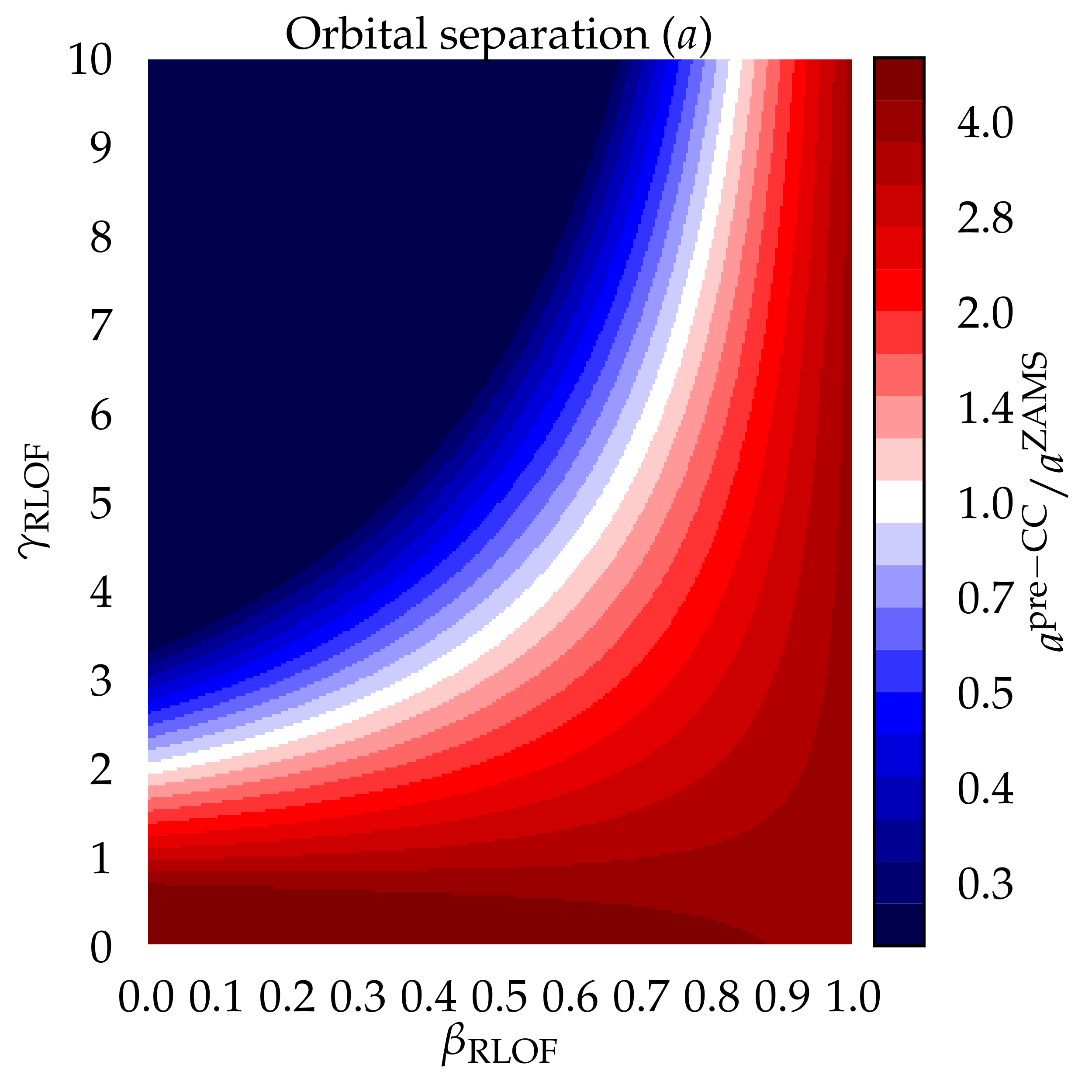}
  \includegraphics[width=0.33\textwidth]{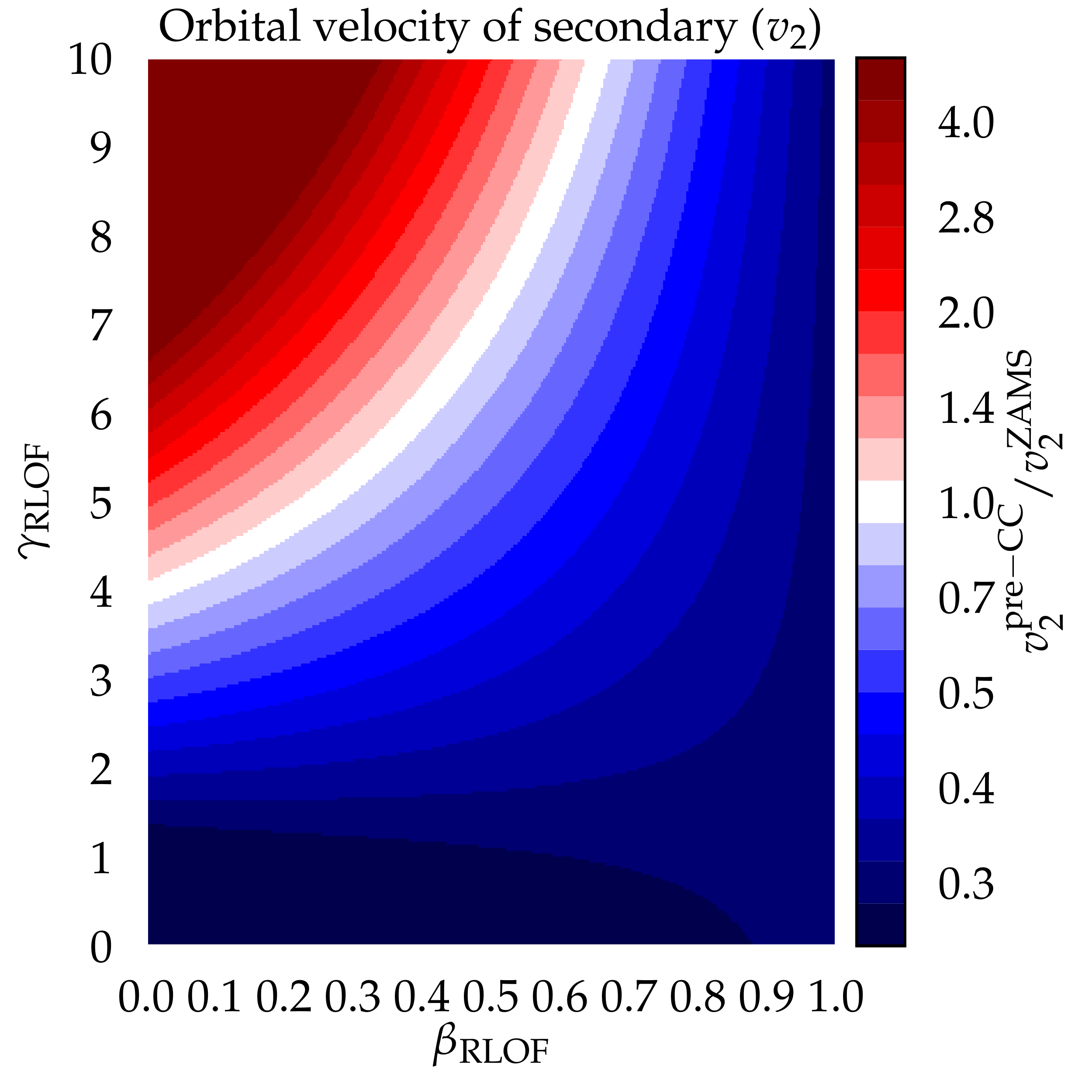}
  \caption{Analytic estimates of the final-to-ZAMS ratio of the
     specific orbital angular momentum $h$ (left panel), the orbital
     separation $a$ (center panel), and orbital velocity of the
     secondary $v_2$ (right panel), assuming $q^\mathrm{ZAMS}=0.75$ and
     $\mu_\mathrm{env}=0.75$, which are representative for the example
     system of \Secref{sec:example}. Blue colors correspond to a
     decrease, red to an increase. The mass transfer efficiency
     $\beta_\mathrm{RLOF}$ and the angular momentum parameter
     $\gamma_\mathrm{RLOF}$ are assumed constant throughout the
     evolution. 
   }
   \label{fig:speedup_factor}
 \end{figure*}

The key quantity to define runaway and walkaway stars is their velocity.
As the example described above demonstrates, the final velocity
obtained at the binary disruption is of the
order of its final (pre-CC) orbital velocity $v_\mathrm{dis}\simeq
v_2^\mathrm{pre-CC}$ \citep[][]{blaauw:61,
  eldridge:11}. Therefore, using a few simplifying assumptions, we can
estimate the ejection velocity for systems that have experienced stable mass transfer, and express it in terms of the initial
parameters of the binary. In particular, we will express the final
orbital velocity as a function of the initial orbital velocity in the
binary, and use it as a proxy for the ejection velocity. This gives
insight on the possible outcomes of the binary evolution since we can estimate the typical velocity at ZAMS from
observational constraints on the binary populations, \citep[e.g.,][]{sana:12}.

Using 
Kepler's third law, we can write the orbital velocity $v_2$ of the
secondary star \newtext{in the frame of the center of mass} at any time during the binary evolution as

\begin{equation}
  \label{eq:v2}
  v_2 =
  \frac{M_1}{M_1+M_2} v_\mathrm{orb} \equiv \frac{M_1}{M_1+M_2} \sqrt{\frac{G(M_1+M_2)}{a}}
  \ \ ,
\end{equation}
where $G$ is the gravitational constant, $a$ is the semimajor axis of
the orbit (assumed to be circular), and $M_1$, $M_2$ are the two
masses.

If we assume that a constant fraction
$\beta_\mathrm{RLOF}=|\dot{M}_\mathrm{acc}|/|\dot{M}_\mathrm{don}|$ of the mass
transferred is accreted by the secondary star, and that the mass not
accreted leaves the binary system with a constant fraction
$\gamma_\mathrm{RLOF}$ of the total
specific orbital angular momentum, we can relate the final and initial
separations with 
\begin{equation}
  \label{eq:af_ai}
  \frac{a_\mathrm{pre-CC}}{a_\mathrm{ZAMS}} =
  \left(\frac{M_1^\mathrm{ZAMS}}{M_1^\mathrm{pre-CC}}\frac{M_2^\mathrm{ZAMS}}{M_2^\mathrm{pre-CC}}\right)^2\left(\frac{M_1^\mathrm{pre-CC}+M_2^\mathrm{pre-CC}}{M_1^\mathrm{ZAMS}+M_2^\mathrm{ZAMS}}\right)^{2\gamma_\mathrm{RLOF}+1}
  \ \ .
\end{equation}

The parameter $\beta_\mathrm{RLOF}$ enters implicitly in the values of
the final pre-CC masses (see \Eqref{eq:DM}). Equation~\ref{eq:af_ai} neglects for simplicity all other processes that can
remove mass from the binary, such as stellar winds \citep[see ][for a
more elaborate expression]{soberman:97}.

The most common type of binary interaction is case B RLOF \citep[][]{kippenhahn:67}.
During \newtext{stable} case B RLOF, the donor star typically loses nearly all its hydrogen-rich envelope
\citep[][]{gotberg:17,yoon:17, gotberg:18}. We parametrize the
fraction of the total mass in the envelope as $\mu_\mathrm{env} =
M_\mathrm{env}/M$, or in other words $1-\mu_\mathrm{env}$ is the fraction of mass in the
helium core. The parameter $\mu_\mathrm{env}$ is dependent on
the overshooting assumed in the underlying stellar evolution
models. The single massive star models from \cite{pols:98} give 
values in the range $0.6\lesssim\mu_\mathrm{env}\lesssim 0.8$ with the larger values
corresponding to less massive (and thus more common) stars. More
recent models \citep[e.g.][]{brott:11} typically adopt larger values
for the overshooting parameter, resulting in smaller values of $\mu_\mathrm{env}$.
Using this parametrization, we can relate the final masses of the two star in
the binary to their ZAMS masses as 
\begin{multline}
  \label{eq:DM}
 M_1^\mathrm{pre-CC} \simeq M_1^\mathrm{ZAMS}-\mu_\mathrm{env} M_1^\mathrm{ZAMS} \
 \ , \\
 M_2^\mathrm{pre-CC} \simeq M_2^\mathrm{ZAMS} +
 \beta_\mathrm{RLOF}\mu_\mathrm{env}M_1^\mathrm{ZAMS}
 \ \ . \hfill
\end{multline}
We can now express the final orbital velocity of the secondary $v_2^\mathrm{pre-CC}$ in
terms of the initial $v_2^\mathrm{ZAMS}$. Introducing $q\equiv q^\mathrm{ZAMS} =
M_2^\mathrm{ZAMS}/M_1^\mathrm{ZAMS}$,
$\beta\equiv\beta_\mathrm{RLOF}$, $\mu\equiv\mu_\mathrm{env}$, and
$\gamma\equiv\gamma_\mathrm{RLOF}$ to simplify the notation, we can write
\begin{equation}
  \label{eq:speedupdown}
  \frac{v_2^\mathrm{pre-CC}}{v_2^\mathrm{ZAMS}}=\left(\frac{q+\beta\mu-\mu
      q-\beta\mu^2}{q}\right)^2
  \left(\frac{1+q}{q+1+\beta\mu-\mu}\right)^{\gamma+1} \ \ .
\end{equation}

 In \Figref{fig:speedup_factor} we visualize the outcome of these
 analytic estimates for different assumptions for the mass transfer
 efficiency $\beta\equiv\beta_\mathrm{RLOF}$ and the angular momentum loss
 parameter $\gamma\equiv\gamma_\mathrm{RLOF}$. To make this figure, we adopt the parameters of the example system of
 \Secref{sec:example} ($q=0.75,\ \mu_\mathrm{env}=0.75$).

 The left panel of \Figref{fig:speedup_factor} shows the relative
 variation in the specific orbital angular momentum \mbox{$h\udef
 J_\mathrm{orb}/(M_1+M_2)=M_1M_2\sqrt{Ga}/(M_1+M_2)^{3/2}$}. For fully conservative evolution
 ($\beta=1$, white region on the right of this panel), $h$ stays
 constant. Also assuming $\gamma=1$ \citep[e.g.,][]{dominik:12} keeps
 the specific angular momentum constant, since this corresponds to
 losing mass and angular momentum at the same relative
 rate. Conversely, low values $\gamma\lesssim1$ result in a
 net increase of the specific orbital angular momentum because mass is
 lost relatively faster than angular momentum, while very large
 $\gamma\gtrsim 1.5$ values result in a loss of specific
 angular momentum.

 The central panel of \Figref{fig:speedup_factor} shows how the
 evolution affects the ratio of the final-to-initial separation (cf.~\Eqref{eq:af_ai}). The
 vast majority of the parameter space results in significant orbital
 widening (red colors), with only rather extreme angular momentum
 losses ($\gamma\gtrsim2$) in non-conservative
 ($\beta\lesssim0.6$) systems result in orbital shrinking.

 The right panel in \Figref{fig:speedup_factor} shows the main
 parameter of interest, the ratio of the final-to-initial orbital
 velocity of the initially less massive star in the system (cf.~\Eqref{eq:speedupdown}). Most
 assumptions for $\beta$ and $\gamma$ result in a decrease in the
 orbital velocity of the secondary (blue colors). This is the combined
 effect of (i) the orbital widening, and (ii) the increase mass of the secondary
 because of mass transfer. Only for very non-conservative systems
 ($\beta\lesssim0.3$) experiencing large
 angular momentum losses ($\gamma\gtrsim4$, corresponding to
 non-accreted mass removing four times the specific orbital angular
 momentum of the binary) the final velocity of the secondary is higher
 than its initial value. 

 In our numerical simulations we adopt physically motivated model for the
 efficiency of mass transfer instead of constant parameters, $\beta$
 and $\gamma$, as we assumed in these analytical considerations. As we
 will argue later, most of the massive unbound companions result from
 nearly conservative mass transfer. Therefore, it is instructive to
 consider the analytical solution for $\beta = 1$. The expression for the final orbital velocity of the secondary (\Eqref{eq:speedupdown}) simplifies to 
 \begin{equation}
   \label{eq:cons}
   \frac{v_2^\mathrm{pre-CC}}{v_2^\mathrm{ZAMS}} =
   \left(\frac{(1-\mu)(q+\mu)}{q}\right)^2 = \left(\frac{3+4q}{16
       q}\right)^2 \ \ ,
 \end{equation}
 where we use as a typical value $\mu=3/4=0.75$ in the last
 step. \Eqref{eq:cons} gives a final-to-initial orbital velocity ratio for the
 secondary always smaller than one for initial mass
 ratios $q>0.25$. In other words, for all systems with initial mass
 ratios of interest evolving through stable and nearly conservative
 mass transfer, the secondary always slows down
 substantially, by up to a factor of 5 for an initially equal mass
 system. We emphasize that we have ignored the additional effect of mass loss by
 stellar winds. 
 These cause the orbit to widen further and lead to a
 further slow down of the secondary star.

\begin{figure*}[tpb]
  \centering
  \tikzset{font=\small,
edge from parent fork down,
level distance=2cm,
sibling distance=3pt,
every node/.style={top color=blue!15,
  bottom color=blue!15,
  rectangle,rounded corners,
  minimum height=0mm,
  align=center,
},
level 4/.style={sibling distance=20pt},
level 5/.style={sibling distance=10pt},
edge from parent/.style={
  draw=blue!50,
  very thick,
},
blank/.style={draw=none},
}

\begin{tikzpicture}
  \Tree [
          .{Binaries with\\$M_1^\mathrm{ZAMS}\geq7.5\,M_\odot$\\ $q^\mathrm{ZAMS}\geq0.1$\\$0.15\leq \log_{10}(P^\mathrm{ZAMS}/\mathrm{days})\leq 5.5$}
               \edge [] node [draw=none, top color=white, bottom
               color=white, above, xshift=-10pt] {$22_{-9}^{+26}\%$};
               [
                   .{Stellar\\mergers}
               ]
               \edge [] node [draw=none, top color=white, bottom
               color=white, above, xshift=10pt] {$78_{-22}^{+9}\%$};
               [
                   .{Non mergers}
                   \edge [] node [draw=none, top color=white, bottom color=white, midway, above] {$1-\mathcal{D}=14_{-10}^{+22}\%$};
                   [
                       .{Non Disrupted\\ {\tiny
                           (cf.~\Secref{sec:bound})}}
                                          \edge [] node [draw=none, top color=white, bottom color=white, midway, left] {$36_{-15}^{+42}\%$};
                       [
                       .{NS companion}
                       ]
                       \edge [] node [draw=none, top color=white, bottom color=white, midway, right] {$64_{-42}^{+15}\%$};
                       [
                       .{BH companion}
                       ]
                   ]                       
                   \edge [] node [draw=none, top color=white, bottom color=white, midway, above] {\textcolor{red}{$\mathcal{D}=86_{-22}^{+10}\%$}};
                   [
                       .{Disrupted}
                       \edge [] node [draw=none, top color=white, bottom color=white, midway, above] {$25_{-10}^{+8}\%$};
                       [
                           .{post-MS companion \\ {\tiny
                               (cf.~\Secref{sec:non-MS-companions})}}
                           \edge [] node [draw=none, top color=white,
                           bottom color=white,  above, xshift=-6pt] {$53_{-46}^{+12}\%$};
                           [
                              .{WD}
                           ]
                           \edge [] node [draw=none, top color=white,
                           bottom color=white,
                           below, yshift=-40pt] {$23_{-20}^{+38}\%$};
                           [
                               .{He star}
                           ]
                           \edge [] node [draw=none, top color=white,
                           bottom color=white,  above, xshift=6pt] {$25_{-6}^{+23}\%$};
                           [
                               .{H-rich}
                           ]                        
                      ]
                      \edge [] node [draw=none, top color=white, bottom color=white, midway, above] {$75_{-8}^{+10}\%$};
                      [
                           .{MS companion \\ {\tiny (cf.~\Secref{sec:MS})}}
                           \edge [] node [draw=none, top color=white,
                           bottom color=white, above, xshift=-3pt] {$5_{-2}^{+14}\%$};
                           [
                              .{Runaway\\$v_\mathrm{dis}\geq30\,\mathrm{km\ s^{-1}}$}
                           ]
                           \edge [] node [draw=none, top color=white,
                           bottom color=white, above, xshift=+3pt] {$95_{-14}^{+2}\%$};
                           [
                               .{Walkaway\\$v_\mathrm{dis}<30\,\mathrm{km\ s^{-1}}$}
                           ]                              
                     ]
                   ]       
               ]
        ]    
\end{tikzpicture}
  \caption{Overview of the binary evolution scenarios up to
    the first CC event. The
    branching ratios shown are from our fiducial simulation, and we
    highlight in red the disruption fraction
    $\mathcal{D}$. The errors on each fraction exclude the run without
  SN kicks ($\sigma_\mathrm{kick}=0\,\mathrm{km\ s^{-1}}$), which
  produces an unrealistically low disruption fraction
  (cf.~\Tabref{tab:parameters} and \Secref{sec:param_variation}).}
  \label{fig:tree}
\end{figure*}
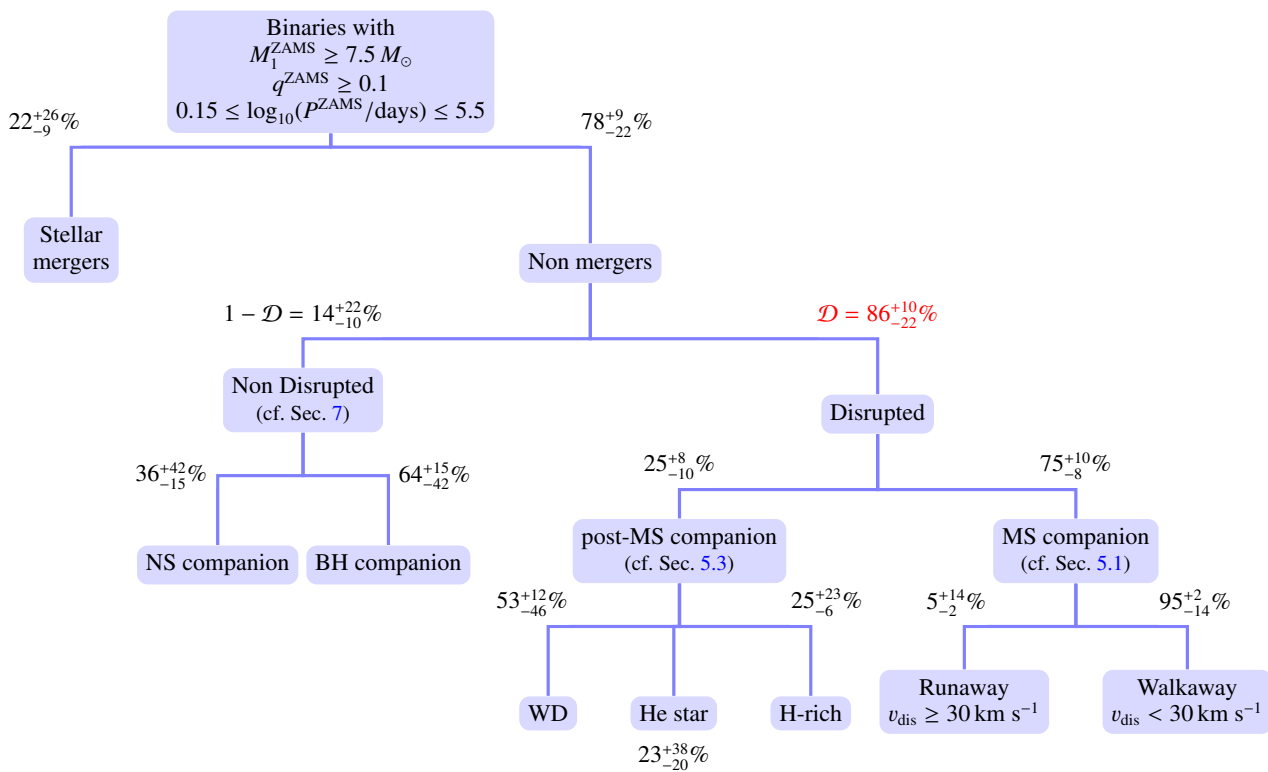

Using the typical values of our example system discussed in
\Secref{sec:example} ($q = 0.75$), we find $v_2^\mathrm{pre-CC}
\simeq 0.25 \, v_2^\mathrm{ZAMS} \simeq  20\,\mathrm{km\ s^{-1}}$.
Our numerical model for this system gives an even lower value,
primarily because of the additional effect of widening as a result of
mass loss through stellar winds. As a sanity check we recomputed the
evolution of the example system, but artificially switching off the
stellar wind mass loss for both stars. This gives
$v_2^\mathrm{pre-CC}\simeq22\,\mathrm{km\ s^{-1}}$, in good agreement
with the analitycal estimate. The remaining difference is due to the fact that the mass transfer is not fully conservative.   

For the analytic estimates presented in this section, we only
considered the case of stable mass transfer. Binary evolution through
unstable mass transfer takes place for a limited range of the initial
distributions \citep[e.g.,][]{soberman:97,schneider:15}. It is
expected to result in a common-envelope phase, followed by substantial
shrinking of the orbit either leading to a merger or the formation of
a compact binary system if the envelope is ejected successfully.
While this is a very interesting pathway to create very fast runaway
stars, the numerical simulations presented later indicate that this
channel produces primarily low mass runaway stars. This is because it typically concerns system with extreme initial mass ratios in which the secondary does not significantly gains mass. This means that the secondaries in these systems are typically not very massive and therefore this channel does not significantly contribute to the production of massive early type runaways, at least in our simulations. We return to this in \Secref{sec:discussion}. 

\section{Ejected companions}
\label{sec:fiducial_res}

In this section, \newtext{we} present the results obtained from our
numerical simulations for a full population computed with our fiducial
assumptions (cf.~\Secref{sec:methods}). We describe the robustness of
our findings against model assumptions in
\Secref{sec:param_variation}. Our main goal is to characterize the
velocity acquired by the companion stars ejected at the time of the first CC
in the binary. Therefore, we ignore all binaries that merge, which constitute $\sim$$22\%$ of our fiducial
population (cf.~\Figref{fig:tree}), in good agreement
with~\cite{sana:12,demink:14,zapartas:17,zapartas:17b}; see also \cite{kochanek:14}. 

Once the merging systems are excluded, the first CC happens in the presence
of a companion. The CC can result in either the disruption of the
binary ($\mathcal{D}=86\%$ in our fiducial run), or the newly formed
compact object remains bound to the companion star. In
\Secref{sec:MS} we focus on the population of ejected MS companions,
which are $75\%$ of all the ejected companions in our fiducial run. We consider companions
that have already evolved off the MS at the time of the first CC in \Secref{sec:non-MS-companions}.

For completeness, we describe the systems
with a MS companion remaining bound in \Secref{sec:bound}.

\subsection{Ejected main sequence companions}
\label{sec:MS}

\begin{figure*}[htp]
  \centering
  \includegraphics[width=\textwidth]{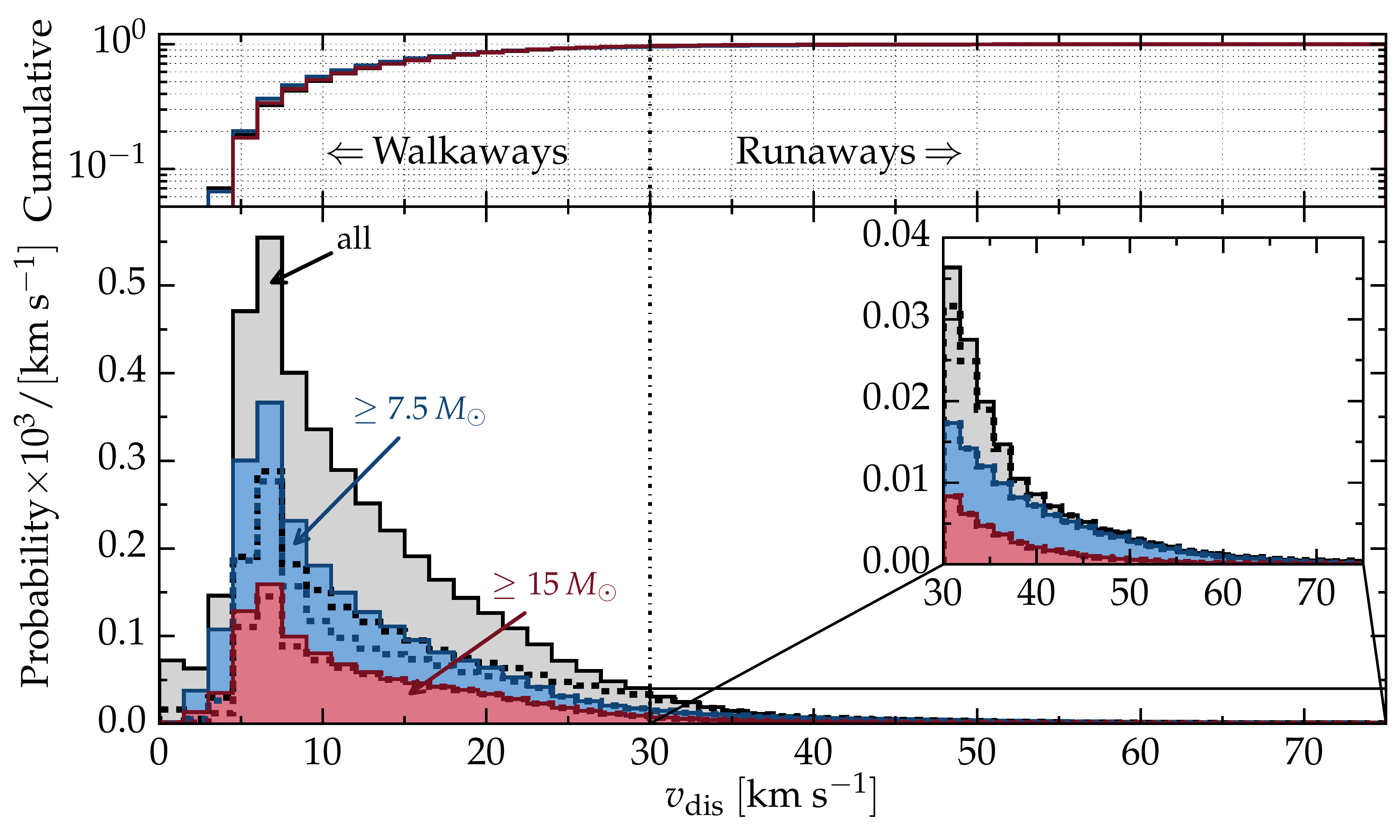}
  \caption{Velocity distribution of MS stars ejected
    from a binary system at the time of the first CC. The top panel shows
    the corresponding cumulative distributions. The 
    walkaways are one order of magnitude more numerous than the
    runaways. The gray, blue, and red histograms show all MS
    secondaries, only MS stars more massive than $7.5\,M_\odot$ (i.e.,
    roughly those that might experience CC), and only MS stars more
    massive than $15\,M_\odot$ (i.e., roughly the O-type stars),
    respectively. The mass considered here is taken right after the CC of the primary star. The dashed
    lines show the distribution of walkaways and runaways that have
    gone through RLOF (or common envelope evolution) before being
    ejected. Almost all massive runaways and walkaways, and the
    majority of the ejected stars at large velocities, have gone
    through RLOF. The inset plot magnifies the runaway regime
    $v_\mathrm{dis}\geq30\,\mathrm{km\ s^{-1}}$. See also
    \Figref{fig:v_dist_log} for a wider velocity range,
    \Figref{fig:v_dist_obs} for a figure accounting for the remaining
    lifetime as walkaway or runaway star, \newtext{and
      \Figref{fig:convolution} for an example of how the velocity
      dispersion of a star forming region affects these distributions}.}
  \label{fig:v_dist}
\end{figure*}
\newtext{The fraction of CC events occurring with a MS companion in
  our fiducial run is $\sim$76\% (and it ranges from $69-90\%$ in our
  parameter variations). The vast majority of these binaries which
  host a MS companion is disrupted at the time of the first CC.} Figure~\ref{fig:v_dist}
shows the distribution of velocities of the ejected companions. The three different colors subdivide the population based on
the mass of the ejected companion. The gray histogram shows all
ejected MS companions. The blue histogram shows only massive MS
companions, $M_2^\mathrm{post-CC}\geq 7.5\,M_\odot$, where
$M_2^\mathrm{post-CC}$ is the mass of the secondary just after the
collapse of the primary. This group roughly corresponds to all
secondaries that will experience CC at the end of their lifetime. The red histogram shows the velocity
distribution of companions with mass larger than $M_2^\mathrm{post-CC}\geq15\,M_\odot$, corresponding roughly to O-type
stars. In the rest of this section, we quote the ratios for the
population of massive stars (i.e.~$M_2^\mathrm{post-CC}\geq 7.5\,M_\odot$) unless
otherwise stated, see also \Tabref{tab:parameters} for more
information.

Although we do produce runaways, the majority of the systems ejects a star slower than $30\,\mathrm{km \
  s^{-1}}$, i.e., a walkaway star. 
We find that the CC of a star
with a MS binary companion is about twenty times more likely to produce a
walkaway than a runaway. In the first line of \Tabref{tab:parameters},
we list the ratio $\mathcal{R}$ of walkaway to runaway stars produced per CC event in a binary. From our
fiducial simulation, we obtain $\mathcal{R}_\mathrm{7.5}\simeq20$ for
massive companions.
As shown in
\Tabref{tab:parameters}, considering progressively larger masses, the number of walkaways
produced per each runaway from a binary increases. \newtext{In
 \Secref{sec:convolution}, we address the impact of the velocity
dispersion within a star forming region on this result.}

Runaways resulting from the disruption of binaries rarely exceed $60\,\mathrm{km \
  s^{-1}}$: about 99.8\% of the massive unbound companions we simulate are slower than
this threshold. We note that higher velocity are not strictly
forbidden, but extremely unlikely (cf.~\Figref{fig:v_dist_log} which
shows the velocity distribution on a logarithmic scale).
Low mass runaways can reach much
tighter pre-CC orbits through common envelope evolution (without
significant accretion), and thus reach higher ejection velocities, but typically
not in excess of $400\,\mathrm{km\ s^{-1}}$. \newtext{Massive} stars significantly
faster than this are likely produced by different mechanisms
(see \citealt{boubert:17b}, and \citealt{brown:15} for a review). However, in extremely rare cases
where a very short pre-CC binary is formed and also successfully disrupted,
much higher velocities can be achieved.  The
absolute maximum velocity we obtain is about $1110\,\mathrm{km\
  s^{-1}}$, in good agreement with \cite{tauris:15}.

The slowest stars come from the widest pre-CC binaries:
the low-velocity drop-off of the distribution in \Figref{fig:v_dist}
is thus an effect of the upper-limit on our period distribution (corresponding
to an upper-limit on the pre-CC period).

The velocity distribution peaks at $v_\mathrm{dis}\simeq6\,\mathrm{km\
  s^{-1}}$. Ejected MS companions with such velocity typically
originate from systems with $q_\mathrm{ZAMS}\gtrsim0.5$ and long initial
periods ($\log_{10}(P^\mathrm{ZAMS}/\mathrm{days})\simeq3$),
regardless of the stellar masses. These systems experience late case B
or case C mass transfer \citep[][]{lauterborn:70}, but owing to the
sufficiently high mass ratio, the mass transfer remains conservative
and the system widens.

\begin{figure}[tbp]
  \centering
  \includegraphics[width=0.5\textwidth]{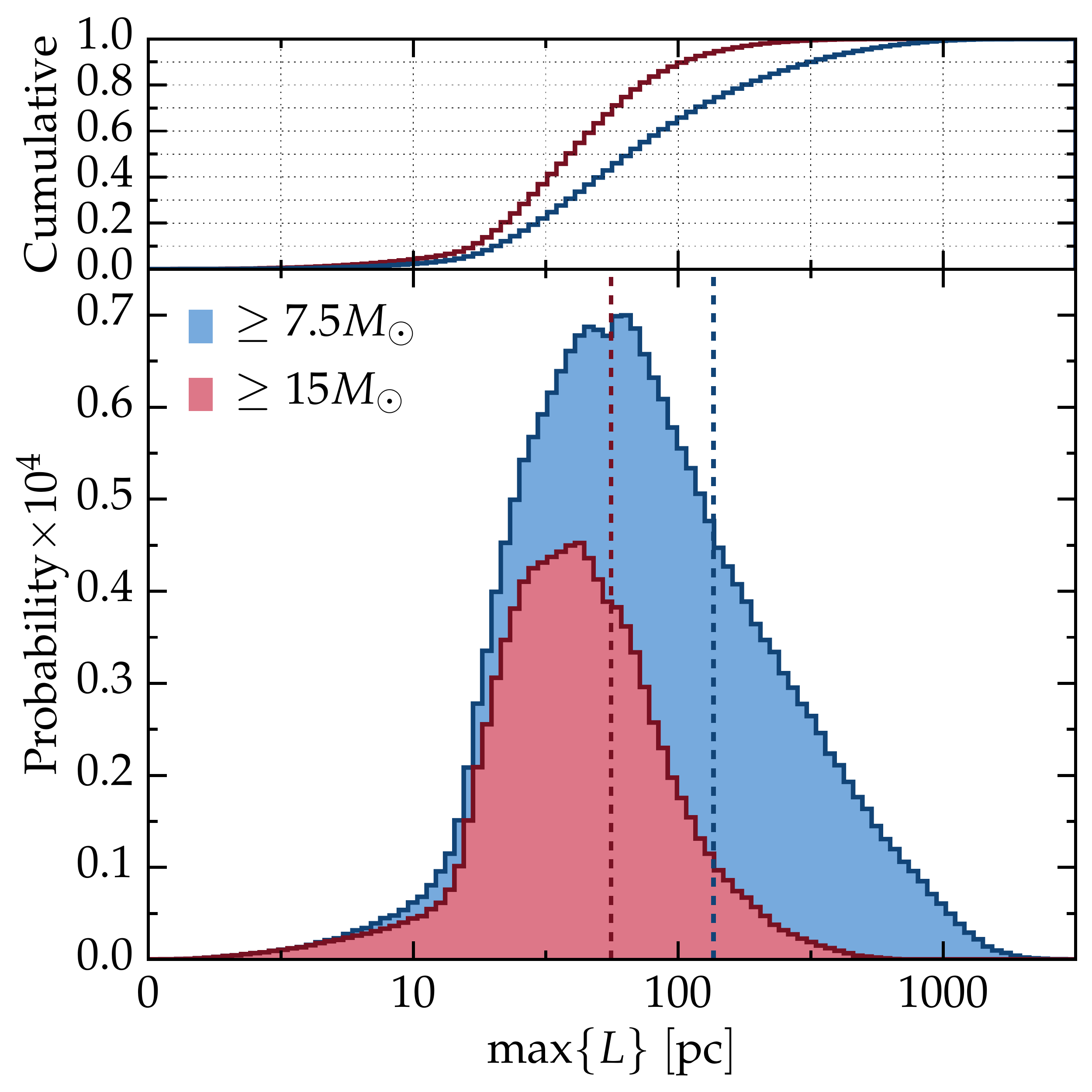}
  \caption{Distribution of the maximum distance $L$ that ejected
    MS secondaries can reach, neglecting any gravitational
    potential. The blue and red histograms are for ejected stars more
    massive than $7.5\,M_\odot$ and $15\,M_\odot$, respectively. The
    vertical dashed lines mark the mean values of
    the distributions.}
  \label{fig:Dhist}
\end{figure}

Among low and intermediate mass walkaways with
$v_\mathrm{dis}\simeq6\,\mathrm{km\ s^{-1}}$, there is also a contribution from
wide systems which do not experience mass transfer. This contribution
is almost negligible for ejected stars more massive than
$15\,M_\odot$. This can be seen from the dashed distributions in
\Figref{fig:v_dist}, which show that the majority
of massive runaways and walkaways have accreted mass from
their companion before its CC. Over the entire velocity range shown
(i.e., considering both runaways and walkaways), $71\%$ of ejected MS
secondaries more massive than $7.5\,M_\odot$, and $91\%$ of those more
massive than $15\,M_\odot$ have accreted from their companion in
the previous evolution. Among low mass ejected companions, $50\%$ have
experienced either RLOF or common envelope evolution before their
ejection from the binary. From the inset in \Figref{fig:v_dist}, where
the blue and red dashed lines representing post-interaction systems
overlap with the full distribution, it is clear that all massive
runaways from binary disruption have experienced mass transfer. 

The majority of the ejected massive MS
companions are not ``usual'' single stars, and this remains true for a
very large fraction of the low mass MS ejected companions. Before the binary disruption,
they have accreted mass from their companion. Mass transfer also causes the convective core of the accretor to grow,
resulting in rejuvenation of the accreting star
\citep[e.g.,][]{hellings:83,schneider:16}. The fact that most ejected
MS stars are accretors is the result of a
combination of assumptions. For disruption to be possible,
the primary needs to be massive enough to collapse. We assume a flat
mass ratio distribution, implying that on average $M_2^\mathrm{ZAMS}\simeq
0.5\,M_1^\mathrm{ZAMS}$, and we also assume that mass transfer is
unstable and leads to mergers if the mass ratio is too extreme
($M_\mathrm{acc}/M_\mathrm{don}<q_\mathrm{crit}$), therefore binaries
that can eject a star have initial mass ratios closer to one than an
average system. Finally, when stable mass transfer occurs, our fiducial
assumption for the accretion efficiency results in rather conservative
mass transfer over most of the period range considered
\citep[][]{schneider:15}.

To make an estimate of the
ratio of walkaways  per runaway {\it existing at a given time}
requires us to
consider the finite MS lifetime of the ejected stars. Assuming that the star formation rate in the Galaxy is
constant and therefore that the Galactic population is in a steady state,
the ratio of walkaways per runaway is
$\sim$13 for masses larger $7.5\,M_\odot$.

Accounting for the durations of each evolutionary phase, we can also
quantify the fraction of MS stars more massive than $15\,M_\odot$ that are
runaway or walkaway stars. In our fiducial simulation, we find a
runaway fraction for masses larger than $15\,M_\odot$ of about $\sim$$0.5\%$ and a
walkaway fraction of $\sim$$10\%$. We only simulate binaries with primaries more
massive than $7.5\,M_\odot$, therefore we do not have in our
population intermediate-mass binaries which could contribute through
mass transfer and/or mergers  to the
normalization factor for the runaway and walkaway fraction at lower
masses \citep[][]{zapartas:17}. We show in \Figref{fig:v_dist_obs} the distribution in
velocity accounting for the lifetime of the stars in each bin. This
can be directly compared to observations \newtext{assuming that} the effects of the
Galactic potential can be neglected.

\begin{figure*}[!htbp]
  \centering
  \includegraphics[width=0.75\textwidth]{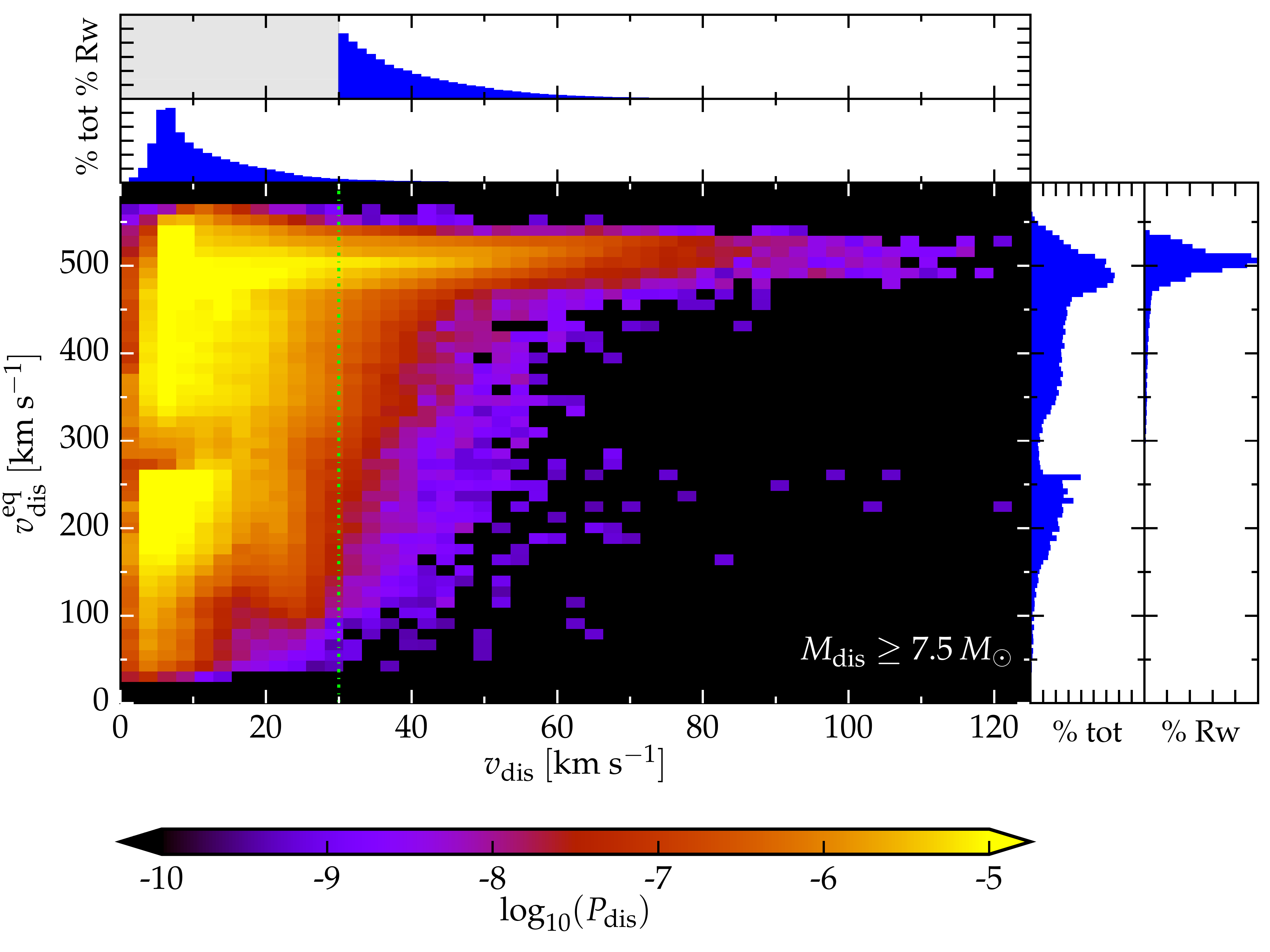}
  \caption{Equatorial velocity at the time of ejection for massive ($M_\mathrm{dis}\geq7.5\,M_\odot$) companions as a function
    of the ejection velocity. Brighter colors indicate location of the
    parameter space more populated by our fiducial simulation. All the massive runaway stars spin with
    an equatorial velocity of $\sim$$500\,\mathrm{km\ s^{-1}}$, close to
  breakup rotation, since they have accreted mass from their companion
  before being ejected. The spread is due to wind spin down before the
  binary disruption. We do not include the effect of projecting on
  the line of sight, or post-ejection wind spin down in this
  plot.  The
    vertical dot-dashed line marks the threshold to define runaway
    stars.}
  \label{fig:vrot_vrw}
\end{figure*}

Figure~\ref{fig:Dhist} shows the distribution of distances traveled by
massive, unbound stars, calculated by multiplying their velocity
$v_\mathrm{dis}$ by their remaining lifetime on the MS, which includes
rejuvenation if the star has accreted mass, plus 10\% of the
MS duration of a star with the same helium core mass, to account for
the duration of the helium core burning phase in the rejuvenated
star\footnote{The duration of the helium core burning depends on the helium core mass, and thus the overshooting
parameter. The value of $10\%$ of the MS duration is typical for overshooting values
larger than what assumed in \cite{pols:98}.}.
This approach neglects the effect of an external potential on the trajectory of the
stars \citep[see][for how this could be done]{boubert:17a,boubert:18}, so it is effectively an upper limit to the distance they can travel. Considering both runaway and walkaway stars with masses
$M_\mathrm{dis} \geq7.5\,M_\odot$, we find
the mean distance $\langle L \rangle$ they travel before experiencing
CC to be 126\,pc, see the blue vertical dashed line in
\Figref{fig:Dhist} and also \Tabref{tab:parameters}. This number
rises to $\langle L_\mathrm{run}\rangle=584$\,pc considering only 
runaways (faster than $30\,\mathrm{km \ s^{-1}}$) more massive than
$7.5\,M_\odot$, but it remains $\langle
L_\mathrm{walk}\rangle=103$\,pc considering massive walkaways only
(slower than $30\,\mathrm{km \ s^{-1}}$). For comparison, the typical
size of OB-associations is of the order of tens of pc,
while the Galactic thin and thick disks have a vertical scale height
of $\sim$300 and $\sim$1500\,pc, respectively
\citep[e.g.,][]{gilmore:83}.

\subsection{Spin of the main sequence ejected companions}

Figure~\ref{fig:vrot_vrw} shows the equatorial rotational velocity
$v_\mathrm{dis}^\mathrm{eq}$ for
the ejected MS companions with $M_\mathrm{dis}\geq7.5\,M_\odot$. The top panels show the normalized one-dimensional
distribution of ejection velocity for the whole sample (the same as the
blue distribution in \Figref{fig:v_dist}), and for the runaways only,
respectively. The side panels show the normalized equatorial
rotational velocity distribution
for the whole sample, and for the runaways only, respectively.
We do not include in \Figref{fig:vrot_vrw} any projection effects due to the inclination
of the stellar spin axis with respect to the line of sight, nor the
wind spin down due to the evolution after the binary disruption. The
combination of these two effects would \newtext{systematically
  decrease the rotational velocities}, allowing any physically possible
\emph{projected} equatorial rotational velocity to be observable.

The main feature at the top of the central panel (also visible in the rightmost panel) indicates that almost all the massive runaways are spinning
nearly at breakup rotation at the time of the ejection, because of the
relatively recent spin-up during the mass transfer phase(s).

The walkaways ($v_\mathrm{dis}<30\,\mathrm{km\ s^{-1}}$) instead can
have a broader range of rotational velocities: this is because a
subset of them comes from wide, non-interacting binaries. The right
panel including both the walkaway and runaway population shows two
peaks for $v_\mathrm{dis}^\mathrm{eq}\simeq200$ and $500\,\mathrm{km\
  s^{-1}}$. The former is due to the combination of the accretion-induced spin up and the wind spin down
before the ejection, which can be more efficient for walkaways
experiencing early mass transfer and with longer delay between the
mass transfer and the disruption of the binary. The former peak is \emph{not} related to tidal locking, which
is efficient only for tight orbits that would result in much larger ejection
velocities $v_\mathrm{dis}$. The latter is populated by secondaries ejected after a
mass transfer phase (analogous to the peak in the equatorial velocity
distribution for the runaways).

\subsection{Ejected post-main sequence companions}
\label{sec:non-MS-companions}

\begin{figure}[htbp]
  \centering
  \includegraphics[width=0.5\textwidth]{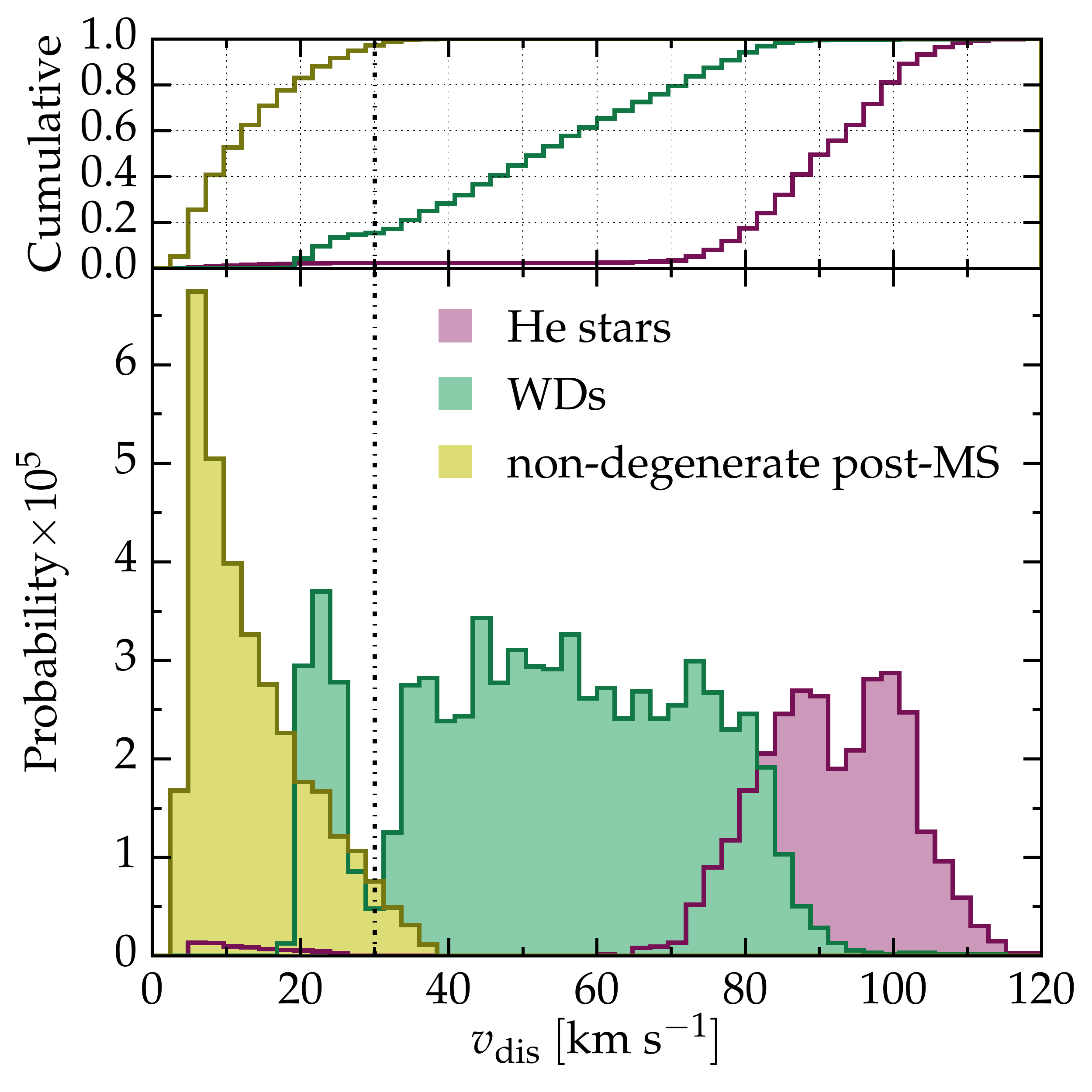}
  \caption{Velocity distribution of companions \newtext{ejected after
      their MS}. Helium
    stars and white dwarfs (WD) can have much shorter pre-CC orbits
    because of their smaller radii, and thus they can reach much
    higher ejection velocities. Conversely, the number of non-degenerate, hydrogen-rich
    post-MS stars is dominated by red supergiants with very large
    radii, which can only exist at large separation from the exploding
    companion, and thus typically have much slower velocities. The top
    panel shows the corresponding cumulative distributions.}
  \label{fig:v2_postMS}
\end{figure}

Figure \ref{fig:v2_postMS} shows the velocity distribution of companion stars which have evolved off
the MS by the time of the first CC in the binary. All
together, these correspond roughly to 25\% of the population of
disrupted binaries (cf.~\Figref{fig:tree}). \newtext{Other
  kind of systems and evolutionary channels not considered here can also produce post-MS
  ejected companions \citep[e.g.,][]{justham:09, zapartas:17}.} We distinguish three sub-populations:
hydrogen-rich non-degenerate secondaries (mostly of red-supergiant
secondaries, corresponding to $\sim$$4\%$
of the entire simulated population, including mergers and
non-disrupted binaries), white dwarfs (WD, $\sim$8\% of
the total), and helium stars which have lost their hydrogen-rich envelope because of previous
binary interactions ($\sim4\%$ of the total).

For the first group, characterized by large stellar radii, the velocity distribution resembles
closely that of the MS ejected companions, with a large number
of walkaways compared to the runaways (roughly corresponding to $96\%$ of the hydrogen-rich
non-degenerate ejected companions).

Ejected WDs and helium stars come from different evolutionary paths. Mass transfer during the primary MS
can reduce the helium core mass thus slowing its evolution, while
accretion of mass speeds up the evolution of the companion. In
$\sim23\%$ of our full population (excluding stellar mergers), this leads
to a reversal of the CC order, with the secondary exploding first \citep[][]{pols:94}. In
these cases the star that is ejected is the initially more massive, which
has become, by the time of the CC, a helium star or a WD.

\newtext{WDs and helium stars are what is left of the initial donor and they are thus less massive than
their companion after the mass transfer phase, therefore they can
reach velocities much larger than (super-)giant stars.
The ejection velocities of
the WDs span} a range  $15 \lesssim v_\mathrm{dis}/\mathrm{[km\
  s^{-1}]}\lesssim 90$, and helium stars are even faster with $70 \lesssim v_\mathrm{dis}/\mathrm{[km\ s^{-1}]}\lesssim
120$. This difference arises because it takes longer for a star to
become a WD than its helium core burning duration. WDs are
ejected by systems where the rejuvenation of the 
secondary is not extreme. This suggests that these systems have
a less conservative mass transfer phase than binaries ejecting a
helium star, which corresponds to less
rejuvenation of the secondary and more orbital widening, and thus
lower velocity of the WD at ejection. Moreover, if
the companion star at pre-CC stage is a WD,  there is a longer time
for the binary to widen, e.g.,~because of winds, than if it is a
helium star.

Conversely, if the ejected star is a non-degenerate helium
star, then the entire evolution of the rejuvenated secondary needs to
be faster than the remaining lifetime of the primary (which now
has a reduced mass): the time delay between the RLOF phase and the
first CC ejecting the helium star is much shorter and there is not enough
time for significant orbital widening.

Figure~\ref{fig:v2_postMS} also shows a small population of slow ($v_\mathrm{dis} <
30\,\mathrm{km\ s^{-1}}$) helium stars. These are very massive
($M_2\gtrsim 30\,M_\odot$ \newtext{after mass accretion}) stars characterized by large Wolf-Rayet
wind mass loss rates, which cause significant orbital widening before
the first CC, resulting in the slow ejection velocity. We emphasize that \Figref{fig:v2_postMS} shows the velocity of
stars ejected while in the corresponding evolutionary stage, and stars
ejected during their MS will then evolve into post-MS hydrogen-rich,
or possibly Wolf-Rayet stars, creating a more complicated velocity distribution of
massive Wolf-Rayet stars \citep[see, e.g.,][]{dray:05, eldridge:11}.

\section{Impact of model variations}
\label{sec:param_variation}

Our predictions depend on a set of parametrized assumptions. We
perform a systematic study of the impact of these uncertainties by
varying the values of the free parameters one-by-one, as we describe in
\Secref{sec:methods}. In each variation, all the other parameters are set to their
fiducial value.

Table \ref{tab:parameters} summarizes our results for binaries with a
collapsing star and a MS companion. If our extreme assumptions
do not change \newtext{significantly} our synthetic population, then the outcome is
robust and independent of the particular physical process. Conversely,
strong dependence on any one parameter indicates that \newtext{this particular} parameter could be constrained by
comparing to observed samples, and therefore allow for physical tests
of the process it represents. Overall, our results are robust and relatively insensitive to
uncertain physical processes, at least within the framework of
parametrizations we assume.

\subsection{Robust  predictions}

The fraction of binaries that are disrupted by the first CC is always larger
than $75\%$, with two notable exceptions. These concern the natal kick
distribution assumed. With a double Maxwellian kick distribution with
$\sigma_\mathrm{kick}=30\,\mathrm{km\ s^{-1}}$ for NSs less massive
than $1.35\,M_\odot$ (and our fiducial kick for more massive
remnants), the disruption fraction
decreases to $\mathcal{D}=65\%$ even though this changes the kick only
for a small \newtext{range of the parameter space}. Low mass NSs are the
typical result of progenitors with small total and core masses, which
are favored by the IMF.

If instead we make the extreme assumption of zero
natal kick in all CC ($\sigma_\mathrm{kick}=0\,\mathrm{km\ s^{-1}}$), then only $\mathcal{D}=16\%$ of the
binaries are disrupted. In these cases the disruption is
only due to the mass loss at the moment of the explosion (Blaauw
kick). Since the ejecta mass must be larger than half of the total
mass of the system to disrupt the binary \citep[][]{blaauw:61}, and since after mass
transfer the primary is typically less massive than the companion, only very wide binaries which did not experience mass transfer
\newtext{might} be disrupted in such a way. Therefore, in this extreme assumption,
the ejected stars are on average
less massive and slower compared to calculations including a natal
kick. We emphasize that this parameter variation is in contrast with
the observed distribution of pulsar velocities, and we
include it for insight only. This variation does not produce any massive
runaways, which is also in contrast with observations. 

We obtain the largest disruption
fraction when we do not rescale the natal kick with the fallback
fraction: in this model, up to $97\%$ of the systems are
disrupted. This assumption is equivalent to assuming a ``BH velocity
kick'' \mbox{\citep[e.g.,][]{belczynski:08,fryer:12}}: BHs and NSs receive
the same kick amplitude at birth, despite the differences in their masses. Therefore, many CC events
resulting in the formation of a massive BH with large fallback
fractions (up to $f_b=1$) still result in the disruption of the binary.

Despite the overall large disruption fraction in all our parameter
variations, the predicted runaway fraction $f_{15}^\mathrm{RW}$ for masses larger than $15\,M_\odot$ is low for all our
parameter variations and never exceeds a few percent. Similarly, the walkaway
fraction $f_{15}^\mathrm{WA}$ is also robust, and is generally around
$10\%$. In other words, the walkaways outnumber the runaways by
a factor of 10-30 (smaller for larger masses), regardless of the
physical assumptions we make. The
most notable exceptions are computations at lower metallicity (see
\Secref{sec:Z}). Another case producing slightly smaller $\mathcal{R}$ is
fully non-conservative mass transfer ($\beta_\mathrm{RLOF}=0$): more
systems are in a very close pre-CC orbits and the average velocity of
an ejected secondary is higher than in our fiducial assumptions.

\newtext{It is worth noting that variations in the common envelope
  efficiency ($\alpha_\mathrm{CE}$) do not have a large influence on
  this result. The most likely reason for this is the combination of
  our assumptions on the stability of late case B and case C
  RLOF and whether massive stars develop deep convective envelopes. The
  models from \cite{pols:98},
  underlying the fitting formulae of \texttt{binary\_c}, barely
  ascend the Hayashi track for $M_\mathrm{ZAMS}\gtrsim15\,M_\odot$, and the
  $q_\mathrm{crit, RSG}$ assumed in our fiducial simulation is rather
  conservative in assessing the stability of late RLOF. Both these
  assumptions contribute in making the parameter space leading to
  common envelope, significant orbital shrinking, and subsequent
  ejection of a fast and massive\footnote{\newtext{The inspiraling star is
    assumed not to accrete significantly during a common
  envelope event, which prevents its growth by mass.}} runaway relatively small. Allowing for less stable late RLOF (higher $q_\mathrm{crit, RSG}$) increases the average speed of ejected companions, and decreases the
  relative number of walkaways. however, extreme variations in this
  parameter (in combination with the common envelope efficiency
  $\alpha_\mathrm{CE}$) are not sufficient to reconcile our
  predictions with commonly-accepted inferences from observations (see
  \Secref{sec:howtofind}).}

The mean value of the mass of the walkaways, runaways, and all ejected
secondaries ($\langle M_2^\mathrm{walk} \rangle$, $\langle
M_2^\mathrm{run} \rangle$, and $\langle M_2 \rangle$, respectively) exceeds $7.5\,M_\odot$ in almost all our parameter variations. With
fully non-conservative mass transfer ($\beta_\mathrm{RLOF}=0$),
the secondary accretes no mass. This makes them on average less
massive than in our fiducial assumptions at the moment of the first CC, and also allows
for less orbital widening (the donor has to lose more mass before the
mass ratio inverts), resulting in a faster population of disrupted
stars with lower masses compared to our fiducial case. If we assume that case B mass transfer is
unstable whenever the accretor is less massive than the donor
($q_\mathrm{crit, B}=1$), or if we take a negative slope of the mass ratio
distribution ($\kappa=-1$), then the fraction of binary systems that
merge after the end of the primary MS increases compared to our
fiducial case. 
When a binary system merges, it cannot eject a walkaway or runaway.
Finally, assuming no natal kicks ($\sigma_\mathrm{kick}=0$), the only
systems disrupted are those where the SN ejecta consist of more than
half of the total mass of the binary at the pre-explosion stage. This
requirement naturally biases the population of ejected companions
towards low masses. If we assume that the kick velocity is not scaled down by the amount
of fallback (``velocity kick'' for BHs), CC events from more massive primaries can also disrupt
the system. On average, this produces more massive ejected
secondaries compared to our fiducial case. Since the remaining lifetimes are shorter, the
distance traveled by the population of ejected secondaries is
also decreased, despite their higher velocity. 

Our assumptions \newtext{about} the stability of case B RLOF (assumed to be stable
if the accretor-to-donor mass ratio is \newtext{larger} than $q_\mathrm{crit,B}$) impact the
velocity and mass distribution of the ejected stars. The sequence from
more stable to more unstable
$q_\mathrm{crit,B}=1,\,0.5,\,0.4$\,(fiducial
value),\,0 shows progressively higher $\mathcal{R}$ indicating
relatively \newtext{fewer} runaways, but higher average velocity $\langle v\rangle$ (and
consequently also a larger $\langle L \rangle$). The increase in velocity and
decrease in average mass of the ejected companions for higher
$q_\mathrm{crit,B}$ indicates an increased importance of evolutionary
channels involving a common-envelope evolution leading to tighter
pre-CC orbits without significant accretion of mass from the companion.

The average distance traveled by walkaways and runaways is also a
robust prediction. The entire population of ejected stars typically
travels slightly farther than $\langle L\rangle\simeq120$\,pc, which
corresponds to an average distance for the runaways of $\langle
L\rangle_\mathrm{run}\gtrsim500$\,pc, while the walkaways alone reach
$\langle L\rangle_\mathrm{run}\gtrsim100$\,pc.

These distances are
significantly shortened only assuming fully conservative mass transfer
($\beta_\mathrm{RLOF}=1$), or assuming higher angular momentum loss
(e.g., through a circumbinary disk, $\gamma_\mathrm{RLOF}=\gamma_\mathrm{disk}$). The former results in
higher masses of the ejected secondaries and thus smaller velocities
and shorter lifetimes,
while the latter results in more orbital shrinking and mergers. Another variation
resulting in small $\langle
L\rangle_\mathrm{walk}\simeq66\,\mathrm{pc}$ is that with zero natal kicks
($\sigma_\mathrm{kick}=0\,\kms$). As mentioned above, only the widest pre-CC binary can be
disrupted in this case, so producing a much slower population of ejected
MS stars.

The predicted rotational velocity of the ejected companions is also a
prediction \newtext{robust against parameter variations}. The vast majority of massive
($M_\mathrm{dis}\geq7.5\,M_\odot$) main sequence runaways accrete mass
from their companions, and thus spin up to critical rotation, before
being ejected. Visual inspection of plots like the one shown in
\Figref{fig:vrot_vrw} for our fiducial simulation show no variation
when changing the initial rotation rate of the stars, confirming that
mass transfer in binaries overwrites the initial rotation of the
accretor. 

\subsection{The mass function of massive runaways as probes for black hole kicks}
\label{sec:bh_kicks_mf_rw}
\begin{figure}[!bp]
  \centering
  \includegraphics[width=.5\textwidth]{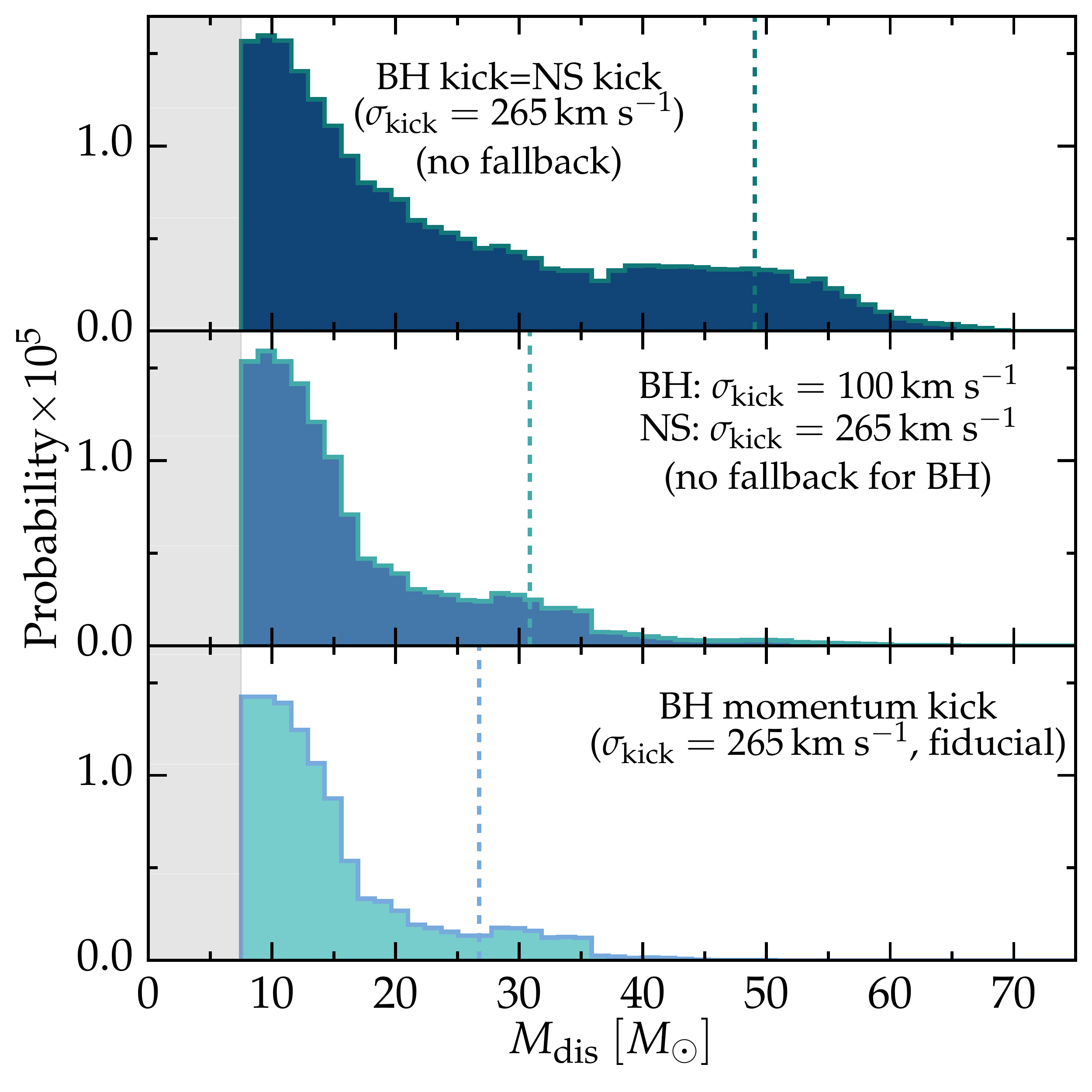}
  \caption{The mass function of massive ($M_\mathrm{dis}\geq7.5\,M_\odot$), runaway
    (faster than $\geq30\,\mathrm{km \ s^{-1}}$) stars depends on the
    assumptions for the BH kick. The bottom panel
    shows our fiducial simulation
    with $\sigma_\mathrm{kick}=265\,\mathrm{km \ s^{-1}}$, including the fallback
    downscaling of the kick both for NSs and BHs. The fallback is an
    important effect only for BHs. The central panel
    shows the mass function for intermediate BH kicks
    ($\sigma_\mathrm{kick}=100\,\mathrm{km\ s^{-1}}$, no fallback
    downscaling) and our fiducial kick for NSs. The top panel shows
    the distribution with our fiducial kick amplitude, but without fallback downscaling (i.e.,~BH
    velocity kick). The dashed lines indicate the $90^\mathrm{th}$
    percentiles of the mass distributions.}
  \label{fig:bh_kicks}
\end{figure}

\begin{figure*}[htbp]
  \centering
   \includegraphics[width=0.33\textwidth]{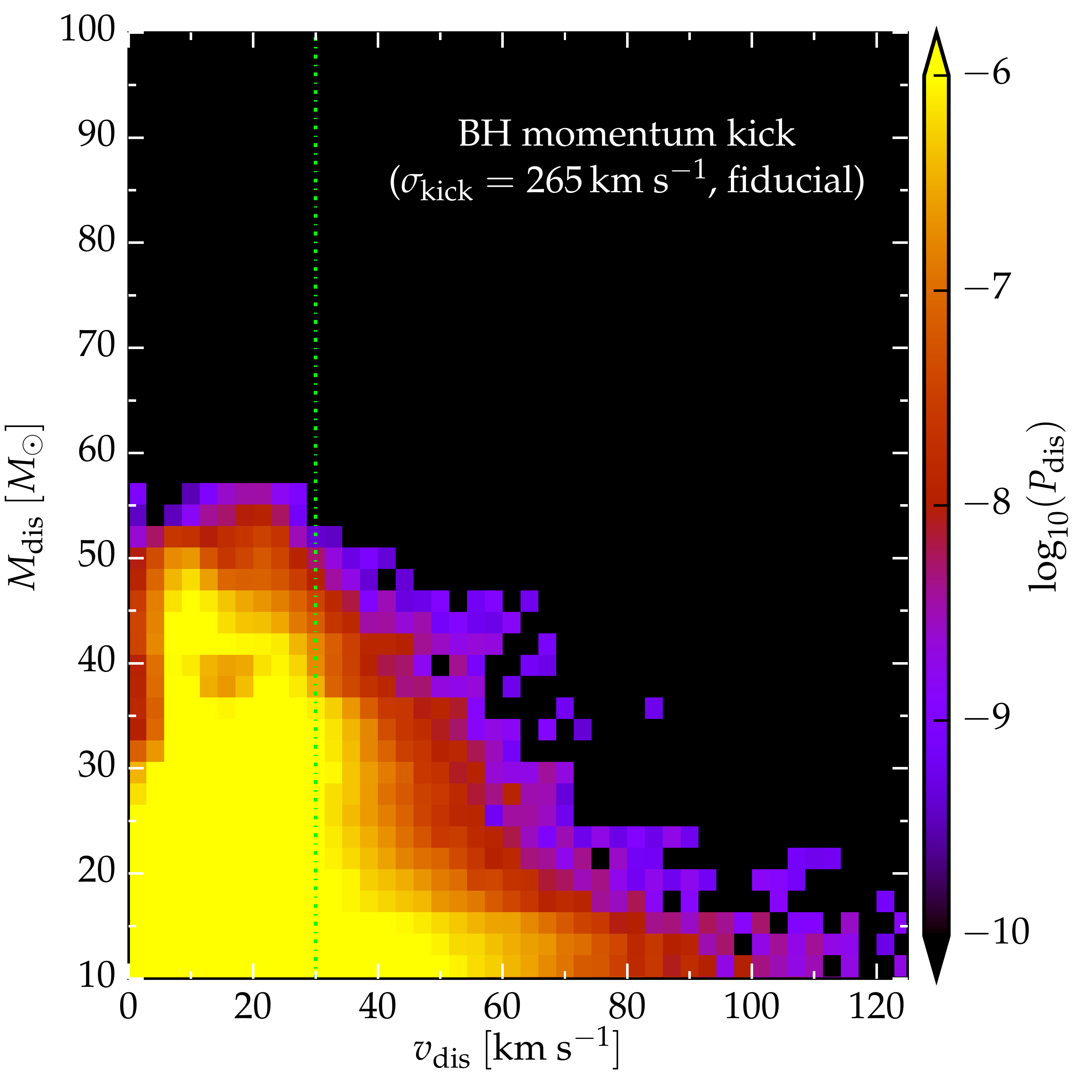}
   \includegraphics[width=0.33\textwidth]{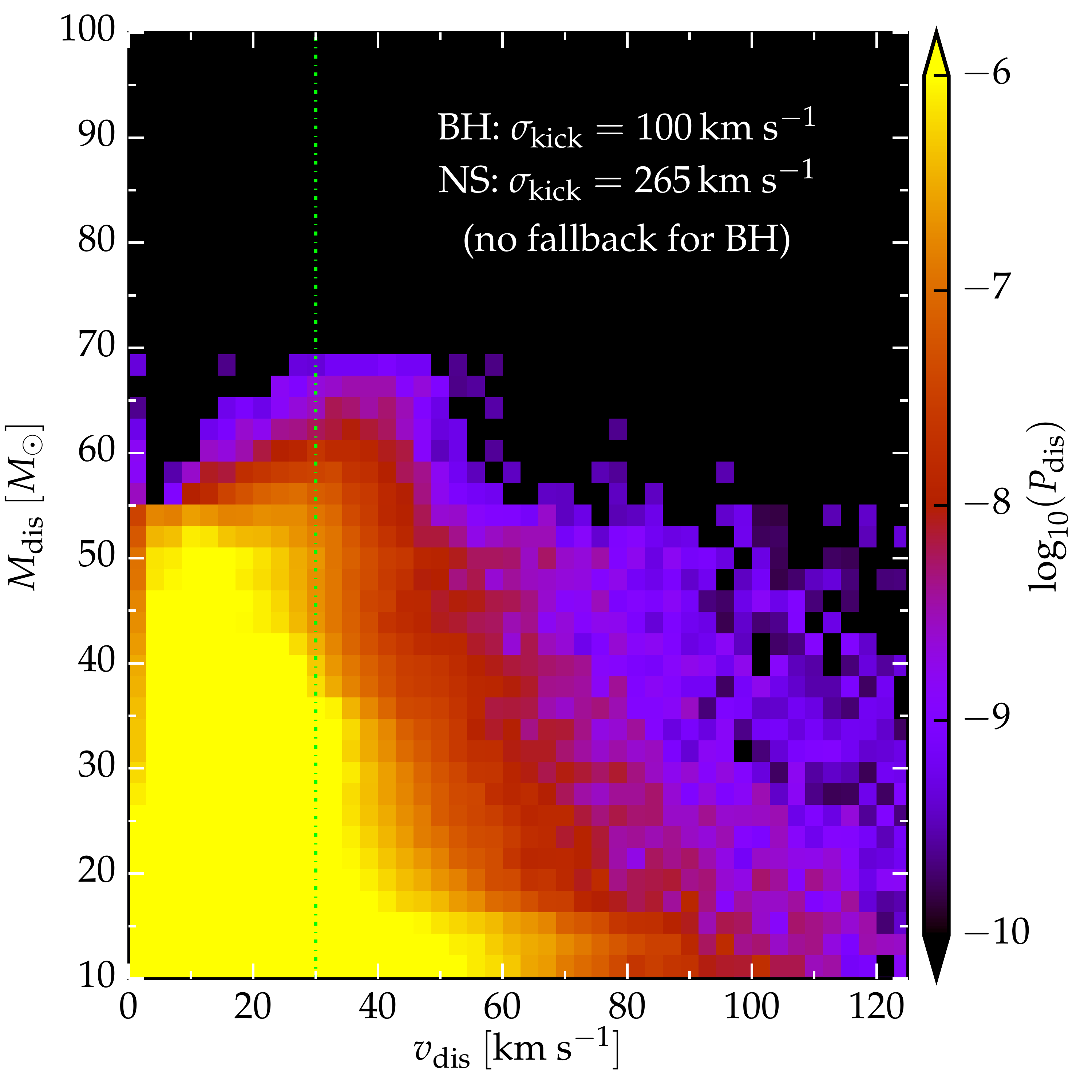}
   \includegraphics[width=0.33\textwidth]{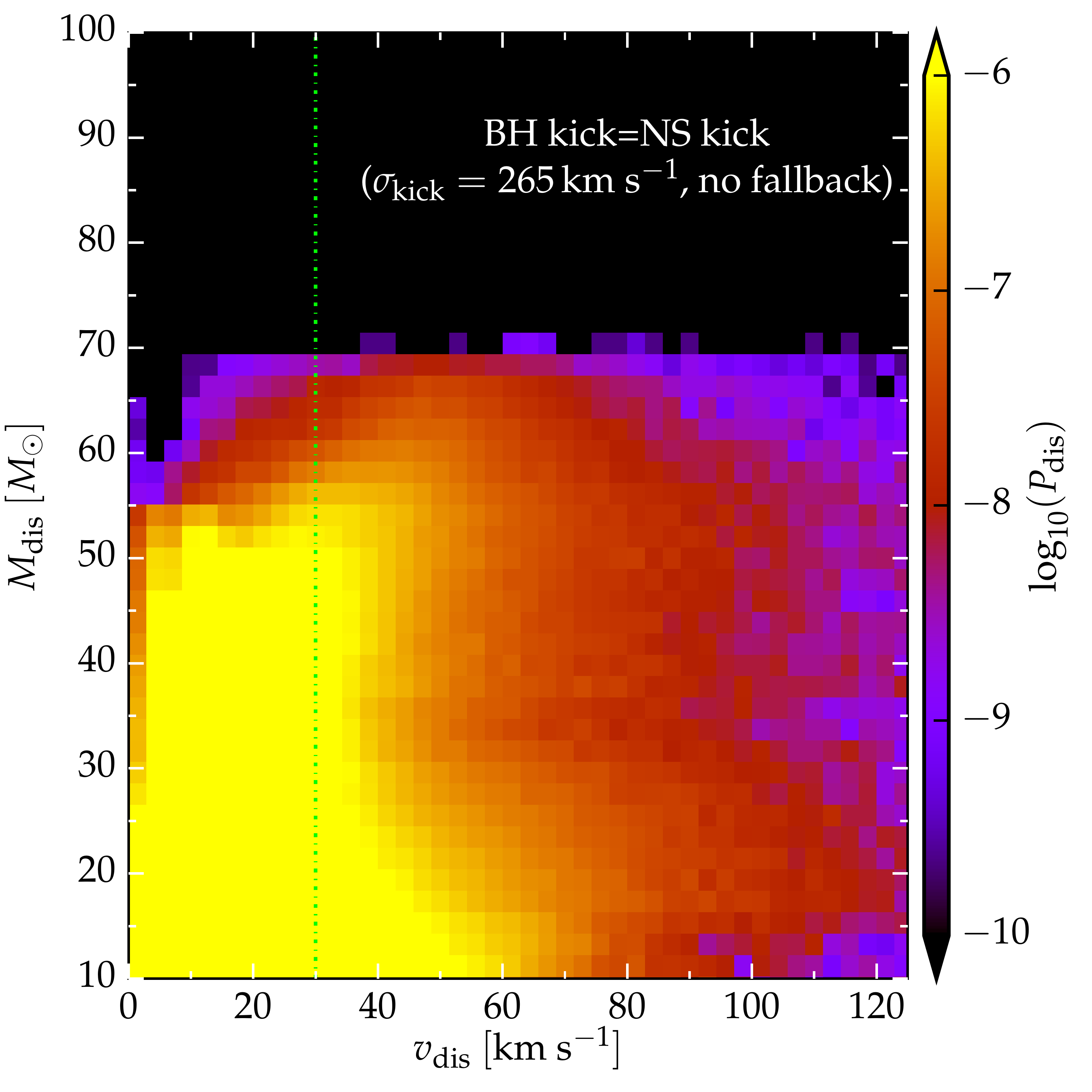}
   \caption{Mass-velocity correlation for the three BH kick variations
     discussed in \Secref{sec:bh_kicks_mf_rw}. The brighter colors indicate
     where \newtext{the probability per system is higher} on this plane, the
     colors have a logarithmic scale. The left panel shows
     our fiducial distribution, the central panel shows the case for
     intermediate BH kicks ($\sigma_\mathrm{kick}=100\,\mathrm{km\
       s^{-1}}$ and no fallback), and the right panel shows large BH
     kicks (same as for NSs). The NS kick is effectively the same in
     all panels, since the amount of fallback is very small for SNe
     producing a NS.}
   \label{fig:M_vrw}
\end{figure*}

\newtext{We find that} the mass distribution of runaways carries
information about the BH kicks. With a sufficiently large sample, it might be used to
discriminate between BH velocity or momentum kicks, at least in a
statistical sense. We focus on the mass function of massive
($M_\mathrm{dis}\geq7.5\,M_\odot$) and fast runaways
($v_\mathrm{dis}\geq30\,\mathrm{km \ s^{-1}}$)
only, i.e., not including the more common walkaways, since we expect
upcoming observational distributions to provide cleaner samples for
these. Ejected companions with high masses are more likely to come from
binaries with an initially very massive primary, so resulting in BH
formation at the end of its evolution.
The constraints on the amplitude of BH kicks from unbound companions are
complimentary to those from binaries surviving the first CC, and \emph{Gaia}
might be able to probe both populations of bound and ejected
companions \citep[][]{breivik:17}. 

\Figref{fig:bh_kicks}
shows the mass distribution of runaways
($v_\mathrm{dis}>30\,\mathrm{km\ s^{-1}}$) more massive than $7.5\,M_\odot$ for three BH
formation scenario. The bottom panel shows our fiducial
simulation, in which natal kicks are damped by fallback (effectively,
this is equivalent to a BH ``momentum kick''). In this case, the $90^\mathrm{th}$
percentile of the massive runaway mass function is $\sim$$25\,M_\odot$.
In this model, the amount of fallback and the amplitude of the BH natal kicks
are degenerate parameters. Because of the large fallback fraction
prescribed by the \cite{fryer:12} algorithm, the runaway mass function
produced by this model is practically indistinguishable from that with
no BH natal kicks: \newtext{almost all the runaways produced come from
NS-forming supernovae.}

The central panel of \Figref{fig:bh_kicks} shows the mass distribution
assuming no fallback downscaling, but a smaller kick amplitude
distribution characterized by $\sigma_\mathrm{kick}=100\,\mathrm{km \
  s^{-1}}$ for the BHs (while we keep the same fiducial assumptions
for NSs). Compared to our fiducial
simulation, BHs receive larger kicks, and thus the runaways produced
are also generally more massive, and the $90^\mathrm{th}$ percentile
of the mass distribution shifts to $\sim$$30\,M_\odot$. 

The histogram in the top panel of \Figref{fig:bh_kicks} instead corresponds
to a scenario in which the natal kick is \emph{not} decreased by the amount of
fallback (e.g., if the natal kick is entirely due to asymmetries in the neutrino
emission). In this case BHs and NSs receive the same kick amplitude
(BH ``velocity kick'') and this results in a larger number of massive
runaways characterized by a mass function skewed towards larger
values. In this scenario, the $90^\mathrm{th}$ percentile of the mass
distribution is as high as $\sim$$50\,M_\odot$.

In summary, the tail of the mass function of massive runaways carries
information about the BH natal kicks: finding runaways of large masses
would support the possibility of large BH natal kicks. We do not
expect large contamination of this tail from dynamical ejection
of runaway stars, since dynamical channels tend to eject the least
massive star among those interacting \citep[although see
also][]{fujii:11, banerjee:12, oh:16}.

Our result allow us to check also for a correlation between the mass
of the ejected companions and its velocity. \Figref{fig:M_vrw} shows
this comparison for the three BH kick variations of
\Figref{fig:bh_kicks}. As expected, there is a correlation between the
maximum velocity that can be achieved because of
the binary disruption, and the mass of the ejected MS companion. This
can be seen by the shape of the colors in each panel. More massive companions have more
inertia, or, in other words, more massive
companion have slower pre-CC orbital velocity because of the mass
dependent factor in \Eqref{eq:v2}.

From left to right, the BH kick amplitude increases (while effectively
keeping the same NS kick, since the fallback fraction is small for
CCSNe resulting in NS formation), and so does the maximum mass of the
ejected companions. There is an sharp drop in the number of ejected
companions for $M_\mathrm{dis}\gtrsim70\,M_\odot$. \newtext{There is a
set of plausible reasons to explain this}. First of all,
stars in this mass range are intrinsically rare because of the
IMF. Also, the lifetimes of very
massive stars become very similar ($\sim3\,\mathrm{Myr}$) above $\sim$$50\,M_\odot$. This
means that by the time the primary collapses, the secondary is also an
evolved star that has finished burning hydrogen in its core, and it
does not appear in the panels of \Figref{fig:M_vrw}. Finally,
these stars tend to have large wind mass loss rates, which can be
enhanced significantly when the star is spun up by the accretion: the
combination of the wind enhancement with already large wind mass loss
rates limits the growth in mass of the companions.

In an observed
population, the short post-ejection lifetime would also make very
massive ejected companions rare. However, the upperlimit of
$\sim$$70\,M_\odot$ is not strict, and it is possible to produce
walkaways in this mass range through binary interactions, although at
a negligible rate compared to less massive walkaways.

\subsection{Effects of metallicity }
\label{sec:Z}

We find that, other than the natal kick distribution, metallicity is the most important parameter
influencing our results. Decreasing the metallicity from super-solar,
$Z=0.03$, to $Z=0.02\simeq Z_\odot$ (fiducial run), to
$Z=0.008$, $Z=0.0047$ to $Z=0.0002\simeq10^{-2}Z_\odot$ a clear trend
emerges from \Tabref{tab:parameters}. At lower metallicity, the
fraction $\mathcal{D}$ of binaries disrupted by the first CC
is lower ($77\%$ for the lowest metallicity, cf. $88\%$ at the highest), but the stars ejected are on average
faster. This can be seen from the average velocity $\langle v\rangle$,
which goes from $11.2\,\mathrm{km\ s^{-1}}$ to $21.6\,\mathrm{km \ s^{-1}}$, but also from the progressive decrease of the ratio of walkaways per runaway
$\mathcal{R}$, regardless of the lower mass cut. Nevertheless,
walkaways outnumber runaways at all the metallicities we explore.

In our lowest metallicity run, the runaways \newtext{($v_\mathrm{dis}\geq30\,\kms$)} travel, on average, as far as 706\,pc from their birth
location, and the walkaways reach, on average, 163\,pc. The mean
distance travelled by population of disrupted stars is 279\,pc for
the lowest metallicity.

The reduced likelihood of disrupting a system and increased velocity
of the stars ejected can be understood in
terms of the effect of metallicity on stellar radii. At lower $Z$
stars of a given mass and evolutionary stage have smaller radii. This means that a given
binary will enter into a mass transfer phase later, resulting in less conservative mass transfer and 
less orbital widening, making the binary harder to
disrupt and the pre-CC orbital velocity of the secondary
faster. This is corroborated by the fact that the average mass of
walkaways and runaways decreases slightly with metallicity, which is indicative of
a less conservative mass transfer. Moreover, at lower metallicity, the post-interaction orbital widening
due to stellar winds is less important.

Another important physical effect, presently not included in
our simulations, is that, at the end of the RLOF, the primary remains more massive at
lower metallicity \citep[owing again to the smaller radii, e.g.][]{gotberg:17,yoon:17}. This
should contribute to the reduction of orbital widening for interacting binaries,
and make the collapsing star more massive and thus the disruption harder.

\newgeometry{margin=1cm}
\begin{landscape}
  \begin{table*}[ht]
    \footnotesize
    \centering
    \caption{Outcome of our model variations. Average distance traveled
      from the ejection to CC for all massive companions ejected, for massive runaways only, and
      massive walkaways only ($\langle L \rangle $, $\langle L_\mathrm{run} \rangle$,
  $\langle L_\mathrm{walk} \rangle $), average and median velocity
  $\langle v\rangle$ and $v_\mathrm{med}$, average
  masses for ejected main sequence companions, runaways only, and
  walkaways only ($\langle M_2 \rangle $, $\langle M_2^\mathrm{run} \rangle$,
  $\langle M_2^\mathrm{walk} \rangle $), produced ratio
  $\mathcal{R}$ of  walkaways to runaways for all masses, above $7.5\,M_\odot$, and
  above $15\,M_\odot$, fraction of all the binaries simulated resulting in a merger
  $\mathcal{M}$, disruption fraction among the non-merging binaries $\mathcal{D}$, and fraction
  of runaways and walkaways with masses larger than $15\,M_\odot$ from the disruption of binaries in a
  steady state population ($f^\mathrm{RW}_{15}$, $f^{WA}_{15}$). \label{tab:parameters}}
\begin{tabular}{lr|c|ccccccccccccccc}\hline\hline 
  \multirow{2}{*}{Physical Assumptions}  & \multirow{2}{*}{Parameter}
  & \multirow{2}{*}{value} & $\langle L \rangle $ & $\langle
                                                    L_\mathrm{run}
                                                    \rangle$ &
                                                               $\langle
                                                               L_\mathrm{walk}
                                                               \rangle
                                                               $ &
                                                                   $\langle
                                                                   v\rangle$
  & $v_\mathrm{med}$ & $\langle M_2 \rangle$ & $\langle
                                               M_2^\mathrm{run}
                                               \rangle$ & $\langle
                                                          M_2^\mathrm{walk}
                                                          \rangle$ &
                                                                     \multirow{2}{*}{$\mathcal{R}_\mathrm{MS}$}
  & \multirow{2}{*}{$\mathcal{R}_{7.5}$} &
                                           \multirow{2}{*}{$\mathcal{R}_{15}$}
  & \multirow{2}{*}{$\mathcal{M}$} & \multirow{2}{*}{$\mathcal{D}$} & $f^\mathrm{RW}_\mathrm{15}$ & $f^\mathrm{WA}_\mathrm{15}$\\

& & & $\mathrm{[pc]}$ & $\mathrm{[pc]}$ & $\mathrm{[pc]}$ &
                                                            $\mathrm{[km\
                                                            s^{-1}]}$
                                                                                                                       &
                                                                                                                         $\mathrm{[km\
                                                                                                                         s^{-1}]}$
                                                                                                                                          &
                                                                                                                                            $[M_\odot]$
                                                                                                                                                                  &
                                                                                                                                                                    $[M_\odot]$
                                                        & $[M_\odot]$
                                                                   &
  & & & & & [\%] & [\%]  \\

\hline
\hline

\multicolumn{2}{l|}{Fiducial population}  &  see \Secref{sec:methods}  & 126 & 584 & 103 & 12.4 & 10.4 & 10.8 & \phantom{0}9.8 & 10.9 &20.1 &19.9 & 26.7& 0.22 & 0.86& 0.5 & 10.1 \\[3pt] 

\hline

\multirow{3}{*}{Mass transfer efficiency} &\multirow{3}{*}{$\beta_\mathrm{RLOF}$}  & 0 & 139 & 493 & \phantom{0}98 & 15.1 & 11.6 & 7.7 & 8.6 &\phantom{0}7.6 & 10.3 &  \phantom{0}8.7 & \phantom{0}5.8& 0.23&0.86 & 0.3 & \phantom{0}1.5\\[3pt]  

 &  &0.5 & 122 & 363 & 111 & 12.8 & 10.5 & 10.2 & 12.0 & 10.1 & 21.7 &20.7& 11.0& 0.22&0.87  & 1.2 & \phantom{0}8.6 \\[3pt]

 & & 1 & \phantom{0}82 & 214 & \phantom{0}80 & 11.2 & \phantom{0}9.8 & 12.1 & 11.9 & 12.1 & 33.0 &  46.9 & 26.6& 0.21&0.87 & 0.7 & 14.7\\[3pt]

\hline

\multirow{2}{*}{Angular momentum loss} &
                                         \multirow{2}{*}{$\gamma_\mathrm{RLOF}$}
                                                                      &
                                                                                               $\gamma_\mathrm{disk}$&
                                                                                                                       \phantom{0}72
                                                  & 544 & \phantom{0}64 & 10.7 &
                                                                      \phantom{0}9.8
                     & 10.5 & \phantom{0}6.4 & 10.7 & 22.6 &  59.0 &
                                                                     53.1& 0.36&0.85 & 0.2 & \phantom{0}7.3 \\[3pt]

 & & 1 & 129 & 592 & 104 & 12.6 & 10.5 & 10.8 & 10.0 & 10.9 & 19.3 &
                                                                     18.5
                                         & 25.3& 0.22&0.86 & 0.6 & \phantom{0}9.9\\[3pt]

\hline

\multirow{2}{*}{Common envelope efficiency} &
                                              \multirow{2}{*}{$\alpha_\mathrm{CE}$}
  &0.1 & 126 & 581 & 103 & 12.4 & 10.3 & 10.9 & 11.3 & 10.9 & 23.8 &
                                                                     19.8
                                         & 26.3& 0.23& 0.86 & 0.5 & 10.1\\[3pt]

 & & 10 & 135 & 730 & 103 & 12.7 & 10.5 & 10.6 & \phantom{0}7.5 & 10.9
                                                                   &
                                                                     12.6 &  18.7 & 26.5& 0.16 &0.84 & 0.5 & 10.0\\[3pt]

\hline

\multirow{2}{*}{Mass ratio for case A merger} &
                                                \multirow{2}{*}{$q_\mathrm{crit,\
                                                A}$} &0.80 & 128 & 593
                                                             & 104 &
                                                                     12.3
                                                                                                                       &
                                                                                                                         10.3 & 10.7 & \phantom{0}9.6 & 10.7 & 20.1 &  19.9 & 27.5& 0.24 & 0.86 & 0.5 & 10.2\\[3pt]

 & & 0.25 & 134 & 637 & 102 & 13.0 & \phantom{0}9.4 & 11.1 & 10.3 &
                                                                    11.1 & 17.5 &  15.9 & 23.0& 0.17&0.86 & 0.6 & \phantom{0}9.4\\[3pt]

\hline

\multirow{3}{*}{Mass ratio for case B merger} & \multirow{3}{*}{$q_\mathrm{crit,\ B}$} &1.0 & 100 & 407 & 100 &\phantom{0}9.5& 10.8 & \phantom{0}8.6 & \phantom{0}3.5 & \phantom{0}8.8& -- &  -- & -- & 0.43& 0.89 & -- & \phantom{0}5.0\\[3pt]
 & & 0.5 & 103 & 717 & 102 & \phantom{0}9.1 & \phantom{0}9.3 & \phantom{0}8.3 & \phantom{0}3.0 & \phantom{0}8.4 & -- & -- &--& 0.48&0.89& --& \phantom{0}5.0\\[3pt]
 & & 0.0 & 167 & 870 & 107 & 13.8 & 10.1 & 10.7 & 9.4 & 10.8 & 12.4 &11.6 & 24.5& 0.16&0.85 &  0.6 & 10.1\\[3pt]
  \hline
Mass ratio for case C merger & $q_\mathrm{crit,\ RSG}$ & 1.0 & 123 &
                                                                     509
                                                             & 82 &
                                                                    14.6
  & 11.4 & 9.9 & 8.1 & 10.2 & 7.9 &  9.5 & 17.0& 0.31 & 0.80 & 0.6 & 6.3 \\[3pt]
\hline
{Mass ratio for case C merger and} & \multirow{2}{*}{$(q_\mathrm{crit,\ RSG},
                \alpha_\mathrm{CE})$} & (1.0, 0.1) & 109 & 626 & 79 &
                                                                      13.3
  & 10.5 & 10.4 & 12.3 & 10.3 & 23.7 &  17.7 & 19.4& 0.41 &0.82 & 0.5
                                                                                                  & 6.4
                                                                   \\[3pt]
{ Common Envelope
  efficiency}  & & (1.0, 10) & 177 & 764 & 91 & 17.9 & 13.2 & 9.6 & 7.9 & 9.9 &
                                                                     6.0
                                         &  6.8 & 12.2& 0.18 &0.78 &
                                                                     0.7
                                                                                                  & 6.4 \\[3pt]                               
\hline

\multirow{3}{*}{Natal kick amplitude} &
                                       \multirow{3}{*}{$\sigma_\mathrm{kick}
                                        \ [\mathrm{km\ s^{-1}}]$}
  &0 & \phantom{0}66 & -- & \phantom{0}66 & \phantom{0}3.8 & 10.2 &
                                                                    \phantom{0}1.6
                                             & -- & \phantom{0}1.6 &
                                                                     --
  &  -- & --& 0.22 & 0.16 & -- & -- \\[3pt]

 & & 300 & 128 & 572 & 103 & 12.7 & 10.4 & 11.0 & 10.3 & 11.0 & 18.7 &
                                                                       18.0
                                         & 22.9& 0.22&0.87 & 0.6 & 10.3 \\[3pt]
  
 & & 1000 & 132 & 490 & 102 & 13.9 & 10.9 & 11.7 & 13.8 & 11.5 & 13.8
  &  12.0 & 11.6& 0.22 & 0.91 & 1.2 & 11.2\\[3pt]

\hline
\multirow{2}{*}{BH kicks
  $\sigma_\mathrm{kick}=\left\{\phantom{\frac{0}{0}}\right.$} &
                                                                \hspace*{-100pt}{$100\,\mathrm{km\
                                                                s^{-1}},\
                                                                (f_b=0$)} &  for
                                                    BH &

\multirow{2}{*}{121} & \multirow{2}{*}{395} & \multirow{2}{*}{97} &
                                                                    \multirow{2}{*}{13.9}
  & \multirow{2}{*}{10.7} & \multirow{2}{*}{12.8} &
                                                    \multirow{2}{*}{18.7}
                                                        &
                                                          \multirow{2}{*}{12.4}
                                                                   &
                                                                     \multirow{2}{*}{13.7}
  &  \multirow{2}{*}{11.3} & \multirow{2}{*}{9.0} & \multirow{2}{*}{0.22} &
                                                    \multirow{2}{*}{0.97} & \multirow{2}{*}{0.1} & \multirow{2}{*}{12.1}\\[3pt]
 & $265\,\mathrm{km\ s^{-1}}$ & for NS & &
                                                             & & & & &
                                                        & & & & & & & \\[3pt]

\hline

\multirow{2}{*}{Double maxwellian with
  $\sigma_\mathrm{kick}=\left\{\phantom{\frac{0}{0}}\right.$} & $30\,\mathrm{km\ s^{-1}}$ &  for
                                                    $M_\mathrm{NS}\leq1.35\,M_\odot$
  & \multirow{2}{*}{105} & \multirow{2}{*}{423} & \multirow{2}{*}{\phantom{0}87} & \multirow{2}{*}{12.2} & \multirow{2}{*}{\phantom{0}9.5} & \multirow{2}{*}{10.2} & \multirow{2}{*}{14.5} &
                                                                      \multirow{2}{*}{10.0}
                                                        & \multirow{2}{*}{30.0} &  \multirow{2}{*}{17.5}
  & \multirow{2}{*}{15.5}& \multirow{2}{*}{0.22}&\multirow{2}{*}{0.65} &  \multirow{2}{*}{0.5} & \multirow{2}{*}{\phantom{0}4.9}\\[3pt]

 & $265\,\mathrm{km\ s^{-1}}$ & for $M_\mathrm{NS}>1.35\,M_\odot$ & &
                                                             & & & & &
                                                        & & & & & & & \\[3pt]

\hline

\multicolumn{2}{l|}{\multirow{2}{*}{Restricted kick directions}}
                                         &$\alpha < 10\,\mathrm{deg}$
  & 127 & 582 & 103 & 12.5 & 10.5 & 10.9 & \phantom{0}9.7 & 11.0 &
                                                                   19.8 &  19.4 & 27.4& 0.22&0.87 & 0.6 & 10.3\\[3pt]

 & & $90\,\mathrm{deg}-\alpha < 45\,\mathrm{deg}$ & 128 & 572 & 102 &
                                                                   12.5
                                                                                                                       &
                                                                                                                         10.4
                     & 10.8 & 10.7 & 10.8 & 18.1 &  17.2 & 19.4& 0.22&0.86 & 0.5 & 10.0\\[3pt]

\hline

Fallback fraction &  $f_b$ &0 & 121
                                                  & 395 & \phantom{0}97 & 13.9 &
                                                                      10.7
                     & 12.8 & 18.7 & 12.4 & 13.7 &  11.3 & 9.0 & 0.22&0.97
                                                                                                                & 1.5 & 12.1 \\[3pt]

\hline





\multirow{2}{*}{Period distribution slope} &  \multirow{2}{*}{$\pi$}
  &-1 & 128 & 581 & 102 & 12.9 & 10.7 & 10.6 & 10.1 & 10.6 & 18.7 &
                                                                    17.6
                                         & 22.8& 0.28&0.89 & 0.6 & \phantom{0}8.9\\[3pt]

 & & 0 & 120 & 593 & 102 & 11.5 & \phantom{0}9.8 & 10.8 &
                                                          \phantom{0}9.0 & 10.9 & 23.8 &  26.4 & 35.9& 0.17&0.85 & 0.4 & 10.7\\[3pt]

\hline

\multirow{2}{*}{Initial mass function slope} &
                                               \multirow{2}{*}{$\alpha$}
  &-1.9 & 125 & 554 & 101 & 12.8 & 10.5 & 11.8 & 10.9 & 11.8 & 18.9 &
                                                                      18.0
                                         & 23.9& 0.24&0.80 & 0.6 & \phantom{0}9.1\\[3pt]

 & & -3 & 125 & 626 & 104 & 11.7 & 10.1 & \phantom{0}9.6 &
                                                           \phantom{0}8.3
                                                        &
                                                          \phantom{0}9.6 & 22.6 &  24.2 & 32.7& 0.20 &0.93 & 0.5 & 11.6\\[3pt]

\hline

\multirow{2}{*}{Mass ratio slope} &  \multirow{2}{*}{$\kappa$} &-1 &
                                                                     163
                                                  & 629 & 130 & 13.2 &
                                                                       11.1
                     & \phantom{0}7.8 & \phantom{0}6.8 &
                                                         \phantom{0}7.9 & 14.5 &  14.5 & 20.7& 0.36&0.86 & 0.4 & \phantom{0}6.3\\[3pt]
 & & 1 & \phantom{0}99 & 526 & \phantom{0}84 & 11.7 & \phantom{0}9.7 &
                                                                       13.0 & 12.7 & 13.1 & 28.3 &  27.4 & 33.2& 0.14&0.87 & 0.5 & 12.4\\[3pt]

\hline

\multirow{4}{*}{Metallicity} &  \multirow{4}{*}{$Z$} &0.0002 & 279 &
                                                                     706
                                                             & 163 &
                                                                     21.6
  & 16.9 & 9.2 & 10.8 & \phantom{0}8.9 & \phantom{0}4.4 &
                                                          \phantom{0}3.7
                                         & 3.0& 0.15&0.77 & 2.6 & \phantom{0}7.7\\[3pt]

 & & 0.0047 & 190 & 718 & 130 & 15.9 & 12.5 & 10.5 & 11.0 & 10.4 &
                                                                   10.7 &  8.8 & 11.1& 0.19&0.84 & 1.2 & 10.3\\[3pt]

                                         & & 0.008 & 164 & 682 & 123 &
                                                                       14.4
  & 11.9
  & 10.5 & \phantom{0}9.9 & 10.6 & 13.7 &  12.8 & 17.4& 0.21&0.84 & 0.1& 11.0\\[3pt]
  
 & & 0.03 & 109 & 576 & \phantom{0}87 & 11.6 & \phantom{0}9.7 & 11.2 &
                                                                       \phantom{0}9.9
                                                        & 11.3 & 22.5
  &  21.6 & 32.6& 0.24 & 0.88 & 0.5 & 10.0\\[3pt]

\hline

\multicolumn{2}{l|}{Initial spin distribution} & \citetalias{ramirez-agudelo:15} & 126 & 582 & 103
                                                             & 12.4 &
                                                                      10.4 & 10.9 & \phantom{0}9.8 & 10.9 & 20.1 &  19.8 & 26.5& 0.22&0.86 & 0.5 & 10.1\\[3pt]

  \hline
  
\end{tabular}
\end{table*}
\end{landscape}
\restoregeometry


\begin{figure*}[!htbp]
  \centering
  \includegraphics[width=\textwidth]{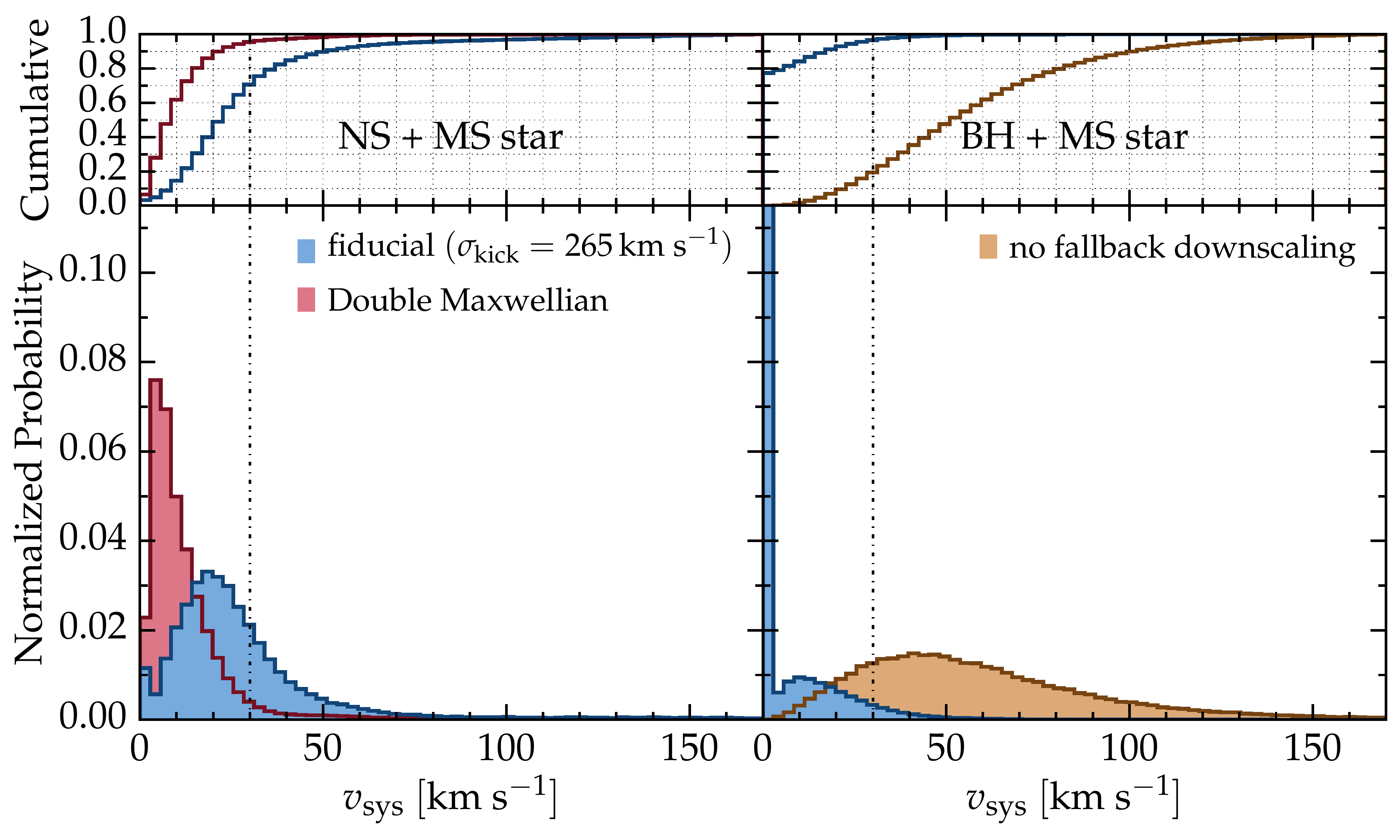}
  \caption{Normalized distribution of systemic velocities of binaries
    with a MS star remaining bound after the CC of the first star. The
    distributions in both panels are normalized to unity. The left panel shows
    the distribution for NS companions, the right panel shows only
    binaries with a BH companion. The blue histogram
    represents our fiducial simulation. The orange solid line gives the
    distribution when the kick is not rescaled because of
    fallback, which changes the BH kicks, but not the NS kicks. The red distribution corresponds to a double Maxwellian
    kick distribution, which changes the NS kicks but not the BH kicks. The top panels show the corresponding cumulative distributions.}
  \label{fig:bound_v_dist}
\end{figure*}

\section{Systemic velocities of NS and BH binaries}
\label{sec:bound}

Our simulations also provide predictions for the systemic velocities
of binary systems that remain bound after the first CC
\citep[e.g.,][]{vanoijen:89}. \newtext{The results of our simulations,
  which will be made available upon publication,
  also provide pre-CC and post-CC separations, eccentricities, and mass
  ratios.}

\newtext{The systems with a MS star and a compact object remaining
  bound are} only \mbox{$1-\mathcal{D} = 14_{-10}^{+14}\%$}
\newtext{of the total} (see \Secref{sec:param_variation} for a
discussion of model variations).  These are of large interest since
they can give rise to X-ray binaries if the compact object starts to
accrete from the companion, \newtext{and later possibly double compact
objects}.  They typically originate from systems where the newly formed compact object received a small natal kick. This bias limits the extent to which these special systems can be used to infer conclusions about the properties of the general population of NS and BH.  For comparison, the average effective natal kick for the compact objects formed in systems that remain bound is  $\sim$$66\,\mathrm{km\ s^{-1}}$, much lower than the average of $\sim$$330\,\mathrm{km\ s^{-1}}$ (as drawn from our standard Maxwellian after rescaling to account for fallback). 

Systems that remain bound also obtain a systemic velocity, because of
mass lost during the supernova explosion
while orbiting around the center of mass.  We define the newly
obtained systemic velocity $v_\mathrm{sys}$ as the velocity of the new
center of mass (of the compact object plus the secondary star) in the
frame of the old center of mass (e.g.,
\citealt{brandt:95}, \citealt{kalogera:96}, \citetalias{tauris:98}).  

Figure~\ref{fig:bound_v_dist} shows the normalized distribution of
systemic velocities $v_\mathrm{sys}$ for different natal kick
assumptions. The left (right) panel shows NS+MS (BH+MS) binaries, with
the cumulative distributions in the top panels, respectively.  We find that the distribution of systemic velocities are very insensitive to the model variations that we considered, except for the natal kicks. We only show the distributions for variations that show large deviations from the fiducial simulation.   

For NS+MS binaries  our fiducial simulation shows a median systemic velocity of $\sim 20\kms$, for nearly all assumptions that we consider. The strongest deviation is found with a run in which we adopt a double Maxwellian distribution, where low mass progenitors are assumed to form neutrons stars with very low kick velocities.  This substantially increases the fraction of systems that remain bound to $1-\mathcal{D}\simeq35$\%. It primarily adds systems with low systemic velocities. The median of the normalized distribution for this simulation lies near $7 \kms$.  

In our fiducial simulations for BH+MS binaries we obtain a bimodal
distribution with a large peak at $0 \kms$ accounting for nearly 80\%
of all systems and a second broader component peaking at $\sim
10\kms$.  The peak near zero results from progenitors that experience
complete fallback ($f_b=1$ according to the algorithm from
\citealt{fryer:12}). They receive no kick at all, and no mass is lost from
the system (we neglect here, as elsewhere, the mass possibly lost to
neutrinos during the CC, \newtext{which is an $\sim$10\% effect}). For these systems we also find no changes in
the separation and/or eccentricity after the CC of the first star.
The sharp peak at $v_\mathrm{sys}\simeq0\,\kms$ would likely be
smeared out by the velocity dispersion of the region where the system
formed.

Nearly all  variations considered give indistinguishable results. However, we find large deviations \newtext{from} the simulation where we used the extreme assumption that compact objects get kicks of a similar velocity amplitude regardless of their mass (no fallback downscaling, orange in \Figref{fig:bound_v_dist}). This effectively results in much larger kicks for black holes and is sometimes referred to as ``velocity kick''. Only $1-\mathcal{D}\simeq3\%$  of systems remain bound in this simulation.  The few resulting BH binaries typically have large systemic velocities: we find a broad distribution with a median at $50\kms$. 

Our simulations also provide predictions for the distributions of further properties of the bound systems.  
If we exclude systems with $f_b=1$, i.e.\ if we consider only NSs and
BHs formed via fallback but still producing ejecta, we find
an anti-correlation between the maximum post-CC separation and systemic
velocity: the wider the system, the slower the maximum systemic
velocity it can reach \citep[see also][]{brandt:95}. 

Most systems remaining bound come from initially 
relatively tight pre-CC orbits. Typically, they
have a reversed mass ratio
(i.e., at the time of the first CC $M_2^\mathrm{pre-CC}>M_1^\mathrm{pre-CC}$)
because of the previous mass transfer phase. In our fiducial run, for the
systems remaining bound the average pre-CC mass ratio is
$\langle q_\mathrm{pre-CC}\rangle\equiv\langle
M_2^\mathrm{pre-CC}/M_1^\mathrm{pre-CC}\rangle\simeq 2.9$.
The majority of the bound systems, about $60\%$ for all variations
with non-zero natal kicks, have separations of less than
about $500\,R_\odot$. This suggests that these systems will
evolve through a phase during which they might be detectable as X-ray
sources, and possibly even through a mass transfer phase. The tail of
the distribution at large separations \newtext{extends to}
post-CC separation as large as $\sim6000\,R_\odot$.

We do not consider cases in which the compact object is shot within the
Roche lobe of the secondary. A prompt collision between the
newly formed compact object and the companion star could lead to the
formation of a Thorne-Zytkow object \citep[if the compact object is a
NS,][]{thorne:75,thorne:77, leonard:94}, or a transient, possibly involving the disintegration of the companion. For a given binary system, the probability of a prompt collision
caused by the natal kick (whose direction is isotropically distributed
in the frame of the collapsing star) can be estimated as the solid
angle subtended by the companion at the position of the collapsing
star. For our fiducial population, excluding systems that do not
receive a natal kick because of the fallback rescaling, we obtain that
roughly 1 out of 10\,000 CC events in a binary would result in a
prompt collision (i.e., a probability per binary system of about
$10^{-4}$). \newtext{For this estimate, w}e exclude from the normalization systems merging before the first CC.

\section{Astrophysical Implications}
\label{sec:discussion}
\subsection{Other possible contributions to the inferred velocity}
\label{sec:vel_contrib}

\begin{figure}[!hbtp]
  \centering
  \includegraphics[width=0.5\textwidth]{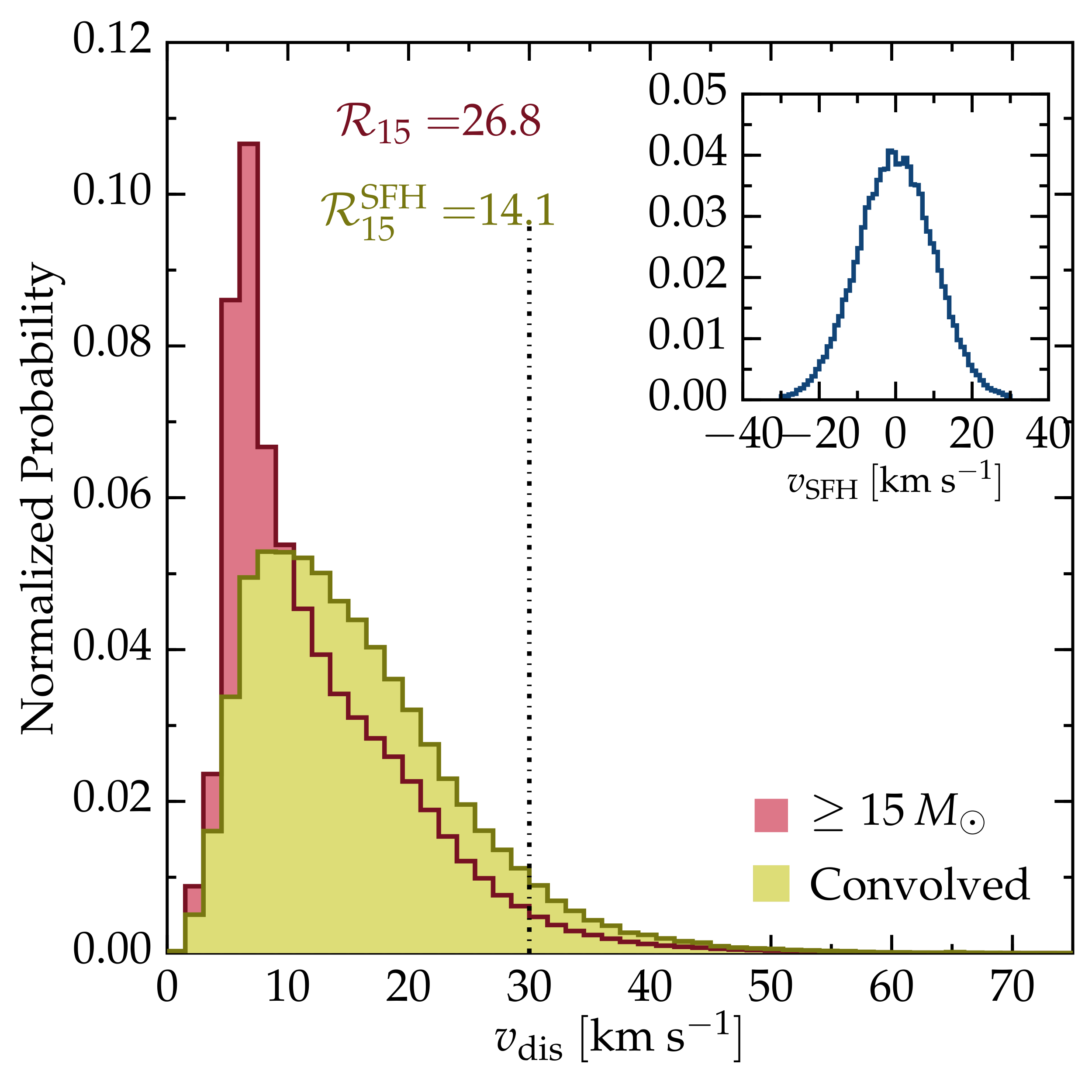}
  \caption{\newtext{The yellow histogram shows the convolution of the distribution of ejected MS
      stars more massive than $15\,M_\odot$ (red histogram,
      cf.~\Figref{fig:v_dist}) with a Gaussian with
      full-width-half-maximum of $10\,\kms$
      representing (an upper limit to) the velocity dispersion of a star forming region
      (shown in the inset plot). In this figure, the red histogram is
      normalized so that its surface area is one.}}
  \label{fig:convolution}
\end{figure}

\newtext{Throughout this study, we have used velocities calculated in
  the rest frame of the initial binaries. In reality, the progenitor
  massive binaries are likely to form in high-mass star forming
  regions with a certain velocity dispersion, typically smaller than
  $10\,\kms$ \citep[][]{debruijne:99,steenbrugge:03, kiminki:18}. Moreover,
  the high-mass star forming regions have been observed to have a systematic lag velocity of about
  $\sim$ $5\,\kms$ compared
  to the rotation of the Galactic disk \citep[][]{reid:14}.

  In an observed sample using the putative parent association to
  define a frame of reference, the ``thermal'' velocity of the progenitor
  binary within the star-forming region will add to the
  velocity resulting from the disruption. To
  illustrate how this affects the velocity
  distribution of \Figref{fig:v_dist}, we numerically convolve the
  distribution for ejected stars more massive than $15\,M_\odot$ (red)
  with a Gaussian distribution
  with full-width-half-maximum of 10\,$\kms$, assumed to be an
  upperlimit of the typical velocity dispersion of OB associations \citep[][]{debruijne:99,steenbrugge:03, kiminki:18}. We
  note that both the orientation of the ``thermal'' velocity of the parent binary,
  and the velocity acquired by the ejected star are randomly
  oriented, and the latter depends on the binary inclination and phase
  at the first CC event.

  Figure~\ref{fig:convolution} shows the re-sampled distribution and the
  result of the convolution with a Gaussian velocity dispersion
  distribution with dispersion of $10\,\kms$, which we consider an
  over-estimate for OB-associations. While adding a dispersion velocity smears
  out the peak of the distribution and increases the contribution of
  the runaways faster than $30\,\kms$, the effect is insufficient to
  reconcile the observed runaway fraction with our results (see \Secref{sec:howtofind}). The ratio of walkaways
to runaways more massive than $15\,M_\odot$ remains
$\mathcal{R}_{15}^\mathrm{SFH}\sim15$ even accounting for the velocity
dispersion of the parent association
(cf.~$\mathcal{R}_{15}\sim27$ before accounting for it).

  For observed samples using the Galactic disk rotation to define a
  reference frame,
  the peculiar lag velocity of the high-mass star-forming region could potentially contribute to overestimating the ejection velocity of the stars.  
  If the lag velocity of the
  parent association is not removed from the measures, it
  would introduce a systematic shift in one component of the velocity
  of the star (corresponding to the direction of the peculiar motion
  of the parent association). This can be modeled with a convolution
  of our distributions with a Dirac's $\delta$-distribution with a velocity
  $\mathbf{v}_\mathrm{lag}=(v_\mathrm{lag}, 0, 0)$ for an appropriate
  choice of the orientation of the (Cartesian) frame. The direction of the lag
  velocity is in this case fixed, and only the ejection velocity is
  randomly oriented. 

  The convolution with a 
  $\delta$-distribution would also shift the peak of \Figref{fig:v_dist}
  to higher velocities, but would smear it out much less. Adopting a
  lag velocity of $v_\mathrm{lag} \simeq 5\,\kms$ \citep{reid:14}
  would barely change the fraction of \emph{apparent} O-type runaways
  we predict. To infer an O-type runaway fraction of $\sim$10\%, the
  true lag velocity of high-mass star forming regions would need to be
  larger than $\gtrsim$$15\,\kms$,
  and would need not to have been corrected when making inferences from observations. 
  
  We note that for O-type runaways with lifetimes of
  $\lesssim15$\,Myr, it is likely possible to locate the parent
  association and correct for its possible peculiar motion. The possible lag velocity of
  high-mass star forming regions only affects stellar velocities
  measured using a frame co-rotating with the Galactic disk.
}

\subsection{On the runaway fraction}
\label{sec:howtofind}
Our results predict a runaway fraction among O type stars of at best a
few percent, in agreement with \cite{eldridge:11}, but in tension with
the observational result that 
$\sim$$10-20\%$ of O type stars are runaways \citep[][Sana et al., in
preparation]{blaauw:61, gies:87, stone:91, tetzlaff:11, maiz-appellaniz:18}, and the claim
that roughly two-thirds of runaways come from the
disruption of binaries \citep[][]{hoogerwerf:01}. The latter has also been
challenged by the more recent observations from \cite{jilinski:10}.
\newtext{As a consistency check we can estimate the runaway fraction with}:
\begin{equation}
  \label{eq:ratios}
  f^\mathrm{RW} \simeq
  f_\mathrm{bin}\times(1-f_\mathrm{mergers})\times\mathcal{D}\times
  (1+f_\mathrm{acc})^{\alpha}\times f_{v>30} \ \ ,  
\end{equation}
where we can assume a binary fraction $f_\mathrm{bin}=1$, reasonable
for the population of O type stars considering initial periods as
large as $10^{5.5}$\,days, the fraction of binaries that do not merge is
$1-f_\mathrm{mergers}\simeq0.8$ \citep[][]{sana:12}, the disruption
fraction is $\mathcal{D}\simeq0.8$ \citep[this study, but also,
e.g.,][]{eldridge:11}, the fraction of material accreted, \newtext{for
which we assume} $f_\mathrm{acc}\simeq 0.2$ \citep[][]{packet:81}, and the
fraction of ejected companions fast enough to be a runaway which we
can estimate\footnote{We neglect here that $\mathcal{D}$ is the
  disruption fraction including also binaries with no MS stars at the
  pre-CC stage} from $\mathcal{D}$ and $\mathcal{R}_{15}$, and is
$f_{v>30}\simeq \mathcal{D}/(1+\mathcal{R}_{15})\simeq0.04$ for our fiducial run. Assuming an IMF slope $\alpha=-2.3$ we obtain a
runaway fraction of $f^\mathrm{RW}\simeq 0.01$, in reasonable
agreement with our results. 

The tension between our results and the observed runaway fraction suggests that either (i)
binaries tend to evolve towards shorter pre-CC orbits, corresponding
to higher ejection velocities, but are still disrupted easily by the
CC event, (ii) the contribution of dynamical ejections to the runaway
population is presently underestimated, (iii) other ejection
mechanisms exists and are presently overlooked, (iv) \newtext{the
fast runaways might be easier to detect since they likely move away
from their gas-rich birth environment} (\newtext{leading to typically lower extinction}, \citealt{maiz-appellaniz:18}), \newtext{(v) the
velocities of observed O type runaways are overestimated
(cf.~\Secref{sec:vel_contrib})}, or any
combination of the previous.

Since
$v_\mathrm{dis}\simeq v_2^\mathrm{pre-CC}\propto
v_\mathrm{orb}^\mathrm{pre-CC}\propto J_\mathrm{orb}$, with
$J_\mathrm{orb}$ orbital angular momentum, to shift the
peak of the velocity distribution in \Figref{fig:v_dist} from
$\sim$$6\,\mathrm{km\ s^{-1}}$ to
$v_\mathrm{dis}\gtrsim30\,\mathrm{km\ s^{-1}}$, the binaries would
need to lose about five times more orbital angular momentum during the evolution
before the collapse of the primary. This discrepancy
deserves to be revisited once the runaway fraction can be evaluated from the
homogeneous \emph{Gaia} datasets. If the contribution from
dynamical ejections can robustly be quantified, the 
population of runaways might provide new constraints on the mass
and angular momentum losses during mass transfer in
binaries.

A mechanism to remove angular momentum and mass from the
binary presently not included in our population are eruptive mass loss
events for which there is growing evidence from early observations of
SNe \citep[e.g.][]{khazov:16}. Such mass loss events might be much
more common than previously thought. One possible physical cause for
these event are gravity waves
excited by late shell burning depositing energy at the base of the
envelope \citep[][]{quataert:12, fuller:17, fuller:18}. It is possible
that the mechanism driving these mass loss events might also result in
pre-CC binary interactions which might significantly change the pre-CC
orbit. SN-impostor events happening earlier in the evolution might
also lead to similar effects on the orbital evolution, which we
presently do not model.

Another \newtext{speculative} possibility to increase the angular momentum losses could be
magnetic braking. If non conservative mass transfer can simultaneously
generate a magnetic field and eject some mass from the system, this
might result in torques on the binary much larger than we consider here.

\subsection{How to identify walkaway stars}
\label{sec:convolution}

The velocity dispersion of OB stars in our Galaxy is
$\lesssim$10\,$\mathrm{km\ s^{-1}}$ (e.g.,
\citealt{blaauw:56}; \citealt{gies:87}; \citetalias{tauris:98};
\citealt{hoogerwerf:01}), and it is typically lower for OB
associations \citep[][]{debruijne:99,steenbrugge:03, kiminki:18}. We emphasize also that the velocity dispersion
increases going from samples of OB stars in clusters, in associations,
and in the field \citep[][]{gies:87}. The average velocity of ejected
secondaries is \newtext{typically} higher than this value
(cf.~\Tabref{tab:parameters}), and if considering only the
post-binary interaction secondaries (i.e.,~removing all the ejected
secondaries from very wide and non-interacting binaries, cf.~dashed
distributions in \Figref{fig:v_dist}) it increases \newtext{further},
especially for $M_\mathrm{dis}\geq7.5\,M_\odot$. Therefore, the
companions ejected after a mass transfer episode might form a distinct
population in the \emph{Gaia} data. Moreover, regardless of their previous history, the walkaway
and runaway stars from binary disruptions do not need to follow the
local Galactic rotation curve, and their motion can \newtext{bring them to
locations in which they are more easily observed}
\citep[e.g.,][]{boubert:18,maiz-appellaniz:18}. 

\citet{tetzlaff:11}, following \cite{stone:91}, proposed to fit the
velocity distribution of young ($\lesssim50\,\mathrm{Myr}$) stars
within 3\,Kpc of the Solar neighborhood using
two Maxwellians: one for the ``low-velocity'' ($v\lesssim30\,\mathrm{km\
  s^{-1}}$) and one for the ``high-velocity'' group. Our findings
suggest that $\sim$$10\%$ of the O-stars in the low velocity group might still be the
result of binary disruptions.

Because of the binary interactions taking
place before the first CC, these stars carry observational signatures which might
make them recognizable even if they do not stand out from a kinematic
point of view. For instance, if there is a mass transfer phase during
the previous binary evolution, the ejected star will be spun up
\citep[e.g.,][]{packet:81, pols:91,boubert:18} and
possibly chemically polluted with He-rich and/or N-rich material \citep[e.g.,][]{blaauw:93}. Mass transfer
might also be responsible for the presence of \newtext{strong} magnetic fields,
\citep[e.g.][]{schneider:16}. These features should in principle leave
observable signatures in the spectra of these stars \citep[e.g.,][]{maiz-appellaniz:18}. However, the
effects of the metallicity gradient in the radial direction of the
Galaxy complicates the spectral identification of binary products, as pointed out by \cite{mcevoy:17}.

Characterizing the population of stars ejected from isolated binary
disruptions is necessary to reduce the number of false positive detections of hypervelocity stars in current and
upcoming astrometric catalogues \citep[e.g.,][]{marchetti:17b}.

Stars ejected by a successful SN explosion recently (i.e., less time
ago than the lifetime of the SN remnant) can be connected to the SN
location, and the associated NS can sometimes also be found. Assuming
a typical visible lifetime of the SN remnant of
$\tau_\mathrm{SNR} \simeq10^4$\,years, the \newtext{typical} distance traveled by
an ejected companion is of order $D\equiv
v_\mathrm{dis}\times\tau_\mathrm{SNR}\simeq
0.1\,\mathrm{pc}$. Therefore runaways, but especially walkaways
are expected to
still reside within the SN remnant while it is still observable. This
strategy has been successfully applied to search for runaways by
\cite{vandenbergh:80,guseinov:05,tetzlaff:13,tetzlaff:14,dincel:15,boubert:17a}. 
\cite{kerzendorf:17} also used the predicted kinematics of disrupted
binaries to search for companions of the star that exploded producing
the supernova remnant Cas A. \cite{tetzlaff:11} also suggested that,
with precise astrometry, it might be possible to relate the ejected
companion to the remnant of the companion up to $\sim$5\,Myr after the
SN explosion, provided that the compact remnant of the former companion is visible (e.g., as a pulsar).

The results of \cite{banerjee:12} and \cite{perets:12} suggest there
might be a spatial distinction between dynamically ejected stars and
post-binary runaways and walkaways. Dynamical interactions in a
cluster are most efficient early in the cluster evolution \citep[e.g.,][]{oh:16},
well before the first stellar CC event happens. Moreover, the median velocity
of cluster ejection is higher than the typical walkaway
velocity: \cite{banerjee:12} derived 
a median velocity of dynamically ejected stars of
$50\,\mathrm{km\ s^{-1}}$ after 3\,Myr of evolution. This value
 is significantly higher than what we find from binary disruptions. 
The combination of these two effects would suggest that
walkaways and runaways from binary disruptions would generally be closer to their parent cluster than
dynamically ejected stars. However, this simplistic prediction is
complicated by the fact that binaries can eject stars from the parent
cluster outskirts, while most dynamical interactions produce runaways
from the center of the cluster. Another complication is the possible
ejection of stars as tidal tails in a cluster merger process, \citep[e.g.,][]{lucas:17}.

\cite{evans:15} and \cite{neugent:18} have reported the first
discovery of extraGalactic post-MS massive runaways. They reported the
observation of a red super giant in M31 and a yellow super giant in the SMC,
respectively. They inferred peculiar space velocities larger than
$\sim$$400\,\mathrm{km\ s^{-1}}$ and $\sim$$150\,\mathrm{km\ s^{-1}}$ for
these stars, and proposed a binary origin at least for the yellow
supergiant in the SMC. Our fiducial population can hardly reach such
high peculiar velocities, unless the velocity of the system as a whole binary itself was already 
high. Even if these stars were ejected from a \newtext{massive} binary during their main
sequence, velocities in excess of a few hundred $\mathrm{km\ s^{-1}}$
are only marginally reached in our simulations
(cf.~\Figref{fig:v_dist},~\Figref{fig:v2_postMS}, and
\Figref{fig:v_dist_log}). \newtext{If our present understanding or
  binary physics is correct, this discrepancy} might suggest a different ejection
mechanism (or combination of mechanisms) for these two stars. \cite{evans:15} also report that
vast majority of O/B stars and red super giant stars in the Milky-Way have radial
velocities lower than $\sim70\,\mathrm{km\ s^{-1}}$, in very good
agreement with our results.

\subsection{X-ray binaries and gravitational wave sources}

Binaries remaining bound after the CC of the primary are a minority.
Selecting from our results only bound systems consisting of a BH and a
MS star, we obtain the systemic velocity distributions for BH 
binaries with different BH kick scenarios (see right panel of
\Figref{fig:bound_v_dist}). \emph{Gaia} will give astrometric constraints on the 19 known BH X-ray binary in the Galaxy: if the
effect of the Galactic potential can be singled out from these data,
the comparison with our distribution might shed light onto the typical
amplitude of BH kicks \citep[e.g.,][]{fragos:09,repetto:12,Fragos+2013a,mandel:16}.

The binaries surviving the first CC can
become X-ray sources if their separation is
short enough for the compact object to accrete during the subsequent evolution. For all our natal kick
assumptions, the post-CC separation peaks at $\sim300\,R_\odot$,
and most bound system have separations smaller than
$\sim500\,R_\odot$, indicating that the majority of the systems
remaining bound will become X-ray sources, and potentially even go
through a common envelope evolution phase because of the large mass
ratio between the compact object and the secondary star (which has
accreted mass, cf.~\Secref{sec:example}). 

However, observed
X-ray sources \newtext{more likely} have relatively short orbital periods, or, in
other words, the energy released at the CC of the first star did not
widen significantly the binary and instead contributed mostly to the
kinetic energy of the system. This could mean that the observed sample
of Galactic BH X-ray binaries is biased towards the high velocity tail
of our distributions.

System that might become X-ray sources have systemic velocities of a few tens of
$\mathrm{km\ s^{-1}}$, and since the secondary star is also an
accretor and rejuvenated, they are able to travel a
distance comparable to $\langle L_\mathrm{walk}\rangle$. This can also \newtext{change} their contribution as feedback engines
\citep[e.g.,][]{justham:12,Fragos+2013a}.

If a bound system is not disrupted by the first CC and does not merge
during the evolution of the secondary star, it is less likely to be
disrupted by the second CC
\citep[e.g.][]{stevenson:17,oshaughnessy:17},  \newtext{and therefore
  more likely to become a gravitational wave source (if the separation is small enough)}. This is because the impact of the kick on the orbit
scales as $v_k/v_\mathrm{orb}$, and the orbit of bound systems is likely to shrink during the evolution after the first CC: the
donor star needs to become less massive than the compact object for
the system to widen as described in
\Secref{sec:example} and \Secref{sec:analytics}. \newtext{We also
  emphasize that typically RLOF has a larger impact than stellar winds on the
  orbital separation.}

\subsection{Runaways and walkaways as feedback engines}
\label{sec:disrupted}

The large majority of binary systems is disrupted at the first CC
(with a few notable exceptions, see also \Secref{sec:param_variation}):
systems that can become X-ray sources and/or gravitational wave
sources are the exception rather than the rule. The constraints on binary evolution coming from the much more
common evolutionary path should be considered together with those
coming from the more rare channels.

\newtext{Because of their motion,} both walkaways
and runaways can reach significant distances from their birth
location, see \Tabref{tab:parameters} and \Figref{fig:Dhist}. The typical distances traveled
by the ejected secondaries (cf.~\Figref{fig:Dhist}) are sufficient to
get massive stars out of their parent cluster or association, but only
the fastest runaways would get beyond the thick disk in our Galaxy,
especially if considering the \newtext{distribution of} relative inclinations between the
trajectories of the stars and the disk itself. As noted by
\cite{boubert:18}, this introduces a bias in observed samples that
favors the detection of the fastest runaways.

The motion of the ejected stars
effectively spreads the massive stars and their (possible) SN explosions in a larger
volume, enhancing their impact on the composition of galaxies (more efficient mixing of
nuclear yields), on the ionization of their surroundings (enhanced
escape fraction, e.g.~\citealt{conroy:12, kimm:14, ma:16}), but it might possibly
decrease the efficiency of their mechanical feedback because of the
lower density of the surrounding gas.
The large number of walkaways
formed per each runaway $\mathcal{R}$ suggests that walkaway stars might
\newtext{change} massive stars feedback more significantly than the faster runaways, although they cannot
reach as far out as the latter because of their lower velocity.

As a consequence of the typical evolutionary scenario
described in \Secref{sec:example}, the first CC event in a \newtext{typical}
binary system will result either in a stripped SN or a failed SN
with little or no electromagnetic signature. The second CC will
instead be from the accretor star, which could again result in a
SNIb/Ic (if the star becomes a Wolf-Rayet because of its own stellar wind),
but, because of the IMF and the possibly non-fully conservative
accretion rate, the secondary is more likely to remain H-rich until
its CC. As already pointed out by
\cite{eldridge:11}, this suggests that \newtext{stripped SNe} should preferentially
happen in star forming regions, while type II SN should show a larger
spatial spread (because of the motion of the ejected star).

In this study, we did not consider the effects of the background
cluster and/or Galactic potential on the motion of the ejected
secondaries. These can in principle be used as test particles to probe
the potential itself, and consequently allow for testing the local
and/or Galactic dark matter distribution \citep[e.g.,][]{rossi:14,marchetti:17}.

While runaways have peculiar kinematic 
characteristics that make them recognizable, walkaway stars might more
easily be mistaken for genuine single stars. Especially \newtext{for} masses larger
than $7.5\,M_\odot$ and velocities larger than $v_\mathrm{dis}\gtrsim20\,\mathrm{km\
  s^{-1}}$, the walkaways and runaways accrete mass before being \newtext{ejected}. Therefore, the disruption
of binaries can
pollute observed samples of
present-day single stars with binary evolution by-products populating the field of the Galaxy
\citep[][]{gvaramadze:12,demink:14}.

As an example, the runaway star $\zeta$ Puppis has been considered to be
a ``canonical'' O-type star \citep[e.g.][]{ramiaramanantsoa:17}, and
it was used to calibrate the free parameters in the \cite{castor:75}
(CAK) theory of stellar winds \citep[][]{pauldrach:94} widely applied
to model single stars. However, this object is
likely to be a accretor ejected from a binary, \citep[][]{vanrensbergen:96,ramiaramanantsoa:17}.

Massive walkaway \newtext{stars} ejected from a binary might also
appear in isolation. \newtext{The origin of massive stars observed in
  isolation is an open problem:} did they form in isolation? or could they
possibly have reached that location, if ejected from a binary? \cite{bestenlehner:11} identify an isolated
$\sim150\,M_\odot$ star (VFTS682) in the 30 Doradus field. Such
massive object is unlikely to be the result of a binary disruption,
but might have been ejected from the massive young cluster
R136. \cite{bestenlehner:11} suggest that if this star has not been
ejected from this cluster, it might be a direct indication of isolated
star formation. Walkaway massive stars can pose similar problems if
not recognized as rejuvenated binary products.

This might also connect to the evolutionary path of luminous blue
variable (LBV) stars. These are massive stars which experience
extreme and yet unexplained outbursts of mass loss. In the classical
picture of massive stellar evolution (so-called ``Conti scenario''),
LBVs are an intermediate evolutionary stage between O-type and Wolf-Rayet
stars, during which the bulk of the H-rich envelope is lost \citep[e.g.][]{conti:75,
  maeder:94,maeder:96}. However, \cite{smith:15}
pointed out that LBV and LBV-candidates are farther from the nearest
O-type star than Wolf-Rayet stars, and suggested that LBVs might be rejuvenated products of binary
evolution\footnote{This idea has also been questioned by
  \cite{humphreys:16,davidson:16}, see also \cite{smith:16}.} (accretors or merger
products). The followup study of \cite{aghakhanlootakanloo:17}
suggested that partial rejuvenation during mass transfer coupled to a walkaway (or
runaway) velocity might be sufficient to explain the isolation
observed by \cite{smith:15}. 

\section{Summary and conclusion}
\label{sec:conclusion}

We have carried out a suite of numerical simulations to provide
predictions for the kinematics and properties of massive stars ejected
from binary systems disrupted by the supernova explosion of the
companion star. Our aim with this study is three-fold:  (i) identify
which theoretical predictions are robust against model uncertainties
and discuss their astrophysical implications (ii)  investigate
theoretically which uncertain physical processes affect the kinematics
of the population of unbound stars in such a way that their imprints
would be observable and  (iii) provide a framework of models that can
be used for comparison against observations or as input for
other simulations. 

We summarize our main findings below. When we quote percentages or
fractions, the error bars given indicate the maximum variations that
we encountered in our model variations for different physical
assumptions (excluding the unphysical variation \newtext{with zero}
natal kicks for \newtext{all} newly born compact objects).

\begin{itemize}

\item[$\bullet$] Nearly a quarter,   $22_{-8}^{+26}$\%, of all binary
  systems with at least one star more massive initially than
  $7.5\Msun$ merges prior to the explosion of the first star (see \Figref{fig:tree} and \Tabref{tab:parameters}).

\item[$\bullet$] The large majority of binary systems is disrupted at the moment of core collapse of the first star, $\mathcal{D}=86_{-9}^{+11}\%$ in our simulations, consistent with earlier studies. This prediction is robust against variations in the treatment of the evolutionary and interaction processes, but it is sensitive to the choices concerning the natal kicks of neutron stars and black holes.  

\item[$\bullet$] The disruption of the binary system produces an
  unbound stellar companion. Remarkably, we find that these events
  only rarely produce runaway stars (i.e. stars with peculiar motions
  larger than $30\,\kms$).  The velocity distribution peaks at
  $\sim6\,\mathrm{km\  s^{-1}}$, with a $90^\mathrm{th}$ percentile
  around $20\kms$ well below the typical threshold for runaway stars.  

\item[$\bullet$] The slow-moving unbound companions, to which we refer
  as walkaway stars, outnumber the runaway stars by at least an order
  of magnitude in nearly all our models. This is a robust outcome of
  the models and is primarily due to the reversal of the mass ratio
  prior to the explosion as we show numerically and analytically.
  This finding has been noted in earlier studies \citep[e.g., ][]{dedonder:97,eldridge:11},  but we believe this result (and its potential implications) is under appreciated.  

\item[$\bullet$] We also produce runaway stars but we rarely find them
  to be faster than $60\,\kms$, which corresponds to
  $98.9^\mathrm{th}$ \newtext{percentile} of \newtext{our fiducial} population. This is much lower than the
  theoretical maximum set by the finite size of the stars and their
  orbit, which is in \newtext{the} range 400--1000$\,\kms$ depending on their masses
  and evolutionary state. We expect that stars with larger peculiar
  velocities result from dynamical ejections from clusters, capture
  from another Galaxy, or most exotically, interaction with a
  intermediate mass or supermassive black hole.

\newtext{\item[$\bullet$] Isolated binary evolution produces a runaway fraction among stars more
  massive than $15\,M_\odot$ between $0.2-2.6\%$. None of the combination of parameters that
  we consider produces a runaway fraction approaching the observed
  value of $\sim$$10\%$ for O-type stars. This result is consistent with previously
  published theoretical studies (see also
  Appendix~\ref{sec:literature}), and might indicate that (i) the observed fraction of
  O-type stars that are runaways is overestimated,
  possibly because of biases favoring the observation of fast and isolated
  massive stars or because of the difficulties in defining the frame
  of reference in which to measure the velocity, or
  (ii) contrary to previous claims, it is not correct that the majority of observed O-type
  runaways come from the disruption of binaries
, or (iii) massive binary evolution models
are lacking some physical process allowing to produce short
pre-collapse orbits which would result in faster ejection velocities
if disrupted.}

\item[$\bullet$] We find a mild trend for runaway stars to be faster
  and more common at lower metallicity, caused by the fact that stars
  are on average smaller and the reduced effects of stellar winds.
  \emph{Gaia} should in principle be able to test these predictions by
  identifying the fastest runaway stars in the Magellanic clouds. In
  our most metal poor simulations for $Z=0.0002$, which may have
  relevance for the earliest stellar populations, we find 5 times more
  \newtext{(massive)} runaway stars, if we keep all other assumptions for the initial distributions the same.
  \newtext{The sensitivity of our results to the assumed metallicity also suggests that binary evolution might
    create systematic trends in the stellar feedback with metallicity and galaxy size.}
  
\item[$\bullet$] Both runaways and walkaways typically accrete mass from their
companion prior to the disruption, This is especially true for those more massive than
$7.5\,M_\odot$ at the time they are ejected. They typically also gain
angular momentum \newtext{and nuclearly processed material} in the process. These features might make walkaway stars recognizable as binary products even if they do not stand out from a kinematic point of view. Their further evolution and structure may also differ from that of non-rotating single stars, as has been speculated in the context of the LBV phenomenon.  

\item[$\bullet$] We suggest that the high mass tail of the runaway
  mass function could provide insight on the formation mechanism of
  black holes from the core-collapse of the primary star, and in particular on whether these receive a ``velocity'' or a ``momentum''
kick. So far, studies have focussed on the surviving X-ray binaries,
but we propose to use their unbound counter parts since they
\newtext{are more common} and should be identified by \emph{Gaia}. 

\item[$\bullet$] By ejecting secondaries, the disruption of binaries can enhance the role of
massive stars as feedback engines. The average distance they travel,
calculated neglecting any external potential, is of order
$\gtrsim100$\,pc, and up to $\gtrsim 500$\,pc if considering only the
faster runaways. These suggests that both runaways and walkaways
can exit the dust cloud embedding their birth location, thus \newtext{changing}
the escape fraction for their ionizing radiation. If they are massive
enough, the ejected star will explode far from other stars and gas
overdensities, with potential implications for the driving of
turbulence and star formation. 

\end{itemize}

Our simulations will be made available upon final publication of this study.  Early requests for access can be addressed to the lead author. 

\begin{acknowledgements}
  We acknowledge for the many helpful discussions
  F.~Broekgaarden, A~.de~Koter, D.~Hendriks, L.~Kaper, N.~Langer, I.~Mandel, C.~J.~Neijssel, Ph.~Podsiadlowski,
  K.~Sakar, F.~N.~Schneider, S.~N.~Shore, E.~van~den~Heuvel, A.~Vigna-Gomez.
  MR is grateful to the KITP for the visiting graduate fellowship during which a large portion of this
  study has been carried out. This research was supported in part by
  the National Science Foundation under Grant No. NSF PHY11-25915. SdM has received funding under the European Unions Horizon 2020 research and innovation programme from the European Research
Council (ERC) (Grant agreement No. 715063).
  RGI thanks the Science and Technology Facilities Council (STFC) for funding his Rutherford fellowship under grant ST/L003910/1.

\end{acknowledgements}

\bibliographystyle{aa}

\appendix

\section{Comparison to selected previous population studies}
\label{sec:literature}

Various earlier studies have simulated the production of binaries
containing a compact object and/or ejecting runaway stars. We briefly discuss a selection that focuses on estimates for unbound companions and comment briefly on how our findings compare. 

\cite{dedonder:97} presented an extensive binary evolution study discussing predictions for the O-type runaways as well as the systems that remain bound. They consider a variety of assumptions for the initial distributions and the uncertain physical parameters concerning the efficiency of mass transfer, angular momentum loss and the common envelope ejection. Generally, we find that our results agree well.  They find that 16--23\% of systems remain bound after the SN explosion of the primary, which is consistent with our findings $1- \mathcal{D} = 14^{+22}_{-10}\%$.  They find that between 6\% and 27\% of the O-type are unbound former companions that are now single, which is also agreement with our walkaway fraction (one but last column in their table 1). 

The find that  2--7\% of the O-type stars have velocities large than $30\,\kms$ (final column in their Table 1), which is slightly larger but still consistent with what we find.  We expect that this difference may be in part the result of a difference the treatment of mass transfer and angular momentum loss, although the combined effect of further differences in our assumptions will also contribute. Their default assumption is that a fixed fraction $\beta_\mathrm{RLOF} = 0.5$ of the material transferred during Roche-lobe overflow leaves the system through the outer L2 Lagrangian point forming a ring around the binary system, which leads to larger angular momentum loss, shrinking the orbit further.  This is somewhat similar to our model variation where we assumed $\gamma_{\rm RLOF} = \gamma_{\rm disk} $ although our results cannot be compared one to one, since we adopt a different, physically motivate, assumption for the mass transfer efficiency. In our simulations, we find that this assumption is not increasing the number of runaway stars, \newtext{because} the fraction of system that merges increases too. 

\cite{eldridge:11} perform population synthesis simulations with a
detailed stellar evolutionary evolutionary code to investigate O and
early B
type runaway stars. Their primary aim is the predict the spatial distribution of different types of CCSN and gamma-ray bursts. They find a disruption fraction, $\mathcal{D}=80\%$, in good agreement with our results, $\mathcal{D} = 86^{+10}_{-22}\%$.  In their study, they use the term runaway to refer to all unbound companions with velocities larger than $5\kms$, which encloses the large majority of what we refer to as walkaway stars, but they also quote estimates for stars faster than $30\kms$ which corresponds to our definition of runaway stars.  They predict a walkaway (runaway) fraction of $2.2\%$ ($0.5\%$) for O type stars
in their simulations for $Z=0.02$. For comparison, for stars more
massive than $15\,\Msun$, which roughly corresponds to O type stars,
we estimate a walkaway (runaway) fraction of  $10^{+4.7}_{-8.5}$\%
($0.5^{+1.0}_{-0.3}$\%).  Our runaway fraction agrees very well.  We
find a somewhat larger walkaway fraction, but we consider this a
fairly good agreement, given the uncertainties and differences in
definitions that we have adopted. Also the distance they estimate that
ejected stars can reach is in good agreement with our predictions and
they also note that the majority unbound companions accrete mass from
their companions before the first core-collapse.  Our predictions for
the velocities of bound systems are slightly lower compared to
\cite{eldridge:11}: the vast majority of \newtext{our} bound
post-core-collapse systems is slower than $30\,\mathrm{km\ s^{-1}}$,
and \newtext{there is} a high-velocity tail barely extending beyond about $100\,\mathrm{km\  s^{-1}}$.


Very recently, \cite{boubert:18} published a study  investigating  the hypothesis that Be stars are products of mass transfer in binary systems \citep[e.g.][]{pols:91, demink:13}.  They 
compare the  kinematics of a flux-limited sample of Galactic Be stars
with binary population synthesis simulations. These simulations were
obtained with a different version of the  binary evolutionary code that we use. Generally, our findings are in agreement, despite the minor differences in the model assumptions. They also find a large fraction of unbound companions, many of which are rapidly rotating and moving at velocities slower than $30\,\kms$ (e.g., their Fig. 6 and 7b), in good agreement with our \Figref{fig:vrot_vrw}.   They also find that the natal kick distribution does not greatly affect the resulting velocity distribution and/or the runaway fraction and that the maximum distance traveled by ejected Be stars is likely to be smaller than the vertical scale height of the thin disk, in agreement with our findings. 

We find some disagreement in the provided explanations of the
theoretical results.  For example, the authors state in their
section~3.2 that that mass transfer shrinks the orbit and accelerates
the secondary star. While this the case initially, upon the onset of
mass transfer, we find that the orbit generally widens after the
reversal of the mass ratio. The widening and the inversion of the mass
ratio both slow down the orbital velocity as we verified both
analytically \Secref{sec:analytics} and by detailed inspection of
representative example systems, e.g. \Secref{sec:example}.  In
their Section~3.3, they state that  ``whether a binary is disrupted by a
supernova is principally determined by whether the primary loses more
than half its mass \citep{blaauw:61}, and the kick on the compact
object is only a second order effect''.  We find instead that mass
loss during the explosion is rarely responsible for unbinding of a
binary system that can produce a runaway star. The amount of mass lost
needs to exceed half of the total mass of the system (and not half of
the primary star), which is rarely achieved in our simulations. This
is because the  CC progenitor loses most of its mass during the
preceding mass transfer phase. We find that the Blaauw kick due to
rapid mass loss is only important of initially very wide binaries, in which the two stars have not exchanged mass prior to CC.  This difference may be in part due to the differences in the assumptions for the range of initial orbital periods. The authors consider systems with initial orbital periods up to  $10^{10}$ days in their simulations, which means that the majority of their progenitors should effectively evolve as single stars. We consider instead systems up to $10^{5.5}$ days, which is more appropriate for the more massive progenitors that we are interested in.  

Many further studies investigated the populations of binaries in the
context of the formation of X-ray binaries and binary neutron stars
and black holes. Providing a complete overview and detailed
comparison is beyond our present scope. We discuss below a comparison
to a limited set of studies.

\cite{brandt:95} focus on binary systems remaining bound at the first CC, with the aim of understanding the
effects of natal kicks on X-ray binaries.  They estimate a disruption
fraction in their calculations of $\mathcal{D}\simeq73-81\%$
(depending on the companion mass), in good agreement with our
results.

\cite{kalogera:96} presented a similar analytic study of the effects of natal kicks on the systemic velocities of X-ray binaries. Our population synthesis results agree in predicting systemic velocities of bound post-CC systems generally lower than the pre-explosion orbital velocity,
except with $\sigma_\mathrm{kick} \gg v_\mathrm{orb}^\mathrm{pre-CC}$ (corresponding to large $\xi$ in the notation used by \citealt{kalogera:96}).

 \cite{fryer:98} investigated the impact of NS natal kicks on the
 formation of NS X-ray binaries and NS-NS binaries, but also present
 results for the unbound companions. Assuming a bimodal kick
 distribution, they find that most O/B-type ejected companion move
 slower than about $50\kms$, consistent with our findings.

 \cite{dray:05} focus on high-mass runaways, which become Wolf-Rayet stars during or before their post-disruption evolution. They argue in favor of significant BH kick amplitudes to explain the rarity of BH-WR binaries and the observed velocity distribution of WR runaways. We reach similar conclusions based on our simulations for different assumptions for the BH kicks. The systemic velocities we find for bound systems are also in reasonable agreement with those found by \cite{dray:05}

Recently, \cite{tauris:17} presented a detailed study of the evolutionary processes leading to the formation of NS NS binary systems. They also find that the majority of binaries hosting a NS after the first CC have systemic velocities smaller than $\sim 30\,\mathrm{km\ s^{-1}}$, in good agreement with the results shown in the left panel of \Figref{fig:bound_v_dist}.

\section{Output files}
\label{app:data}

The outcome of our population synthesis calculations will be made
available upon publication. For early inquiries, please contact the
lead author.
Each file corresponds to one parameter variation (see
\Secref{sec:methods} and \Secref{sec:param_variation}), and logs the
following information for each binary system where a star goes
CC\footnote{We stop our computations at the first CC event.}.
\begin{itemize}
 \item primary ZAMS, pre-CC, and post-CC (corresponding to the NS or
   BH mass if the primary is the star collapsing) masses in $M_\odot$ units:
   \texttt{M1zams}, \texttt{M1preCC}, \texttt{M1postCC};
 \item secondary ZAMS, pre-CC, and post-CC masses in $M_\odot$ units:
   \texttt{M2zams}, \texttt{M2preCC}, \texttt{M2postCC};
 \item fallback fraction $f_b$ for each star (set to zero for the star
   that is not collapsing):
   \texttt{fb1}, \texttt{fb2}
 \item evolutionary stage before and after CC (stellar types listed
   according to \citealt{hurley:00}):
   \texttt{type1preCC}, \texttt{type1postCC},
   \texttt{type2preCC}, \texttt{type2postCC}
 \item post-CC velocities of the 2 stars in the original frame, in
   $\mathrm{km\ s^{-1}}$: \texttt{v1postCC}, \texttt{v2postCC};
 \item pre-CC and post-CC eccentricity (the latter is -1 for mergers
   and disrupted systems): \texttt{e\_preCC}, \texttt{e\_postCC};
 \item pre-CC and post-CC separation in $R_\odot$ units (the latter
   is set to 0 for mergers and disrupted systems):
   \texttt{a\_preCC}, \texttt{a\_postCC};
 \item ZAMS, pre-CC, and post-CC periods in
   days: \texttt{Pzams}, \texttt{PpreCC}, \texttt{PpostCC};
 \item kick amplitude in $\mathrm{km\ s^{-1}}$: \texttt{v\_kick};
 \item kick direction, with $\theta$ angle between the collapsing star
   orbital velocity and the kick (see also \citetalias{tauris:98} for notation):
   \texttt{theta}, \texttt{phi};
 \item systemic velocity in $\mathrm{km\ s^{-1}}$: \texttt{v\_sys};
 \item age of the system at the time of CC in Myr: \texttt{t\_explosion};
 \item time left in the current evolutionary stage in Myr:
   \texttt{t\_remaining};
 \item \newtext{time spent by the system with at least one star more massive
   than 15\,$M_\odot$: \texttt{duration\_*};}
 \item system probability (see below): \texttt{Prob}.
\end{itemize}

The system probability corresponds to the hyper-volume of the initial
parameter space ($M_1^\mathrm{ZAMS}$, $q^\mathrm{ZAMS}$,
$P^\mathrm{ZAMS}$) represented by each binary system in our model
grid. In other words, the probability of each system is the
statistical weight of the system seen as sampling point for the initial
distributions. To construct distributions of the output quantities (e.g.~those presented in
\Figref{fig:v_dist}-\ref{fig:bound_v_dist}), the properties of each
system should be weigh with the corresponding probability. Similarly,
the mean value of a quantity $\langle x \rangle$ (e.g.~$\langle v \rangle$ in
\Tabref{tab:parameters}) should be calculated using:

\begin{equation}
  \label{eq:mean}
  \langle x \rangle = \frac{\int x P(x)\,dx}{\int P(x)\,dx} \equiv
  \frac{\sum_i^{}  x_i P_i}{\sum_i^{} P_i} \ \ ,
\end{equation}
where $P$ is the probability, and the index $i$ runs over all the
binary systems in a population.

To calculate the average distance traveled by stars
ejected by the binary disruption, we use:

\begin{equation}
  \begin{aligned}
    \label{eq:meanD}
\langle L \rangle \equiv \langle v\times \Delta t\rangle = \\
  \frac{\sum_i \texttt{v2postCC}\times [\texttt{t\_remaining}+0.1\times\tau_\mathrm{MS}(M_2)]
    \times P_i}{\sum_i P_i} \ \ ,
  \end{aligned}
\end{equation}


where the second term in squared brackets accounts, albeit in a
simplified way, for the helium core burning duration of the
rejuvenated star.

\section{Observable velocity distribution}

\begin{figure*}[htbp]
  \includegraphics[width=\textwidth]{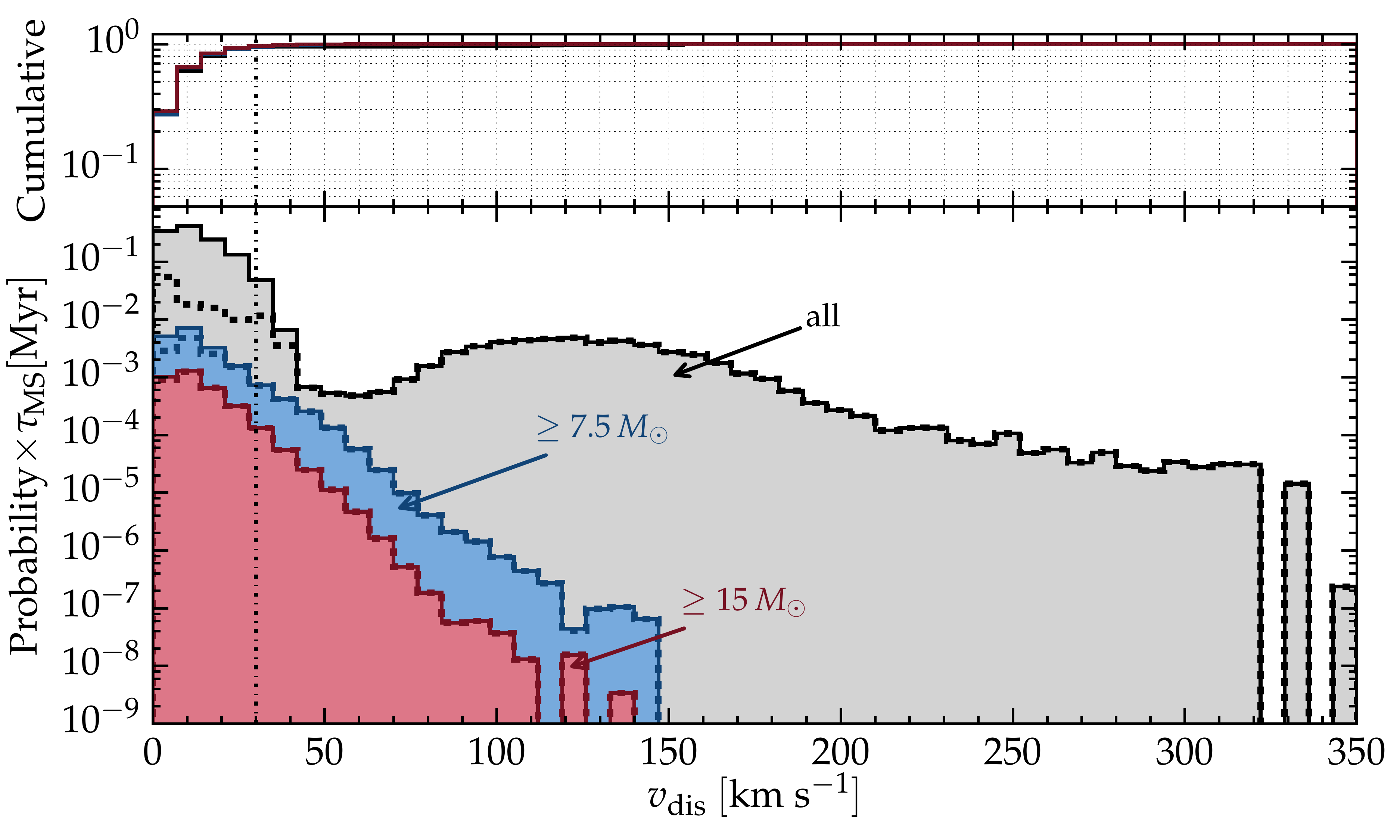}
  \caption{Velocity distribution of ejected stars, including the
    finite MS lifetime to populate the bins (see also
    \Figref{fig:v_dist}). }
  \label{fig:v_dist_obs}
\end{figure*}

Figure~\ref{fig:v_dist_obs} shows the velocity distribution of MS
stars ejected by the disruption of binaries that can
be directly compared to observations provided that (i) the
contribution of dynamical ejection can be separated in the observed
sample and (ii) the effects of the Galactic potential can be
neglected, i.e.\ effectively each ejected star moves in a straight
line at constant velocity for the remaining duration of its MS. This is the same information of
\Figref{fig:v_dist}, but each bin is populated considering also the
remaining MS lifetime of the ejected star
($\tau_\mathrm{MS}$).

\begin{figure*}[bp]
  \includegraphics[width=\textwidth]{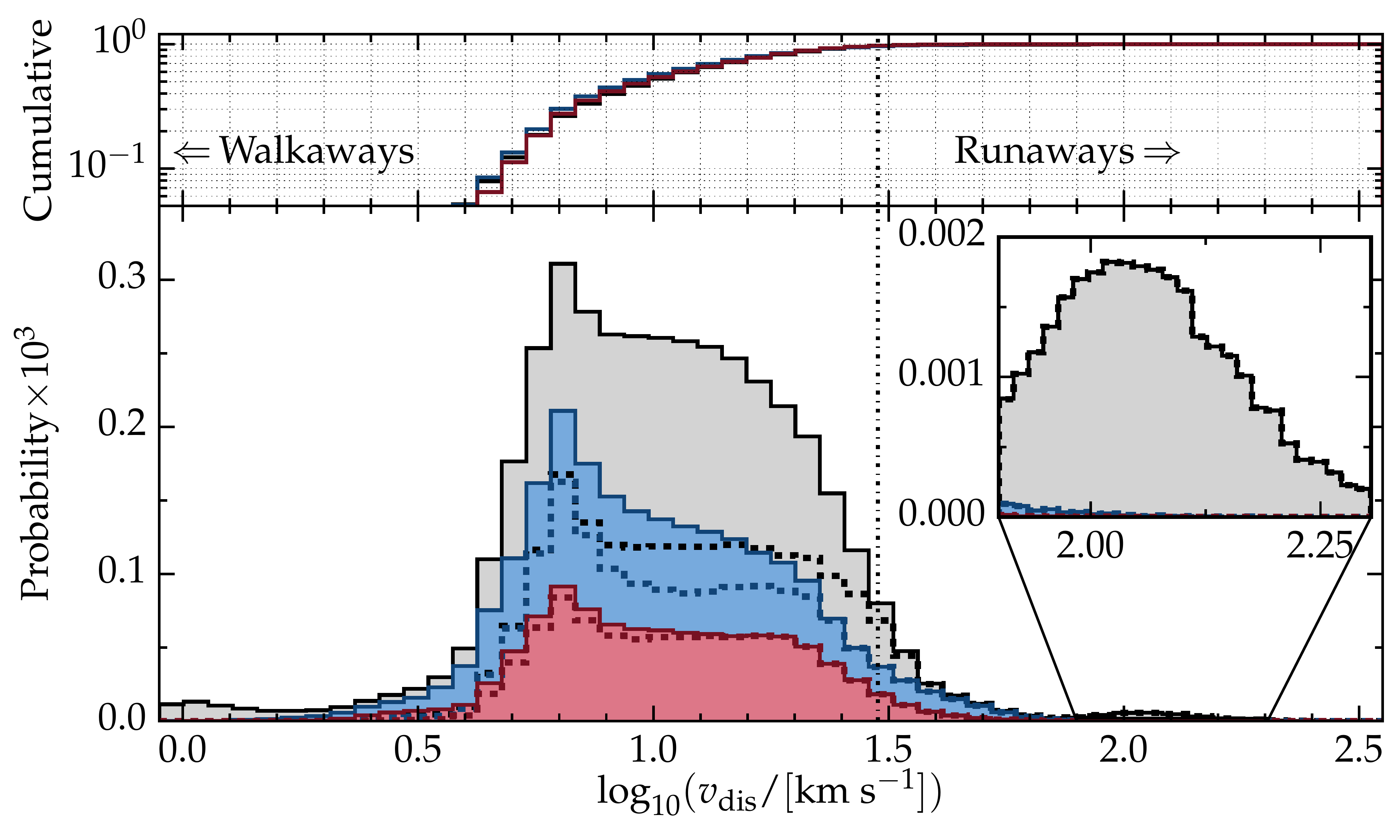}
  \caption{Same as \Figref{fig:v_dist}, but using a logarithmic scale
    for the velocity. The use of a logarithmic scale allows for the
    display of a wider range of velocities. A minor peak in the grey
    histogram can be seen between $100\lesssim
    v_\mathrm{dis}/\mathrm{km \ s^{-1}} \lesssim
    400$, but is absent in the histograms for massive ejected
    stars. Such high ejection velocities are reached through a common
    envelope evolution without accreting mass.}
  \label{fig:v_dist_log}
\end{figure*}

\section{Pre-collapse distributions}

We present in this appendix the pre-CC distribution in separation,
mass of the collapsing star, and mass of the companions for all the
binaries with a MS companion to the collapsing star in our fiducial
simulation. Similar distributions can be derived for all our parameter
variations from the data files that will be made
available.
These distributions can inform studies of the interaction of the
SN shock with the companion star
\citep[e.g.][]{wheeler:75,liu:15,rimoldi:16, hirai:18}.

\begin{figure}[htbp]
  \includegraphics[width=0.5\textwidth]{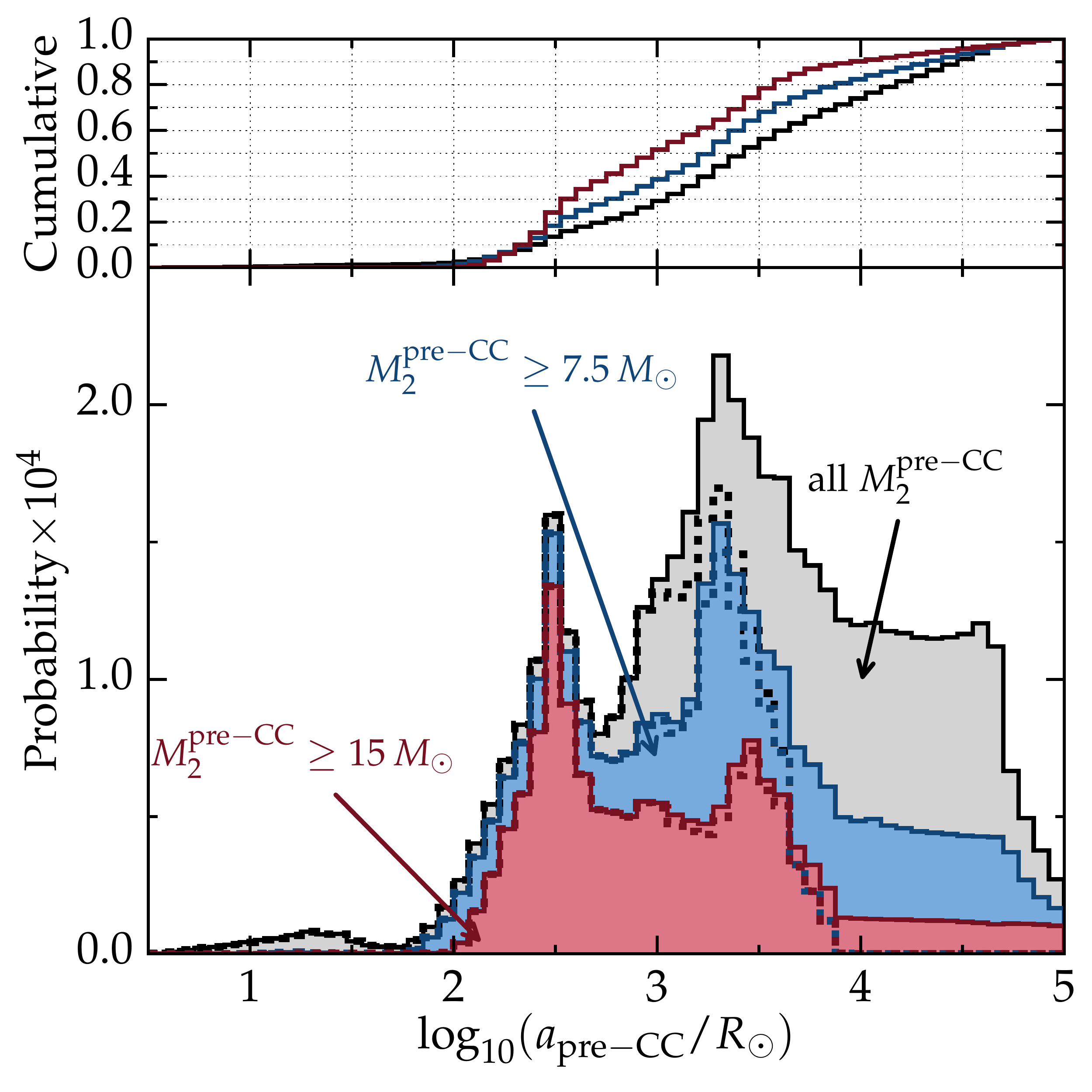}
  \caption{Pre-CC separation distribution for binaries with a
    collapsing star and a main sequence companion. Colors indicate the
    pre-CC mass of the MS companion according to the legend. Dashed
    histograms indicate post-interaction (RLOF or common envelope)
    binaries. The top panel shows the corresponding cumulative
    distributions.}
  \label{fig:preCC_sep_dist}
\end{figure}

We show in \Figref{fig:preCC_sep_dist} the pre-CC separation distribution. The colors indicate indicate the minimum mass of the MS
companion (i.e., not of the collapsing star). Roughly speaking, the
two peaks shown in \Figref{fig:preCC_sep_dist} correspond to the
orbital widening due to conservative (case A and early case B) and
non-conservative (late case B and case C) mass transfer phase. For all
pre-CC separations shorter than $10^3\,R_\odot$, the dashed and solid
histograms coincide, indicating that all these binaries have
experienced a direct interaction previously during the evolution.

\begin{figure}[htbp]
  \includegraphics[width=0.5\textwidth]{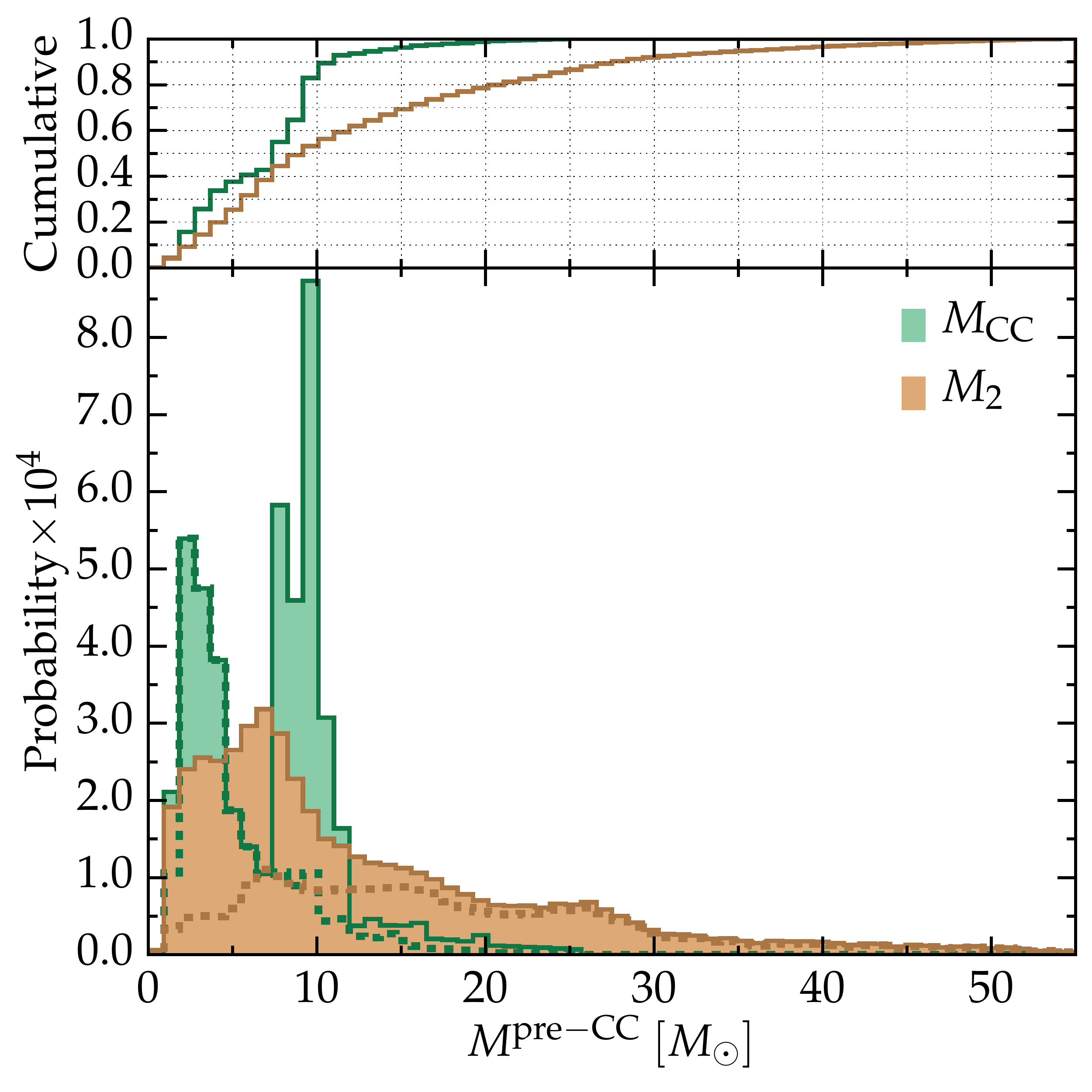}
  \caption{Pre-CC mass distribution for the exploding star
    ($M_\mathrm{CC}$) and the companion \newtext{($M_2$). We plot all systems
    where the companion is a MS star at the time of the explosion,
    regardless of whether the binary is disrupted or not.} Dashed
    histograms indicate post-interaction (RLOF or common envelope)
    binaries. The top panel shows the corresponding cumulative
    distributions.}
  \label{fig:preCC_mass_dist}
\end{figure}

Figure \ref{fig:preCC_mass_dist} shows the mass distribution for the
exploding star and the MS companion ($M_\mathrm{CC}$, and $M_2$,
respectively), at the pre-CC stage. The combination of the
distributions shown in \Figref{fig:preCC_mass_dist} and
\Figref{fig:preCC_sep_dist}, together with the effects of the natal
kick distribution results in the ejection velocities in
\Figref{fig:v_dist} which is our main result.

\end{document}